\documentclass[12pt,letterpaper]{article}
\pdfoutput=1
\usepackage{jheppub}
\usepackage[english]{babel}
\usepackage[latin1]{inputenc}
\usepackage{amsfonts}
\usepackage{graphicx}
\usepackage{amssymb,amsmath,mathabx}
\usepackage{framed}

\allowdisplaybreaks[1]

\title{Four-point amplitudes in \boldmath$\mathcal{N}=2$ SCQCD}

\author[]{Marta Leoni$^{\ast, \hash}$,} 
\author[\dag]{Andrea Mauri}
\author[\hash]{and Alberto Santambrogio}

\affiliation[\ast]{Dipartimento di Fisica dell'Universit\`a degli studi di Milano, via Celoria 16, I-20133 Milano, Italy}
\affiliation[\hash]{INFN, Sezione di Milano, via Celoria 16, I-20133 Milano, Italy}
\affiliation[\dag]{Dipartimento di Fisica, Universit\`a degli studi di Milano--Bicocca, Piazza della Scienza 3, I-20126 Milano, Italy}

\emailAdd{marta.leoni@mi.infn.it}
\emailAdd{andrea.mauri@mi.infn.it}
\emailAdd{alberto.santambrogio@mi.infn.it} 

\abstract{We compute four--point scattering amplitudes in $\mathcal{N}=2$ SCQCD 
with general external matter configurations using $\mathcal{N}=1$ superspace Feynman diagrams, 
at one loop in the general case and up to two loops in the fundamental sector. 
In the pure adjoint sector at one loop we confirm exact agreement with the corresponding 
amplitudes in $\mathcal{N}=4$ SYM theory, supporting the idea that a closed subsector of 
the SCQCD  might be exactly integrable. External matter in the fundamental representation 
breaks dual conformal invariance already at one loop and also the maximum transcendentality 
principle at two loops.}

\preprint{June 2014 \\ \vspace{-1cm} \begin{flushright} IFUM-1029-FT \end{flushright}}


\newcommand {\non}{\nonumber}
\newcommand{\Li}[2]{{\mbox{Li}}_{#1}\left(#2\right)}

\newcommand{\mc}{\mathcal}

\def\tr{\mathrm{Tr}}
\def\d{\mathrm{d}}
\def\Q{\tilde{Q}}

\def\Tr{\mathrm{Tr}}
\def\ln{\text{ln}}
\def\th{\theta}

\def\bseq{\begin{subequation}}  
\def\eseq{\end{subequation}}
\def\bsea{\begin{subeqnarray}}  
\def\esea{\end{subeqnarray}}

\newcommand{\beq}{\begin{equation}}
\newcommand{\bea}{\begin{eqnarray}}
\newcommand{\eea}{\end{eqnarray}}
\newcommand{\eeq}{\end{equation}}

\renewcommand{\a}{\alpha}
\renewcommand{\b}{\beta}

\newcommand{\pa}{\partial}

\newcommand{\G}{\Gamma}

\newcommand{\e}{\epsilon}

\newcommand{\Db}{\overline{D}}

\newcommand{\thb}{\overline{\theta}}

\begin{document}

\maketitle 

\flushbottom

\section{Introduction}

The study of scattering amplitudes in supersymmetric field theories has recently unveiled the existence of hidden symmetries and unexpected properties. Once again $\mc N=4$ SYM theory played a pivotal role and turned out to be the perfect playground to provide important insights into quantum field theory. 

One of the most surprising novelty is that planar MHV scattering amplitudes of $\mc N=4$ SYM theory enjoy an additional dynamical symmetry, which is not present in the Lagrangian formulation and which constrains the form of the amplitudes to be much simpler than a naive analysis might suggest \cite{Drummond:2007aua,Brandhuber:2007yx}.  This hidden symmetry, called dual conformal invariance, can be related to a duality between planar MHV amplitudes and light--like polygonal Wilson loops and was first suggested in the strong coupling string description \cite{Alday:2007hr}.

As a consequence of dual conformal symmetry, the four and five--point MHV gluon amplitudes were shown to be completely fixed \cite{Drummond:2007cf,Drummond:2007au} in a form that matches the exponential BDS ansatz  \cite{Bern:2005iz}.  Starting from six external particles, dual conformal invariance constrains the amplitudes only up to an undetermined function of the conformal cross ratios which violates the BDS exponentiation \cite{Alday:2007he,Drummond:2007bm}. Nevertheless the duality with Wilson loops was shown to be preserved  \cite{Bern:2008ap,Drummond:2008aq}.

One more remarkable property of $\mc N=4$ SYM amplitudes is that they exhibit uniform and maximal transcendentality weight. In the dimensional reduction scheme, assigning transcendentality $-1$ to the dimensional regularization parameter $\e$, one obtains $L$-loop corrections with uniform degree of transcendentality $2L$. This maximal transcendentality property was first observed for the anomalous dimension of twist--2 operators \cite{Kotikov:2001sc,Kotikov:2004er,Kotikov:2006ts} and then was found it is surprisingly enjoyed by all the known observables of the theory.  It is still unclear whether this property has to be ascribed to the special diagrammatics \cite{Kotikov:2012ac} associated to either (dual)conformal symmetry or supersymmetry, or if it is a unique feature of the model. 

The investigation on the origin of such  properties has led to study theories with less amount of supersymmetry.
A quite extensive analysis has been performed in the case of three-dimensional ABJM theory \cite{Aharony:2008ug}. Dual superconformal symmetry and Yangian invariance were first found at tree level \cite{Bargheer:2010hn,Lee:2010du,Huang:2010qy}. Then it was shown \cite{Chen:2011vv,Bianchi:2011dg} that the two--loop four--point amplitude is given by a dual conformal maximally transcendental expression in perfect match with the corresponding light-like Wilson loop expectation value \cite{Henn:2010ps,Bianchi:2013pva}. This result was generalized  \cite{Bianchi:2011fc} to the less symmetric ABJ model \cite{Aharony:2008gk} and evidence for an exponentiation {\it \`a la} BDS for the four--point ABJM amplitude was given at three loops \cite{Bianchi:2011aa,Bianchi:2014iia}. Nevertheless, the situation becomes more intricate outside the four--point case: a careful supersymmetric extension of the standard Wilson loop is needed and the amplitudes exhibit a pattern of conformal anomalies \cite{Bianchi:2012cq,Bargheer:2012cp}. However all the ABJM  observables computed so far, including form factors \cite{Brandhuber:2013gda,Young:2013hda,Bianchi:2013iha} and finite N amplitudes \cite{Bianchi:2013iha}, have been shown to respect the maximal transcendentality principle.  

The aim of this paper is the investigation of the properties of scattering amplitudes in $\mc N=2$ superconformal QCD theory. This model is an  $\mc N=2$ supersymmetric Yang-Mills theory with gauge group $SU(N)$ coupled to $N_f=2N$ fundamental hypermultiplets. The condition on the number of flavour of the fundamental fields is necessary to ensure exact conformal invariance.

Several aspects of $\mc N=2$ SCQCD have been analyzed in the past few years. In the context of integrability, the dilatation operator at one loop was constructed first in the sector of operators made of elementary scalar fields \cite{Gadde:2010zi} and then for the full theory \cite{Liendo:2011xb}.  Later on, through a diagrammatic analysis, the dilatation operator of the scalar sector was shown to deviate from the one of  $\mc N=4$ SYM at three loops \cite{Pomoni:2011jj}.  After some first promising clues,  it was definitely demonstrated  that the Hamiltonian for the full theory is not integrable \cite{Gadde:2012rv}.  However the possibility that the closed $SU(2, 1|2)$ subsector built only with adjoint fields is exactly integrable still remains open.  In \cite{Pomoni:2013poa} it  was claimed that in this subsector, present in all $\mc N=2$ superconformal models, the integrable structure becomes exactly the one of $\mc N=4$ SYM, by substituting the $\mc N=4$ coupling with an effective coupling. A weak coupling expansion of the $\mc N=2$ SCQCD effective coupling was presented in \cite{Mitev:2014yba}.

Integrability from the perspective of the scattering amplitudes/WL duality has been far less analyzed. Expectation values of Wilson loops have been studied at weak coupling by taking the diagrammatic difference with $\mc N=4$ SYM \cite{Andree:2010na}.  It was shown that light--like polygonal Wilson loops (actually any closed WL) start deviating from the corresponding $\mc N=4$ SYM results at three--loop order, confirming the prediction coming from the localization matrix model construction of \cite{Pestun:2007rz}.  A more general analysis including the strong coupling behaviour of the matrix model was performed in \cite{Passerini:2011fe}. Scattering amplitudes in $\mathcal{N}=2$ SCQCD have been computed at one--loop order only in the adjoint sector using unitarity \cite{Glover:2008tu}.  It was shown that in this sector the results match that of  $\mc N=4$ SYM and thus consist of a dual conformal invariant and maximal transcendental expression. Nothing is known so far about amplitudes in more general sectors of the theory and at higher--loop order.

In this paper we begin an analysis of planar four--point scattering amplitudes in $\mathcal{N}=2$ SCQCD. We work in $\mc N=1$ superspace formalism and perform direct super Feynman diagram computations within dimensional reduction scheme.  At one--loop order we provide a complete classification of the amplitudes, which can be divided in three independent sectors according to the color representation of the external particles. The pure adjoint sector consists of amplitudes with external fields belonging to the $\mathcal{N}=2$ vector multiplet. In this sector we confirm the results of the previous work \cite{Glover:2008tu}, since we obtain exactly the same expressions of the corresponding $\mc N=4$ SYM amplitudes, demonstrating the presence of dual conformal symmetry and maximal transcendentality.  This agrees with the conjectured integrability of the closed subsector  $SU(2, 1|2)$. As a byproduct, we also provide a direct Feynman diagram derivation of the $\mc N=4$ SYM result first derived long ago by stringy arguments \cite{Green:1982sw}. 

Outside the adjoint sector there is no reason to expect amplitudes to be dual conformal invariant. We show that in the mixed and fundamental sectors, with external fields in the fundamental representation, even if dual conformal invariance is broken, the results still exhibit maximal transcendentality weight. We thus show that at one-loop order the maximal transcendentality property of the amplitudes is not a consequence of dual conformal invariance.

In order to check these properties beyond the one--loop perturbative order,  we computed the simplest two--loop amplitude in the fundamental sector. We end up with a result that does not exhibit maximal transcendentality and it is not dual conformal invariant. 
We provide also an analysis of the transcendentality properties of every diagrams which contribute to the two--loop amplitude. 
A very non trivial check of our two--loop result is the fact that it reproduces the expected factorized structure of the infrared divergences predicted for general scattering of massless particles \cite{Catani:1998bh,Sterman:2002qn}. 

The paper is organized as follows. In Section 2 we introduce the model and our notations. In Section 3 we discuss the general features of four--point scattering amplitudes and we summarize the computational steps which are needed to find the results.  In Section 4 we present the one--loop amplitudes in the three independent sectors, while in Section 5 we perform the computation of the two--loop amplitude in the pure fundamental sector. Several technical aspects such as superspace conventions, Feynman rules and  properties of the  integrals are collected in the Appendices. \\

\noindent {\bf Note added} \\
Due to a mistake while converting the conventions used in \cite{Pomoni:2011jj} into our notations, in the published version of the paper the two--loop correction to the chiral superfield propagators presented in equation (\ref{2loopschir}) was not correct. By taking this into account, the two--loop amplitude (\ref{2loopsfinal}) gets an extra term which breaks the maximal transcendentality. Therefore  the remarks about transcendentality at two loops have been accordingly modified.

\section{$\mathcal{N}=2$ superconformal QCD}

The field content of $\mathcal{N}=2$ SCQCD can be conveniently expressed in terms of $\mathcal{N}=1$ superfields. With the superspace conventions of \cite{Gates:1983nr}, which we summarize in Appendix \ref{appena}, the Euclidean action reads
\begin{align}\label{action}
S  = & \int\d^4x \d^4\theta \bigg[  \tr\big(e^{-g V} \bar{\Phi}e^{g V}\Phi\big) + \bar{Q}^{ I}e^{g V}Q_{ I}  + \tilde{Q}^{ I}e^{-g V}\bar{\tilde{Q}}_{ I} \bigg]  + \frac{1}{g^2} \int\d^4x\d^2\theta \   \tr\big(W^{\alpha}W_{\alpha}\big) + \non \\
&+i g \int\d^4x\d^2\theta  \ \tilde{Q}^{ I} \Phi Q_{ I}
-i g \int\d^4x\d^2\bar\theta  \  \bar Q^{ I}\bar\Phi \bar{\tilde{Q}}_{ I} 
\end{align}
where $W_\alpha = i \bar{D}^2(e^{-g V}D_{\alpha}e^{g V})$ is the superfield strength of the vector superfield $V$. The gauge group is $SU(N)$; there is a global symmetry group $U(N_f)\times SU(2)_R \times U(1)_R$, where $U(N_f)$ is the flavour symmetry and $SU(2)_R \times U(1)_R$ the R-symmetry group. If the number of flavours is tuned to be $N_f=2 N$ the theory becomes exactly superconformal. 
The superfield $V$ contains the component gauge field and transforms in the adjoint representation of the gauge group $SU(N)$. The $\mathcal{N}=1$ chiral superfield $\Phi$ also transforms in the adjoint representation of $SU(N)$  and combines with the superfield $V$ into an $\mathcal{N}=2$ vector multiplet. The rest of the matter is described in terms of the quark chiral scalar superfields $Q_{ I}$ and $\tilde{Q}^{ I}$ with  $I=1,\dots,N_f$, which transform respectively in the fundamental and antifundamental representation of $SU(N)$ and together form an  $\mathcal{N}=2$ hypermultiplet. A summary of the field content of the theory is given in Table 1.

At strong coupling the dual string description of $\mc N=2$ SCQCD seems much more problematic than that of  $\mc N=4$ SYM . There are some proposal for the dual string/supergravity background which turn out to be either singular \cite{Gaiotto:2009gz,ReidEdwards:2010qs,Colgain:2011hb}  or related to non critical models \cite{Gadde:2009dj}. Any advancement on the field theory side might help claryfing  the correct properties of the gravitational description.

The $\mc N=2$ SCQCD theory can be quantized in Euclidean space by path integration $\int\mathcal{D} \psi \,e^{S[\psi]}$ over all the fields $\psi$ after performing gauge fixing in $\mathcal{N}=1$  superspace. 
The standard procedure is described in details in \cite{Gates:1983nr} and results in adding to the action (\ref{action}) the following gauge fixing term which involves the anticommuting ghost scalar superfields
\begin{align}
S_{gf} = & \int d^4x d^4\theta \ \Tr\left(-\frac{1}{\alpha} (D^2V) (\bar{D}^2V) +(c'+\bar{c}')\textrm{L}_{\frac{g V}{2}}\big[c+\bar{c}+\coth \textrm{L}_{\frac{g V}{2}}(c-\bar{c}) \big]\right) 
\end{align}
where $\textrm{L}_{\frac{g V}{2}} X = \frac{g}{2}[V,X]$. The gauge propagator can then be extracted from the gauge fixed action and we will choose to work in the supersymmetric Fermi-Feynman gauge $\alpha=1$. After gauge fixing we are left with the set of Feynman rules described in Appendix \ref{appenb}.  In Appendix \ref{appenc} we discuss the perturbative corrections to the propagators and superpotential vertices up to two--loops and we show that if $N_f=2 N$ the coupling $\beta$-function identically vanishes. 

We will compute  scattering amplitudes in perturbation theory in the planar limit $N\rightarrow \infty$ with the 't Hooft coupling $\lambda= \frac{g^2 N}{(4 \pi)^2}$ kept finite. More precisely, in order to preserve conformal invariance the number of flavours will be also sent to infinity and thus the model will be studied in the so called Veneziano limit with $N_f=2 N$.  

 \begin{table}  \begin{minipage}[c]{0.9\textwidth} \centering
 \begin{tabular}{|c|c|c|c|} \hline
 field & $SU(N)$ & $U(N_f)$ & $U(1)_R$ \\ \hline
 $V$ & Adj & 1 & 0 \\ 
 $\Phi$ & Adj &  1  & 1\\ 
 $\bar\Phi$ & Adj & 1 & -1 \\ \hline
 $Q$ & $\Box$ & $\Box$  & 0 \\
 $\bar\Q$ & $\Box$ & $\Box$ & 0 \\
 $\bar Q$ & $\bar{\Box}$ & $\bar{\Box}$ & 0\\
 $\Q$ & $\bar{\Box}$ & $\bar{\Box}$ & 0\\ \hline
 \end{tabular} \vspace{0.1cm}
 \caption{Field content of $\mc N=2$ SCQCD in terms of $\mc N=1$ superfields. The global symmetry $SU(2)_R$ is not manifest in the $\mc N = 1$ superspace formulation.}
 \end{minipage}
 \end{table}

\section{Four-point scattering}

The complete set of four--point amplitudes of the theory can be obtained by means of supersymmetry transformations from superamplitudes involving only the chiral scalar superfields  $\Phi$ and $Q$ as external particles. In fact,  in the $\mathcal{N}=1$ superfields language, supersymmetry rotates the $\Phi$ and $V$ superfield components inside the $\mathcal{N}=2$ vector multiplet and the $Q$ and $\tilde{Q}$ ones in the $\mathcal{N}=2$ hypermultiplet.  We thus can classify the four--point superamplitudes into three independent sectors according to the color representation of the external superfields: four adjoint scalar superfields, two adjoint scalars and a quark/antiquark pair and finally two quark/antiquark pairs. Different amplitudes inside each sector are related by supersymmetry.

We perform standard perturbative computations directly with the $\mathcal{N}=1$ off--shell Lagrangian (\ref{action}) so that the super Feynman diagrams give rise to expressions which are power series in the superspace Grassmann variable $\theta$. From the full superamplitude we choose to extract the lowest component of the expansion, thus presenting the explicit results for the scattering of four scalar component fields.  The other component amplitudes can be easily obtained by choosing different projections of the superspace results.

In the next sections we present our results according to the following scheme. For each sector, after selecting  a suitable process, we first discuss the partial amplitudes color decomposition. We then present the loop results for the subamplitudes, obtained performing the following steps:
\begin{itemize}
\item At first we read the contributions to the partial amplitude by considering the effective action of the model. More precisely, we draw super Feynman diagrams contributing to the four--point scalar supervertex associated to the chosen external configuration, where the diagrams have to be suitably chosen to respect the color ordering. 
\item We then perform D-algebra on the selected superdiagrams. In order to extract the four--point component amplitude with scalar fields as external particles we perform the projection $\int \d^4x\ \d^4\theta \dots =\int \d^4x\ \bar D^2 D^2\ \dots |_{\theta=0}$ on the superspace results.  
\item For each diagram, we are then left with a linear combination of standard bosonic integrals with numerators, which can be simplified by completion of squares and using on-shell symmetries. 
\item The contributions of the different diagrams is then summed up and the final result is expressed, using the integration by part reduction technique, as a linear combination of master integrals (see \cite{Smirnov} for details).
\item Finally, each master integral is expanded in terms of the dimensional regularization parameter $\e=2-d/2$ and the total result is presented as a series in the infrared divergences poles. 
\end{itemize}

\section{One-loop amplitudes}

At one--loop order we provide a complete classification of the four--point scattering amplitudes. In general we define with $(ABCD)$ a process where we treat all the  particles as outgoing so that 
$$
0 \rightarrow A(p_1)+B(p_2)+C(p_3)+D(p_4) 
$$
with light-like momentum assignments as in parentheses and momentum conservation given by $p_1+p_2+p_3+p_4=0$. We define Euclidean Mandelstam variables as
$$
s=(p_1+p_2)^2 \qquad t=(p_2+p_3)^2 \qquad u=-t-s
$$
Diagrammatically we start with the A particle in the upper left corner and proceed with the ordering counterclockwise. 

\subsection{Adjoint subsector}

In the purely adjoint sector we first focus on the process  $(\Phi \bar \Phi \Phi \bar \Phi)$. Since we deal with adjoint external particles of a $SU(N)$ gauge theory, the color decomposition of the planar amplitude is the same as in the four--gluon scattering (see \cite{Dixon:1996wi} for details)
\begin{equation}
 A^{(L)}(\{p_i,a_i\}) =  \sum_{\sigma \in S} Tr(T^{a_{\sigma_1}}T^{a_{\sigma_2}}T^{a_{\sigma_3}}T^{a_{\sigma_4}})\mc A^{(L)}(\sigma_1 \sigma_2 \sigma_3 \sigma_4) 
\end{equation}
where the sum is performed over non-cyclic permutations inside the trace. This gives rise to six {\it a priori} independent color ordered subamplitudes which might be further reduced by exploiting the symmetries of the process. These subamplitudes only receives contributions from diagrams with the specified ordering of the external particles. In any case, all the different subamplitudes divided by the corresponding tree-level contributions are expected to yield the same result since they can be mapped by $\mathcal{N}=2$ supersymmetry to proper gluon MHV amplitudes, which in turns do not depend on the gluon ordering inside the trace. We therefore expect that different orderings produce identical results. 

As we will explain below, from a diagrammatic point of view it is instructive to compute two non trivial orderings  $\mc A (\Phi(1) \bar \Phi (2)\Phi(3) \bar \Phi(4))$ and $\mc A (\Phi(1) \bar \Phi (2)\bar{\Phi}(3) \Phi(4) )$.

\subsubsection{Process $\Phi \bar \Phi \Phi \bar \Phi$} \label{adjalt}

\begin{figure}
\centering
\begin{minipage}{5cm}
\includegraphics[scale=0.28]{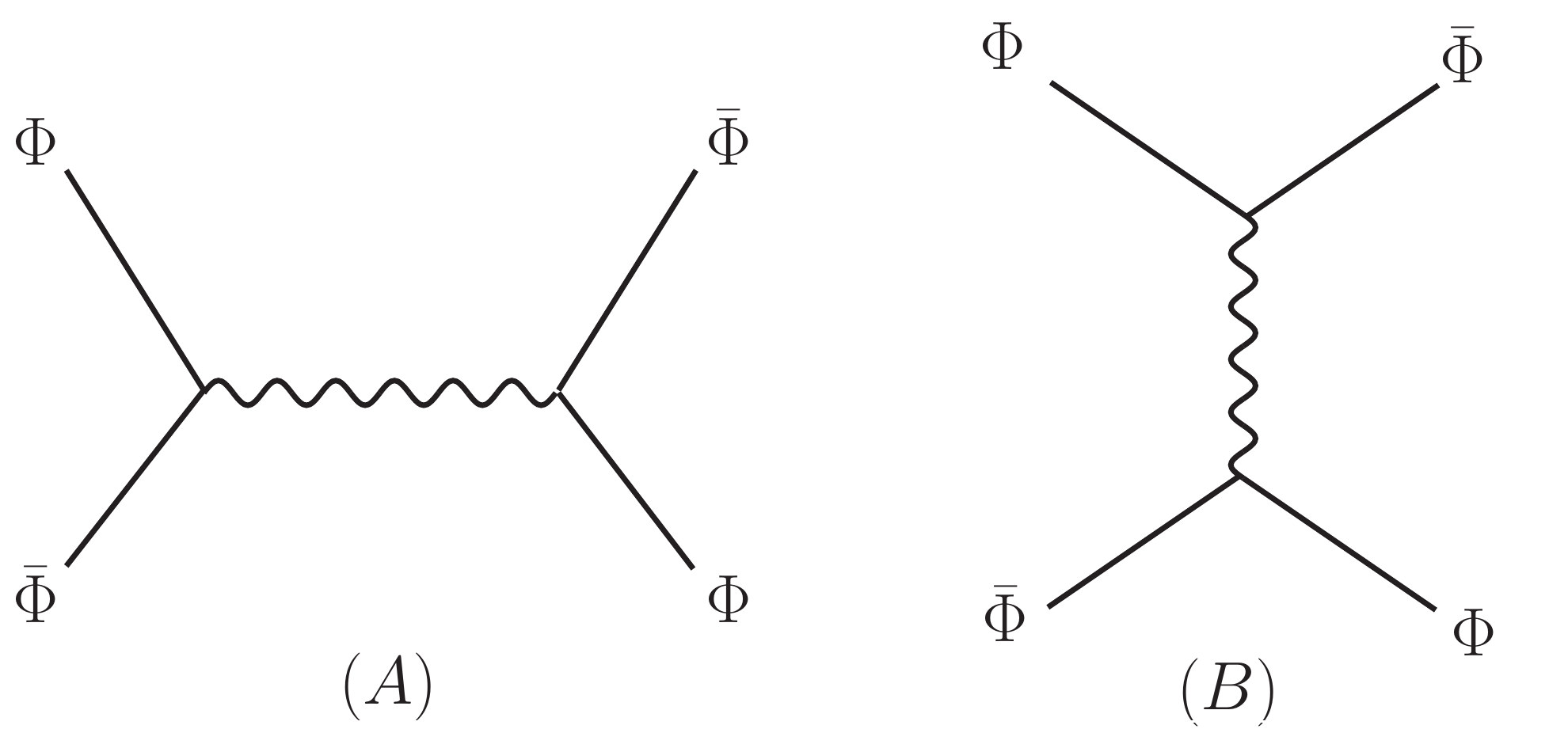}
\end{minipage} \\ \hspace{-2.7cm}
\begin{minipage}{12cm}
\centering \vspace{0.3cm}
\includegraphics[scale=0.37]{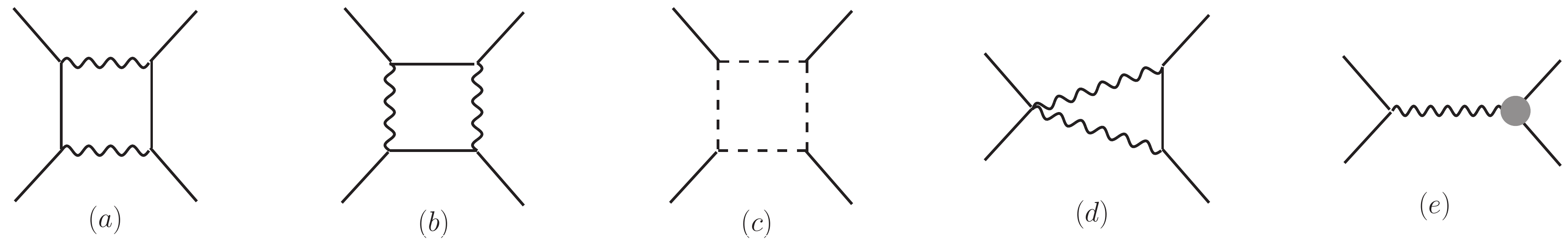}
\end{minipage}
\caption{Tree level and one--loop non--vanishing planar diagrams contributing to the process  $(\Phi \bar \Phi \Phi\bar \Phi )$. The grey bullet in diagram (e) stands for the one--loop vertex insertion. } 
\label{Fig1Loop}
\end{figure}

The color ordered subamplitude $\mc A (\Phi(1) \bar \Phi (2)\Phi(3) \bar \Phi(4))$
receives contributions at tree level from the processes depicted in Fig.\ref{Fig1Loop}(A) and (B). The contribution of the diagrams to the four scalar superfield vertex of the effective action is
\begin{equation}
\mc S^{(0)}= - g^2\ \int \d^4 p_i\ \d^4 \theta \left(\frac{1}{s}+\frac{1}{t} \right) \tr(\Phi(p_1) \bar\Phi(p_2) \Phi(p_3) \bar\Phi(p_4)) \non \end{equation}
From these we extract the relevant color structure and, after projection to the purely scalar component of the superamplitude, we can read the tree level contribution
\begin{equation}
\mc A^{(0)} (\phi(1) \bar \phi (2)\phi(3) \bar \phi(4))= g^2\ \left( \frac{u}{s}+\frac{u}{t} \right)  \label{eTree} \end{equation}
\noindent We now consider the planar one--loop corrections. The diagrams which contribute are listed in Fig.\ref{Fig1Loop}(a)--(e). For each diagram we find first the contribution to the effective action by performing the D--algebra with on--shell conditions. For the diagram $(a)$ we get
\begin{align}
\mc S^{(a)}  = & -  g^4 N\ \tr(T^a T^b T^c T^d)\,  \int \d^4 p_i\, \d^4\theta\, I_{\mathrm{triangle}}(s)\ 
\Phi_a(p_1) \bar\Phi_b(p_2) \Phi_c(p_3) \bar\Phi_d(p_4) \,\, +  \non \\ &
 -  g^4 N\ \tr(T^a T^b T^c T^d) \int \! \d^4 p_i \,\d^4\theta\, I_{\mathrm{box}}^{\alpha \dot\beta} \
\Phi_a(p_1) \bar\Phi_b(p_2) D_{\alpha} \Phi_c(p_3) \bar D_{\dot\beta}\bar\Phi_d(p_4) \non \end{align}
where $I_{\mathrm{triangle}}(s)$ and $I_{\mathrm{box}}^{\alpha \dot\beta}$ are defined in eq. (\ref{IntTriangle}) and (\ref{IntVecBox}) of Appendix \ref{append}. At this point we can project down to the four--scalar component and directly read the contribution to the color ordered amplitude 
\begin{align} \label{a4phi}
(a) & =  g^4 N \ \,\,\begin{minipage}{50px} \includegraphics[width=1.3cm]{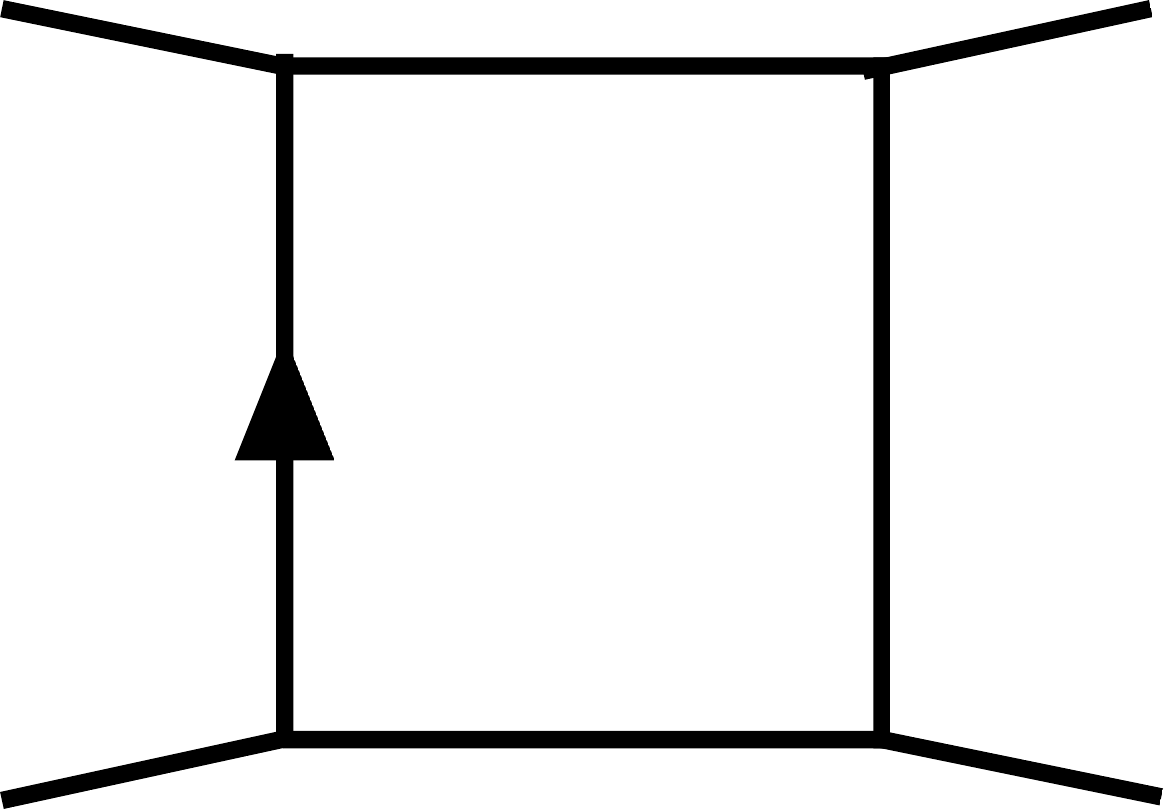} \end{minipage} \!\! \bigg(  u k^2  -  \Tr(k p_4 p_1 p_3) \bigg)
\end{align}
The numerator of the Feynman integral in (\ref{a4phi}) is spelled out explicitly whereas the denominator is represented pictorially together with an arrow indicating the integration variable $k$. Expanding the trace and completing the squares we can cast the final contribution in terms of a linear combination of scalar integrals
\begin{align} \label{adjstart}
(a) & = g^4 N \left[ - (s +2t)\,\, \begin{minipage}{50px} \includegraphics[width=1.3cm]{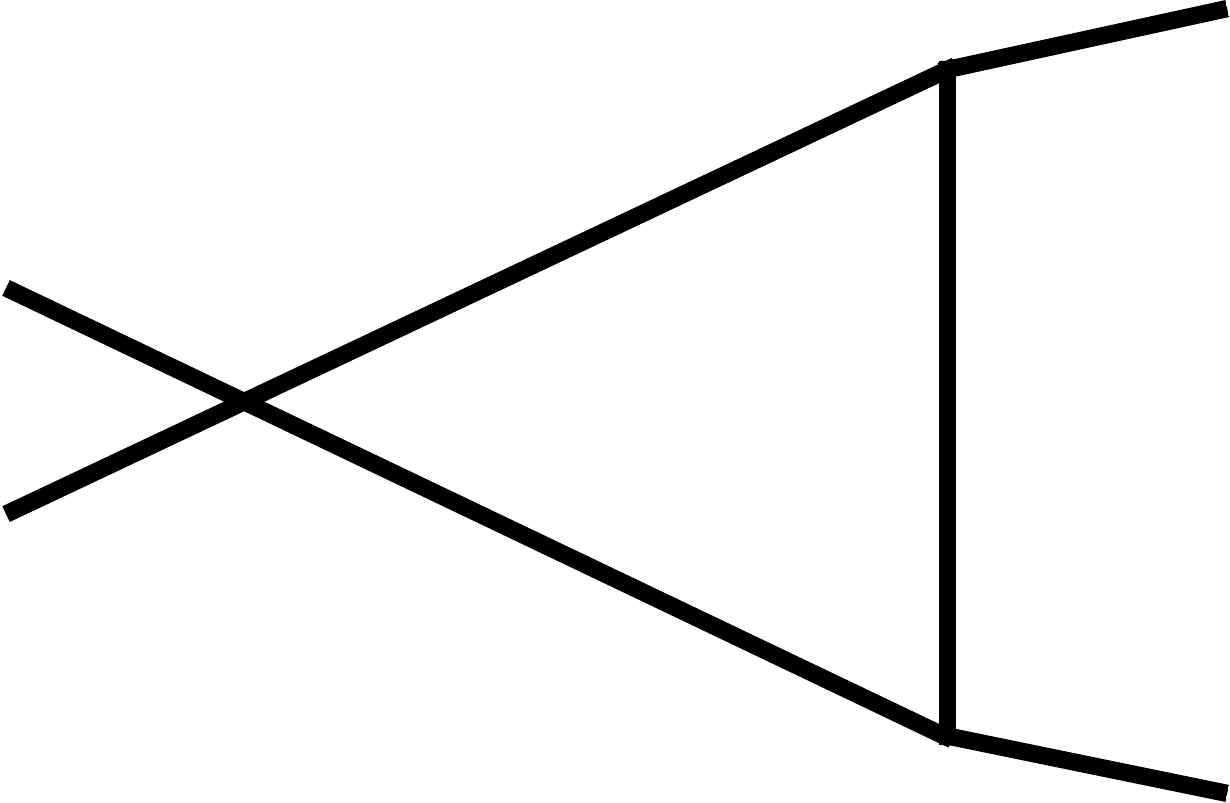}\end{minipage} \hspace{-0.5cm}\,\,\,+ t \,\,\,\begin{minipage}{50px} \includegraphics[angle=270, width=0.75cm]{Itria1.pdf}\end{minipage} \hspace{-0.8cm} +\left( t^2 + \frac{s t}{2}\right) \,\,\,\begin{minipage}{50px} \includegraphics[width=1.3cm]{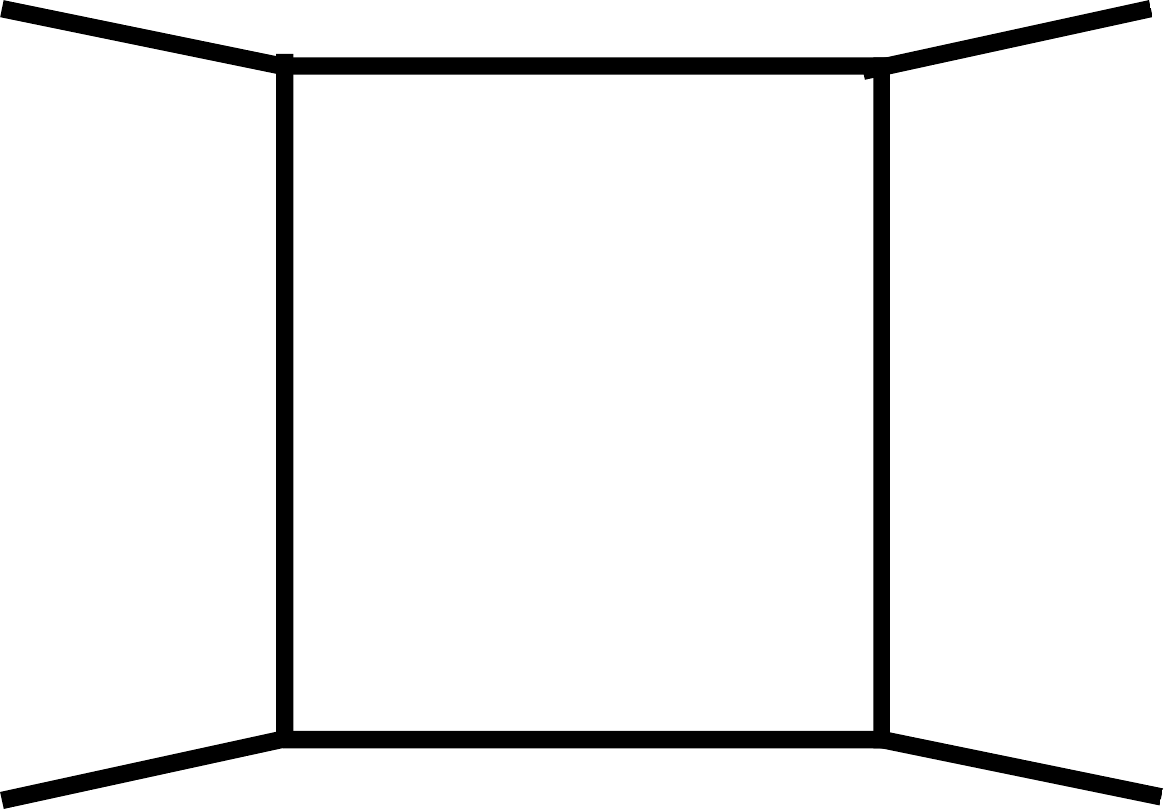}   \end{minipage} \right] 
\end{align}
The contribution of diagram $(b)$ can be immediately obtained from the one of diagram (a) by exchanging $s\leftrightarrow t$
\begin{align}
(b) &  = g^4 N \left[ s \,\, \begin{minipage}{50px} \includegraphics[width=1.3cm]{Itria1.pdf}\end{minipage} \hspace{-0.5cm}\,\,\,-(t+2s) \,\,\,\begin{minipage}{50px} \includegraphics[angle=270, width=0.75cm]{Itria1.pdf}\end{minipage} \hspace{-0.8cm}+\left( s^2 + \frac{s t}{2}\right)\,\,\,\begin{minipage}{50px} \includegraphics[width=1.3cm]{Ibox1.pdf}  \end{minipage}\right]
\end{align}
We proceed similarly for the remaining diagrams, performing D-algebra, component projection and reduction to scalar integrals. For the scalar box diagram $(c)$ we obtain
\begin{align}
(c) & = g^4 N_f \left[ \,- t \,\,\,\begin{minipage}{50px} \includegraphics[angle=270, width=0.75cm]{Itria1.pdf}\end{minipage} \hspace{-0.8cm}- s \,\, \begin{minipage}{50px} \includegraphics[width=1.3cm]{Itria1.pdf}\end{minipage}+ \frac{s t}{2} \,\,\,\begin{minipage}{50px} \includegraphics[width=1.3cm]{Ibox1.pdf}  \end{minipage} \right]
\end{align}
with $N_f=2 N$. For diagrams of type $(d)$ we need to consider the four possible ways to draw the graph, which combine to
\begin{align}
(d) & =  g^4 N \left[  (s+t) \,\, \begin{minipage}{50px} \includegraphics[width=1.3cm]{Itria1.pdf}\end{minipage} \hspace{-0.3cm}+ \,\, (s+t) \,\, \begin{minipage}{50px} \includegraphics[angle=270, width=0.75cm]{Itria1.pdf}\end{minipage} \hspace{-0.8cm}\right]
\end{align}
The diagram $(e)$ represents the one--loop correction to the vertex. The vertex correction insertions are described in Appendix \ref{appenc}. After taking into account the four possible insertions in the s-- and t--channel diagrams we get an overall
\begin{align} \label{adjend}
(e) & =   g^4 N \left[  (s+t) \,\, \begin{minipage}{50px} \includegraphics[width=1.3cm]{Itria1.pdf}\end{minipage} \hspace{-0.3cm}+ \,\, (s+t) \,\, \begin{minipage}{50px} \includegraphics[angle=270, width=0.75cm]{Itria1.pdf}\end{minipage} \hspace{-0.8cm}\right] 
\end{align}
Summing over all the contributions (\ref{adjstart})--(\ref{adjend}) it is easy to see that triangle integrals cancel out, leaving a final result which is proportional to the box integral 
\begin{align}
\mc A^{(1)}&(\phi(1) \bar \phi (2)\phi(3) \bar \phi(4)) \,\,\,=\,\,\, g^4 N\ (s+t)^2 \,\,\,\begin{minipage}{50 px} \includegraphics[width=1.3cm]{Ibox1.pdf}   \end{minipage}\,\,\, = \non\\
& = 2 \ \frac{g^4 N}{(4 \pi)^2}\ \left( \frac{u}{s}+\frac{u}{t} \right) \left\{ -\frac{1}{\e^2}\left(\frac{\mu}{s}\right)^{\e} 
-\frac{1}{\e^2}\left(\frac{\mu}{t}\right)^{\e} + \frac{2}{3} \pi^2 + \frac{1}{2} \ln^2\frac{t}{s} + \mc O(\e) \right\} \label{e1Loop}
\end{align}
where $\mu=4 \pi e^{-\gamma_\e}\nu$, and $\nu$ is the IR scale of dimensional regularization.
The reduced amplitude is then defined as the ratio between the one--loop (\ref{e1Loop}) and the tree-level one (\ref{eTree})\vspace{-0.4cm}
\begin{framed} \vspace{-0.3cm}
\begin{equation} \label{adj1}  \mc M^{(1)}(\phi(1) \bar \phi (2)\phi(3) \bar \phi(4)) = 2 \lambda \left\{ -\frac{1}{\e^2}\left(\frac{\mu}{s}\right)^{\e}  
-\frac{1}{\e^2}\left(\frac{\mu}{t}\right)^{\e} + \frac{2}{3} \pi^2 + \frac{1}{2} \ln^2\frac{t}{s} \right\}  \vspace{-0.3cm}
\end{equation} \end{framed}\vspace{-0.1cm}
\noindent where $\lambda= \frac{g^2 N}{(4 \pi)^2} $. This confirms the result of \cite{Glover:2008tu} obtained via unitarity cuts method and it shows that in this sector the one--loop amplitudes are identical to the corresponding $\mathcal{N}=4$ SYM  ones. Therefore the amplitude in (\ref{adj1}) is completely captured by a dual conformal invariant integral and respects the maximum transcendentality principle.

From a diagrammatic point of view the matching with  $\mathcal{N}=4$ SYM can be understood as follows. We could consider in $\mathcal{N}=4$ SYM  a four--point amplitude of adjoint scalar superfields with equal flavours $(\Phi_1 \bar{\Phi}_1 \Phi_1 \bar{\Phi}_1)$.  We note that diagrams (a), (b), (d) and (e) of Fig.\ref{Fig1Loop} can be drawn also for this process and are identical to the ones computed in $\mathcal{N}=2$ SCQCD. In  $\mathcal{N}=4$ SYM  diagram (c) is substituted with an analouge diagram with adjoint scalars circulating into the loop. This exactly reproduces the contribution of the fundamental loop of $\mathcal{N}=2$ SCQCD  when $N_f=2 N$. Therefore it would have been easy in this case to work taking the diagrammatic difference between the two models and to show that it is vanishing. It is a general feature of $\mathcal{N}=2$  SCQCD diagrams that fundamental matter loops give the same results of $\mathcal{N}=4$ SYM scalar adjoint loops.

\subsubsection{Process $\Phi \bar \Phi \bar \Phi \Phi $}

\begin{figure} 
\centering
\begin{minipage}{3,3cm}
\includegraphics[scale=0.3]{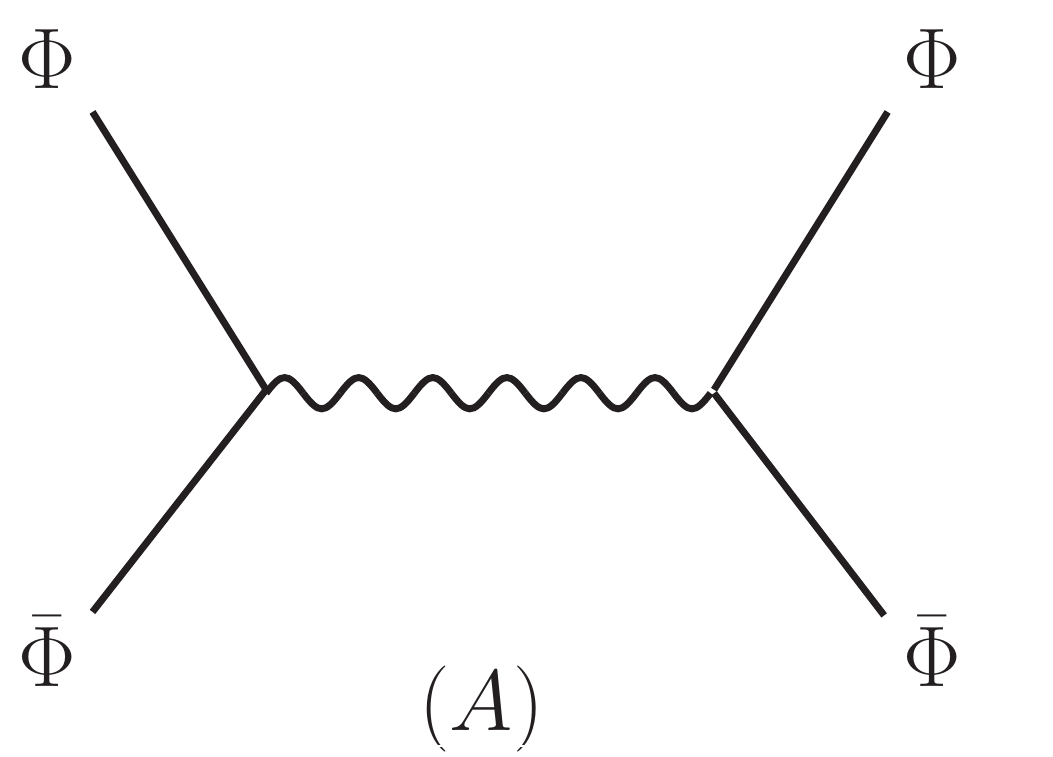}
\end{minipage} \qquad
\begin{minipage}{9cm}
\includegraphics[scale=0.4]{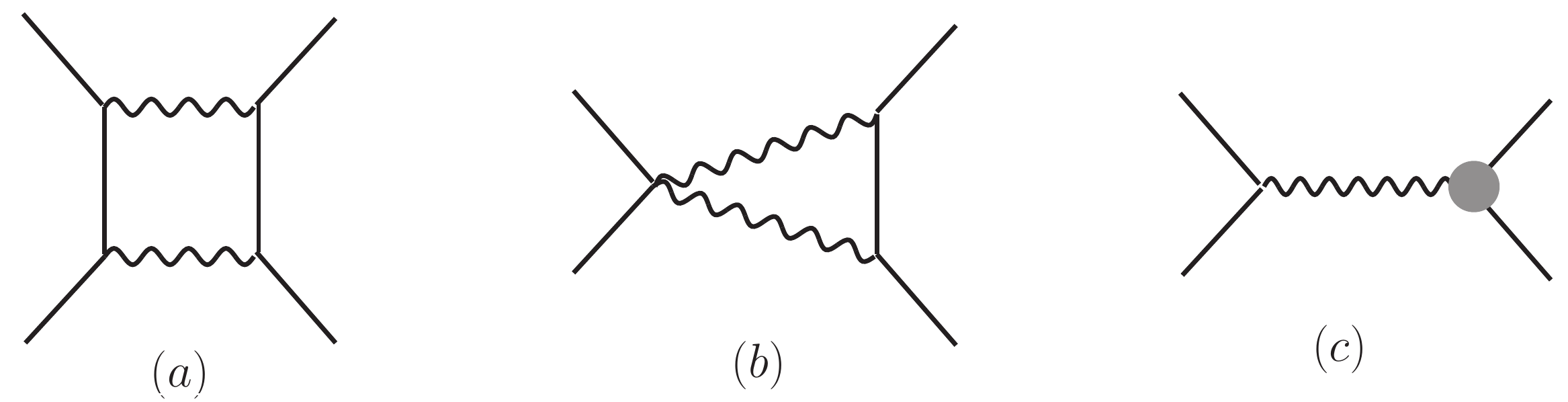}
\end{minipage} \vspace{0.1cm}
\caption{Tree level and one--loop diagrams contributing to $(\Phi \bar \Phi \bar\Phi \Phi)$ process.} \label{Figadj2}
\end{figure}

We now focus on the color ordered subamplitude $\mc A (\Phi(1) \bar \Phi (2)\bar \Phi(3) \Phi(4))$ for the process $(\Phi \bar \Phi \bar\Phi \Phi)$. From a diagrammatic point of view this is equivalent to consider the color ordered subamplitude $\mc A (\Phi(1) \bar \Phi (2)\bar \Phi(4) \Phi(3))$ for the process considered above $(\Phi \bar \Phi \Phi \bar\Phi)$.  At tree level only the diagram in Fig.\ref{Figadj2}(A) contributes according to color ordered rules. 
After projection to the scalar component we obtain
\begin{equation} \label{adj2tree}
\mc A^{(0)}(\phi(1) \bar \phi (2)  \bar \phi(3)\phi(4))= - g^2\ \frac{t}{s} \end{equation}
The non-vanishing planar one--loop diagrams are listed in Fig.\ref{Figadj2}(a)--(c). 
For each diagram we perform D--algebra, component projections and master integrals expansion as detailed above and obtain
\begin{align}
(a) & = g^4 N\ t^2  \,\,\,\begin{minipage}{50px} \includegraphics[width=1.3cm]{Ibox1.pdf}  \end{minipage}  \\
(b) & = - g^4 N  \, t \,\,\,\begin{minipage}{50px} \includegraphics[width=1.3cm]{Itria1.pdf}\end{minipage}   \\
(c) & =  g^4 N \, t \,\,\,\begin{minipage}{50px} \includegraphics[width=1.3cm]{Itria1.pdf}\end{minipage} 
\end{align}
Note that diagrams (b) and (c) now contribute in two ways, which can be obtained from the drawn diagrams by left/right reflection. 
The full amplitude then simply reads
\begin{align}
\mc A^{(1)} (&\phi(1) \bar \phi (2) \bar \phi(3) \phi(4)) \,\,= \,\, g^4 N \ t^2 \,\,\,\begin{minipage}{50px} \includegraphics[width=1.3cm]{Ibox1.pdf}   \end{minipage} \,\, = \non \\
& = 2  \ \frac{g^4 N}{(4 \pi)^2}\ \frac{t}{s} \left\{ \frac{1}{\e^2}\left(\frac{\mu}{s}\right)^{\e} 
+\frac{1}{\e^2}\left(\frac{\mu}{t}\right)^{\e} - \frac{2}{3} \pi^2 - \frac{1}{2} \ln^2\frac{t}{s} + \mc O(\e) \right\} 
\end{align}
Taking the ratio with the tree-level amplitude (\ref{adj2tree}) we immediately get
\begin{equation}
\mc M^{(1)}(\phi(1) \bar \phi (2) \bar \phi(3) \phi(4)) =  \mc M^{(1)}(\phi(1) \bar \phi (2)\phi(3) \bar \phi(4)) 
\end{equation}
as expected. We note that this ordering of the external fields gives rise to a smaller number of diagrams with respect to the ordering of section \ref{adjalt}. 
Moreover, all the diagrams of Fig.\ref{Figadj2} display a corresponding diagram for the analogue process in $\mathcal{N}=4$ SYM  yielding the same result. 
It is then straightforward in this case to predict the final result. 
With this respect, since fundamental matter interaction does not play any role, our computation can be seen as a direct standard Feynman diagram confirmation 
of the $\mathcal{N}=4$ SYM result, computed long ago by taking a low energy limit of a superstring \cite{Green:1982sw} and then readily reproduced by unitarity methods.


\subsection{Mixed adjoint/fundamental sector}

We now consider amplitudes with two external fields in the fundamental/antifundamental representation of the gauge group $SU(N)$.  Focusing on the process $(Q \bar Q \Phi \bar \Phi)$, the color decomposition of planar amplitudes is given by
\begin{equation} \label{mixord}
 A^{(L)}(Q \bar Q \Phi \bar \Phi)= \sum_{\sigma \in S} (T^{\sigma_3}T^{\sigma_4})^{i}_{\phantom{i} i} \mc A^{(L)}(Q(1) \bar Q(2) \sigma_3 \sigma_4)
\end{equation}
There are two non trivial color structures given by strings of color indices starting with the antifundamental index of the $\bar{Q}$ field and ending with the fundamental index of the field $Q$.  The two structures differs by a permutation of the color matrices of the adjoint fields.  Once again we expect to obtain the same result for all the ordering of the reduced subamplitudes.
Concerning the flavour structure of the amplitudes, it is easy to see that we only have non vanishing results for the quark $Q_I$ and antiquark $\bar Q^J$ fields with equal flavours $I=J$. We therefore can omit the flavour indices in our expressions.

\subsubsection{Process $Q \bar Q \Phi \bar{\Phi}$}

\begin{figure}
\begin{minipage}{2.5cm}
\vspace{0.17cm} \includegraphics[scale=0.39]{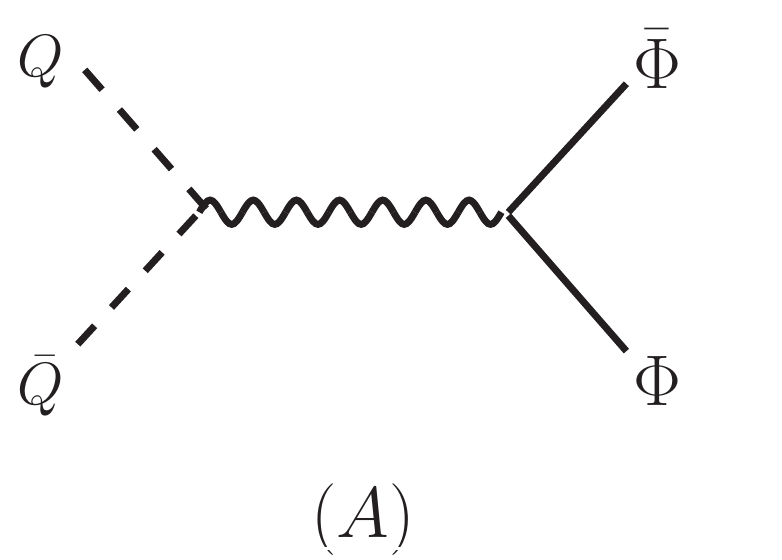}
\end{minipage} \qquad \!\!\! \,
\begin{minipage}{8.8cm}
\includegraphics[scale=0.39]{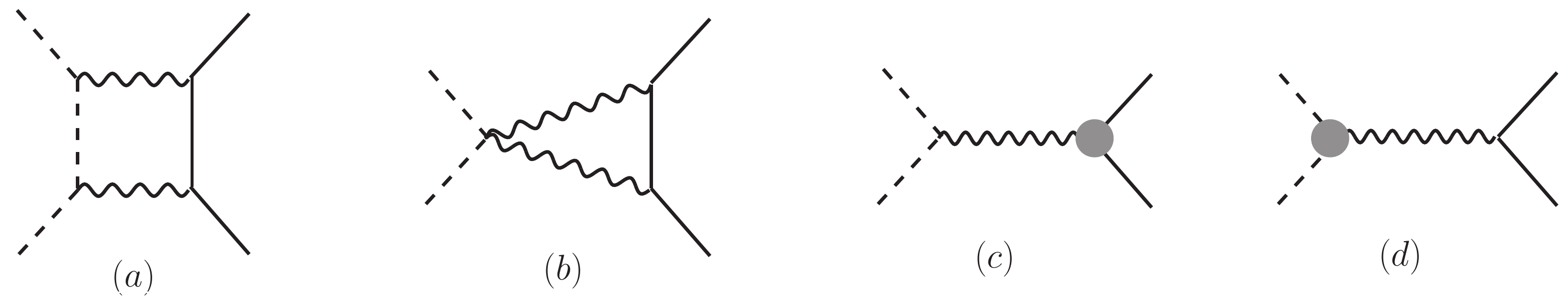}
\end{minipage} \vspace{0.1cm}
\caption{Tree level and one--loop diagrams for the process $(Q \bar Q \Phi \bar \Phi)$.} 
\label{Fig4}
\end{figure}

We consider  first the subamplitude $\mc A(Q(1) \bar Q(2) \Phi(3)\bar{\Phi}(4)) $. At tree level only the process in Fig.\ref{Fig4}(A) contributes and after projection we get
\begin{equation}
\mc A^{(0)}(q(1) \bar q(2) \phi(3)\bar{\phi}(4)) = g^2\ \frac{u}{s} \end{equation}
The diagrams giving non vanishing planar one--loop corrections are listed in Fig.\ref{Fig4}(a)--(d). 
These evaluate to
\begin{align}
(a) & = g^4 N \left[ \,- (2 t + s) \,\,\,\begin{minipage}{50px} \includegraphics[width=1.3cm]{Itria1.pdf}\end{minipage} + t \,\, \begin{minipage}{50px} \includegraphics[angle=270, width=0.75cm]{Itria1.pdf}\end{minipage} \hspace{-0.8cm} + \left( t^2 + \frac{s t}{2}\right) \,\,\,\begin{minipage}{50px} \includegraphics[width=1.3cm]{Ibox1.pdf}  \end{minipage} \right] \\
(b) & = g^4 N  \,\, (s + t) \,\,\,\,\,\begin{minipage}{50px} \includegraphics[width=1.3cm]{Itria1.pdf}\end{minipage} \label{mix1} \\
(c) & = g^4 N  \, \,\frac{(s + t)}{2} \,\,\,\,\,\begin{minipage}{50px} \includegraphics[width=1.3cm]{Itria1.pdf}\end{minipage} \\
(d) & = - g^4 N \,\, \frac{(s + t)}{2} \,\,\,\,\,\begin{minipage}{50px} \includegraphics[width=1.3cm]{Itria1.pdf}\end{minipage} 
\end{align}
where we already combined in (\ref{mix1}) the two possible permutations for diagrams of type (b). Summing over all the partial contributions we get
\begin{align}
\mc A^{(1)}&(q(1) \bar q(2) \phi(3)\bar{\phi}(4))   = g^4 N\ \left[ -t  \,\,\,\begin{minipage}{50px} \includegraphics[width=1.3cm]{Itria1.pdf}\end{minipage}  + t  \,\, \begin{minipage}{50px} \includegraphics[angle=270, width=0.75cm]{Itria1.pdf}\end{minipage} \hspace{-0.8cm} + \left( t^2 + \frac{s t}{2}\right) \,\,\,\begin{minipage}{50px} \includegraphics[width=1.3cm]{Ibox1.pdf}  \end{minipage} \right] = \non \\
& \hspace{-0.5cm} = \ \frac{g^4 N}{(4 \pi)^2}\ \left\{\frac{u}{s} \left[ -\frac{2}{\e^2}\left(\frac{\mu}{t}\right)^{\e}  
-\frac{1}{\e^2}\left(\frac{\mu}{s}\right)^{\e} + \frac{3}{4} \pi^2 + \frac{1}{2} \ln^2\frac{t}{s} \right] -
\frac{t}{s} \left[ \frac{\pi^2}{2} + \frac{1}{2} \ln^2\frac{t}{s} \right] \right\} \, \label{bifu2} \end{align} \vspace{0.2cm}
The reduced amplitude then reads\vspace{-0.6cm}
\begin{framed}\vspace{-0.35cm}
\begin{align} \label{bifu} &
\hspace{-0.2cm} \mc M^{(1)}(q(1) \bar q(2) \phi(3)\bar{\phi}(4)) = \lambda \left\{ -\frac{2}{\e^2}\left(\frac{\mu}{t}\right)^{\e} -\frac{1}{\e^2}\left(\frac{\mu}{s}\right)^{\e}  
	 + \frac{3}{4} \pi^2 + \frac{1}{2} \ln^2\frac{t}{s} - \frac{t}{u} \left[\frac{\pi^2}{2} + \frac{1}{2} \ln^2\frac{t}{s}\right]\right\}  \non \\[0.1cm] &\vspace{-0.7cm}\end{align}\vspace{-0.95cm}\end{framed}\vspace{-0.3cm}
\noindent  We first note that the dual conformal invariance which was present in the pure adjoint sector is lost. This is best seen by looking at the scalar integrals contributing to the amplitude in equation (\ref{bifu2}). Together with the dual conformal box,  triangle integrals survive, inevitably breaking the dual conformal symmetry.  Nevertheless we notice that the result in (\ref{bifu}) respects the maximal transcendentality principle. This explicitly shows that dual conformal invariance and maximal transcendentality are independent properties of the amplitudes at one-loop order.  

We further notice that for the chosen process different channels contribute asymmetrically. We might have considered the process with cyclically rotated fields $(\bar{q} \phi \bar{\phi} q)$. This would produce a result given by (\ref{bifu}) with $s \leftrightarrow t$. It is amusing to note that if we had to sum over the two processes for the given subamplitude the result in  (\ref{bifu}) would be symmetrized in $s$ and $t$ giving an expression proportional to  (\ref{adj1}).

\subsubsection{Process $Q \bar Q \bar{\Phi} \Phi$}

\begin{figure}
\centering
\begin{minipage}{5cm}
\includegraphics[scale=0.4]{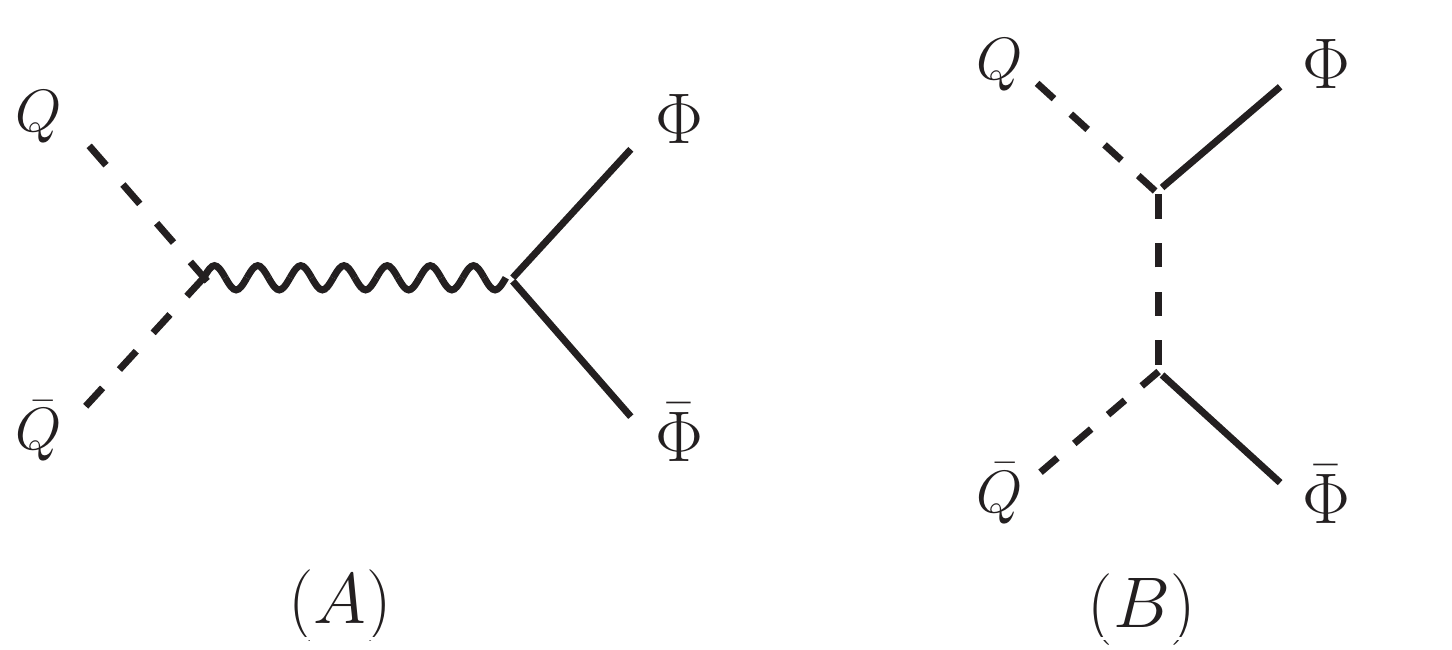}
\end{minipage}\\ \hspace{-3.8cm}
\begin{minipage}{12cm} \vspace{0.35cm}
\includegraphics[scale=0.39]{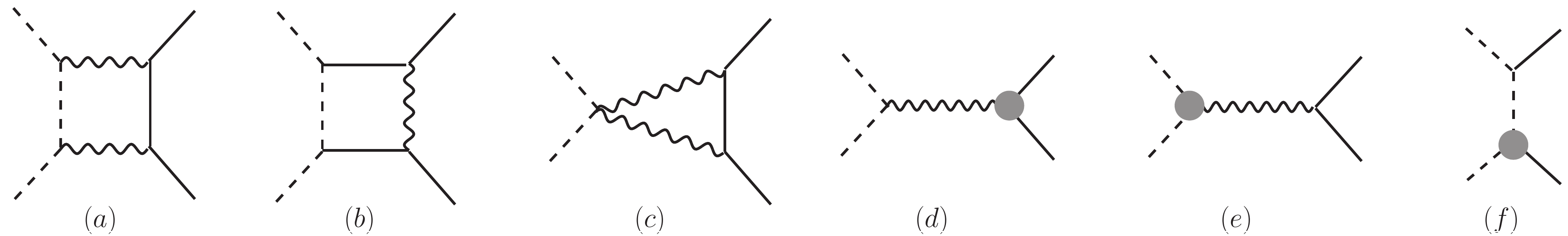}
\end{minipage}\vspace{0.2cm}
\caption{Tree level and one--loop diagrams for  the process $(Q \bar{Q}\bar{\Phi}\Phi)$.} 
\label{Fig5}
\end{figure}

As a check of our computation we analyze the subamplitude $\mc A(Q(1) \bar Q(2) \bar{\Phi}(3) \Phi(4))$ for the process  $(Q \bar Q \bar{\Phi} \Phi )$. 
This subamplitude is diagrammatically identical to the color ordered subamplitude $\mc A(Q(1) \bar Q(2) \bar{\Phi}(4) \Phi(3))$ for the process $(Q \bar{Q}\Phi\bar{\Phi})$. 
At tree level the diagrams in Fig.\ref{Fig5}(A) and (B) contributes to the scalar projection
\begin{equation}
\mc A^{(0)}(q(1) \bar q(2) \bar{\phi}(3) \phi(4))= g^2\ \frac{u}{s} 
\end{equation}
We now consider the planar one--loop corrections to the tree level amplitude. The diagrams which contribute are listed in Fig.\ref{Fig5}(a)--(f) and give
\begin{align}
(a) & = g^4 N\, t^2   \,\,\,\,\,\begin{minipage}{50px} \includegraphics[width=1.3cm]{Ibox1.pdf}  \end{minipage}  \\
(b) & = g^4 N \left[- t  \begin{minipage}{50px} \includegraphics[angle=270, width=0.75cm]{Itria1.pdf}\end{minipage} \hspace{-0.8cm} + \frac{s t}{2}  \,\,\,\,\,\begin{minipage}{50px} \includegraphics[width=1.3cm]{Ibox1.pdf}  \end{minipage}\right] \\
(c) & = - g^4 N  \, t \,\,\,\, \begin{minipage}{50px} \includegraphics[width=1.3cm]{Itria1.pdf}\end{minipage} \\
(d) & =  g^4 N \, \frac{t}{2} \,\,\,\,\,\begin{minipage}{50px} \includegraphics[width=1.3cm]{Itria1.pdf}\end{minipage} \\
(e) & =  - g^4 N  \, \frac{t}{2} \,\,\,\,\begin{minipage}{50px} \includegraphics[width=1.3cm]{Itria1.pdf}\end{minipage} \\
(f) & = g^4 N\  2 t \,\,\,\, \begin{minipage}{50px} \includegraphics[angle=270, width=0.75cm]{Itria1.pdf}\end{minipage}  
\end{align}
with diagrams of type (c) and (f) summed over the two possible choices. Summing over all the partial contributions we find 
\begin{align}
\mc A^{(1)}(q(1) \bar q(2) \bar{\phi}(3) \phi(4)) & = g^4 N\ \left[ -t  \,\,\,\begin{minipage}{50px} \includegraphics[width=1.3cm]{Itria1.pdf}\end{minipage}  + t  \,\, \begin{minipage}{50px} \includegraphics[angle=270, width=0.75cm]{Itria1.pdf}\end{minipage} \hspace{-0.8cm} + \left( t^2 + \frac{s t}{2}\right) \,\,\,\begin{minipage}{50px} \includegraphics[width=1.3cm]{Ibox1.pdf}  \end{minipage} \right] \non \\
\end{align}
This is exactly the result we found for in (\ref{bifu2}). By supersymmetry the same result holds for amplitudes involving $\tilde{Q}$ and $\bar{\tilde{Q}}$.


\subsection{Fundamental sector}

We now consider amplitudes with two pairs of quark/anti-quark superfields as external particles. We describe the color structure for the process $(Q \bar Q Q\bar Q)$ and we remind that a similar description holds when substituting  $\bar{Q}$ with $\tilde{Q}$ and/or $Q$ with $\bar{\tilde{Q}}$. The planar amplitude can be decomposed as follows
\begin{align} \label{fundord}
 A^{(L)}(Q_I \bar Q^J  Q_K \bar Q^M )  = & \,\,\, \delta_I^J \delta_K^M \,Q(1)_{j} \bar{Q}(4)^{j}  \, Q(3)_{i} \bar{Q}(2)^{i} \,  \mc A^{(L)}_1(q \bar q q \bar q) \non \\  
 &+ \delta_I^M \delta_K^J \, Q(1)_{i} \bar{Q}(2)^{i}\,  Q(3)_{j} \bar{Q}(4)^{j}\, \mc A^{(L)}_2(q \bar q q \bar q)  
\end{align}
We thus have two independent color structures corresponding to the two possible ways of contracting the pairs of fundamental and antifundamental indices.  For each color structure we only have a unique choice of flavour flow displayed in equation (\ref{fundord}).  We will omit the flavour indices in what follows. 

\subsubsection{Process $Q \bar Q Q \bar Q$}

\begin{figure}
\begin{minipage}{3cm}
\vspace{0.2cm}\includegraphics[scale=0.39]{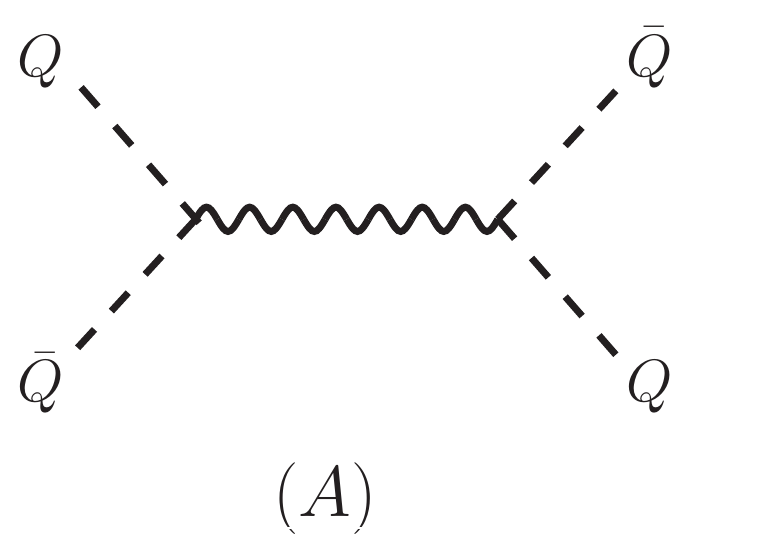}
\end{minipage} \qquad \!\!\!\!\!
\begin{minipage}{9cm}
\includegraphics[scale=0.39]{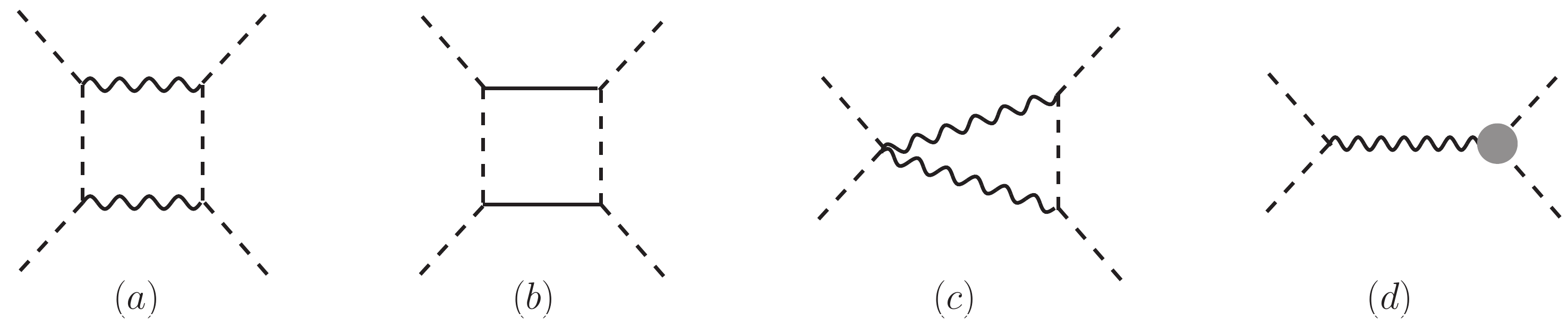}
\end{minipage} \vspace{0.1cm}
\caption{Tree level and one--loop diagrams for $(Q \bar Q Q\bar Q)$. } 
\label{Fig6}
\end{figure}

We compute the partial amplitude $\mc A_1(q \bar q q \bar q)$. At tree level  only the diagram depicted in Fig.\ref{Fig6}(A) gives a contribution
\begin{equation}
\mc A^{(0)}_1(q \bar q q \bar q)= g^2\ \frac{u}{s} \end{equation}
We now consider the planar one--loop corrections to the tree level amplitude. The diagrams which contribute are listed in Fig. \ref{Fig6}(a)--(d) and give the following results
\begin{align}
(a) & = g^4 N \left[ - (s +2t)\,\, \begin{minipage}{50px} \includegraphics[width=1.3cm]{Itria1.pdf}\end{minipage} \hspace{-0.5cm}\,\,\,+ t \,\,\,\begin{minipage}{50px} \includegraphics[angle=270, width=0.75cm]{Itria1.pdf}\end{minipage} \hspace{-0.8cm} +\left( t^2 + \frac{s t}{2}\right) \,\,\,\begin{minipage}{50px} \includegraphics[width=1.3cm]{Ibox1.pdf}   \end{minipage} \right] \\
(b) & = g^4 N \left[ \,- t \,\,\,\begin{minipage}{50px} \includegraphics[angle=270, width=0.75cm]{Itria1.pdf}\end{minipage} \hspace{-0.8cm}- s \,\, \begin{minipage}{50px} \includegraphics[width=1.3cm]{Itria1.pdf}\end{minipage}+ \frac{s t}{2} \,\,\,\begin{minipage}{50px} \includegraphics[width=1.3cm]{Ibox1.pdf}  \end{minipage} \right] \\
(c) & =  g^4 N \, (s + t) \,\,\,\begin{minipage}{50px} \includegraphics[width=1.3cm]{Itria1.pdf}\end{minipage} \\
(d) & =  - g^4 N \, (s + t) \,\,\,\begin{minipage}{50px} \includegraphics[width=1.3cm]{Itria1.pdf}\end{minipage} 
\end{align}
where again we summed over the two left/right reflected diagrams of type (c) and (d). Summing over all partial contributions we find 
\begin{align}
\mc A^{(1)}_1(q \bar q q \bar q)&  = g^4 N\ \left[ - 2 (t + s)  \,\,\,\begin{minipage}{50px} \includegraphics[width=1.3cm]{Itria1.pdf}\end{minipage} + \left( t^2 + s t \right) \,\,\,\begin{minipage}{50px} \includegraphics[width=1.3cm]{Ibox1.pdf}  \end{minipage} \right] = \non \\
& = 2 \ \frac{g^4 N}{(4 \pi)^2}\ \frac{u}{s} \left\{ -\frac{1}{\e^2}\left(\frac{\mu}{t}\right)^{\e} + \frac{7}{12} \pi^2 
+ \frac{1}{2} \ln^2\frac{t}{s} + \mc O(\e) \right\} \end{align}
The ratio between the one--loop amplitude and the tree-level one is \vspace{-0.2cm}
\begin{framed}
\begin{equation} \label{1fund}
\mc M^{(1)}_1 (q \bar q q \bar q) \,\, = \,\, 2 \lambda\ \left\{ -\frac{1}{\e^2}\left(\frac{\mu}{t}\right)^{\e} + \frac{7}{12} \pi^2 
+ \frac{1}{2} \ln^2\frac{t}{s} + \mc O(\e)\right\} \vspace{-0.2cm}\end{equation} \end{framed}\vspace{-0.2cm}
Once again we see that the result does not display dual conformal invariance whereas it respects maximal transcendentality.  
It is possible to show that the partial amplitude $\mc A^{(1)}_2$ is equal to  $\mc A^{(1)}_1$ with the exchange $s \leftrightarrow t$.

\subsubsection{Process $Q \tilde{Q} \bar{\tilde{Q}}\bar Q$} \label{prox}

\begin{figure}
\centering
\begin{minipage}{4cm}
\vspace{0.35cm}\includegraphics[scale=0.4]{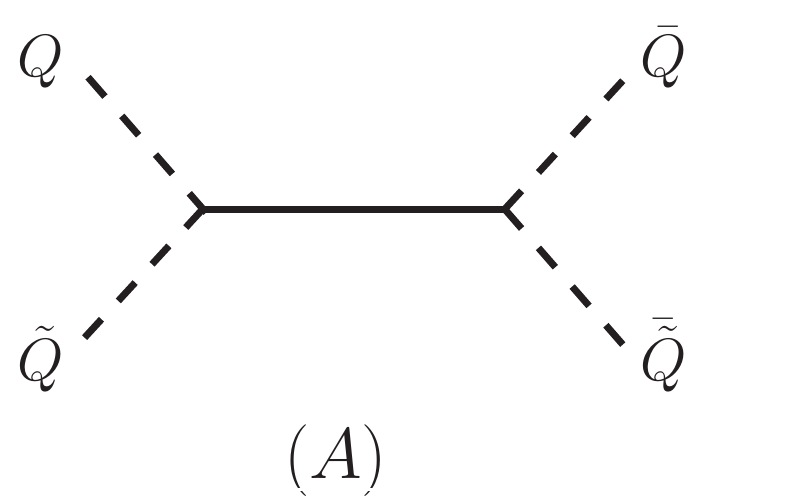}
\end{minipage}\quad
\begin{minipage}{4cm}
\includegraphics[scale=0.4]{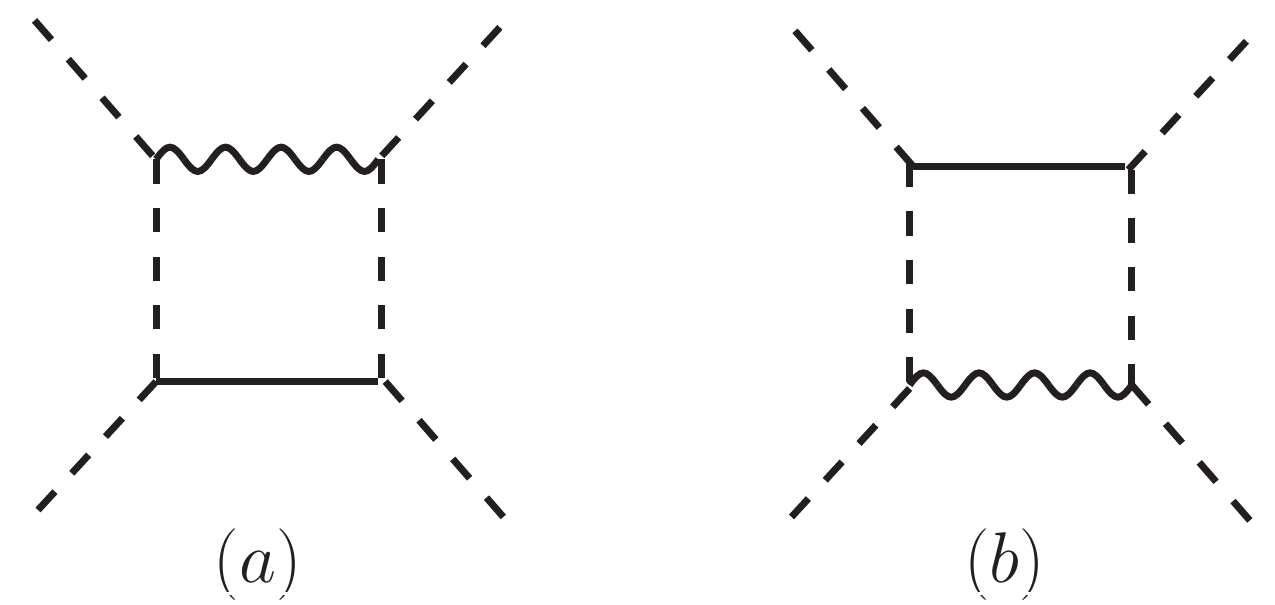}
\end{minipage}
\vspace{0.1cm}
\caption{Tree level and one--loop diagrams for $(Q \tilde{Q} \bar{\tilde{Q}}\bar Q)$ process. } 
\label{Fig3}
\end{figure}

As a check of our result (\ref{1fund}) we consider the process $(Q \tilde{Q} \bar{\tilde{Q}}\bar Q)$, which is expected to provide an identical expression because of supersymmetry. 
We consider the color structure $Q(1)_{j} \bar{Q}(4)^{j} \, \bar{\tilde Q}(3)_{i} \tilde Q(2)^{i}$, which is the analogue of the one considered for the previous process.
The amplitude corresponding to the tree level diagram depicted in Fig.\ref{Fig3}(A) is the following
\begin{align}
\mc A^{(0)}_{1}(q \tilde{q} \bar{\tilde{q}} \bar{q}) & = - g^2  \end{align}
We now consider the planar one--loop corrections to the tree level amplitude. The relevant diagrams are listed in Fig.\ref{Fig3}(a) and (b).
The contributions of diagram $(a)$ is
\begin{align}
(a) & = g^4 N\ \left[ - s \,\,\,\begin{minipage}{50px} \includegraphics[width=1.3cm]{Itria1.pdf}\end{minipage}  + \frac{s t}{2}  \,\,\,\begin{minipage}{50px} \includegraphics[width=1.3cm]{Ibox1.pdf}  \end{minipage}\right] 
\end{align}
The contribution of diagram $(b)$ is equal to diagram $(a)$. 
\vspace{4pt}
Summing the two diagrams above we find
\begin{align}
\mc A^{(1)}_1 (q \tilde{q} \bar{\tilde{q}} \bar{q})& = g^4 N\ \left[ -2 s  \,\,\,\begin{minipage}{50px} \includegraphics[width=1.3cm]{Itria1.pdf}\end{minipage} + s t  \,\,\,\begin{minipage}{50px} \includegraphics[width=1.3cm]{Ibox1.pdf}  \end{minipage} \right] =  \non \\
& = 2 \ \frac{g^4 N}{(4\pi)^2}\  \left\{ \frac{1}{\e^2}\left(\frac{\mu}{t}\right)^{\e} -\frac{7}{12}\pi^2 
-\frac{1}{2} \ln^2\frac{t}{s} + \mathcal{O}(\e) \right\}   \end{align}
Taking the ratio with the tree level result we obtain again the result in (\ref{1fund})
\begin{equation}
\mc M^{(1)}_1(q \tilde{q} \bar{\tilde{q}} \bar{q}) = \mc M^{(1)}_1 (q \bar{q} q \bar q)
\end{equation}
This process turns out to be the simplest from the computational point of view and it will then be chosen for the two--loop analysis in the next Section. Once again, one might want to consider the other color ordering or also reshuffled processes whose results can be obtained by  suitable permutations of the Mandelstam variables.

\section{Two-loop amplitudes}

At two--loops the supergraph computation starts becoming cumbersome because of the increasing number of diagrams contributing to each process.  There are some indications based on Feynman diagrammatics and integrability arguments that in the pure adjoint sector at two--loops the amplitude should be identical to that of $\mathcal{N}=4$ SYM. In fact, in \cite{Pomoni:2011jj} the dilatation operator of the theory has been found to coincide with that of $\mathcal{N}=4$ SYM up to two--loops in the purely scalar sector. Moreover, in \cite{Pomoni:2013poa} it has been argued  that the sector built only with adjoint letters should be exactly integrable.  If dual conformal invariance  and the duality with light--like  Wilson loops are a consequence of integrability, we then expect from the Wilson loop computation in \cite{Andree:2010na} to obtain a result that deviates from the $\mathcal{N}=4$ SYM result only at three loop order. A diagrammatic check of this claim is in progress \cite{progress}.

In the other two sectors nothing is known a priori and we expect a behaviour which is qualitative different from the $\mathcal{N}=4$ SYM case. From our one--loop detailed analysis it is easy to see that inside each sector the degree of complexity for different processes is very variable. It is therefore advisable to choose the special amplitude giving rise to less contributions. We present here the full result for the computationally easiest choice, the process of section \ref{prox} in the pure fundamental sector. We will see that the result is not dual conformal invariant and the maximal transcendentality principle is not respected at two--loop order. 

\subsection{Fundamental sector} \label{2loopscomput}

\begin{figure} 
\includegraphics[width=15.5cm]{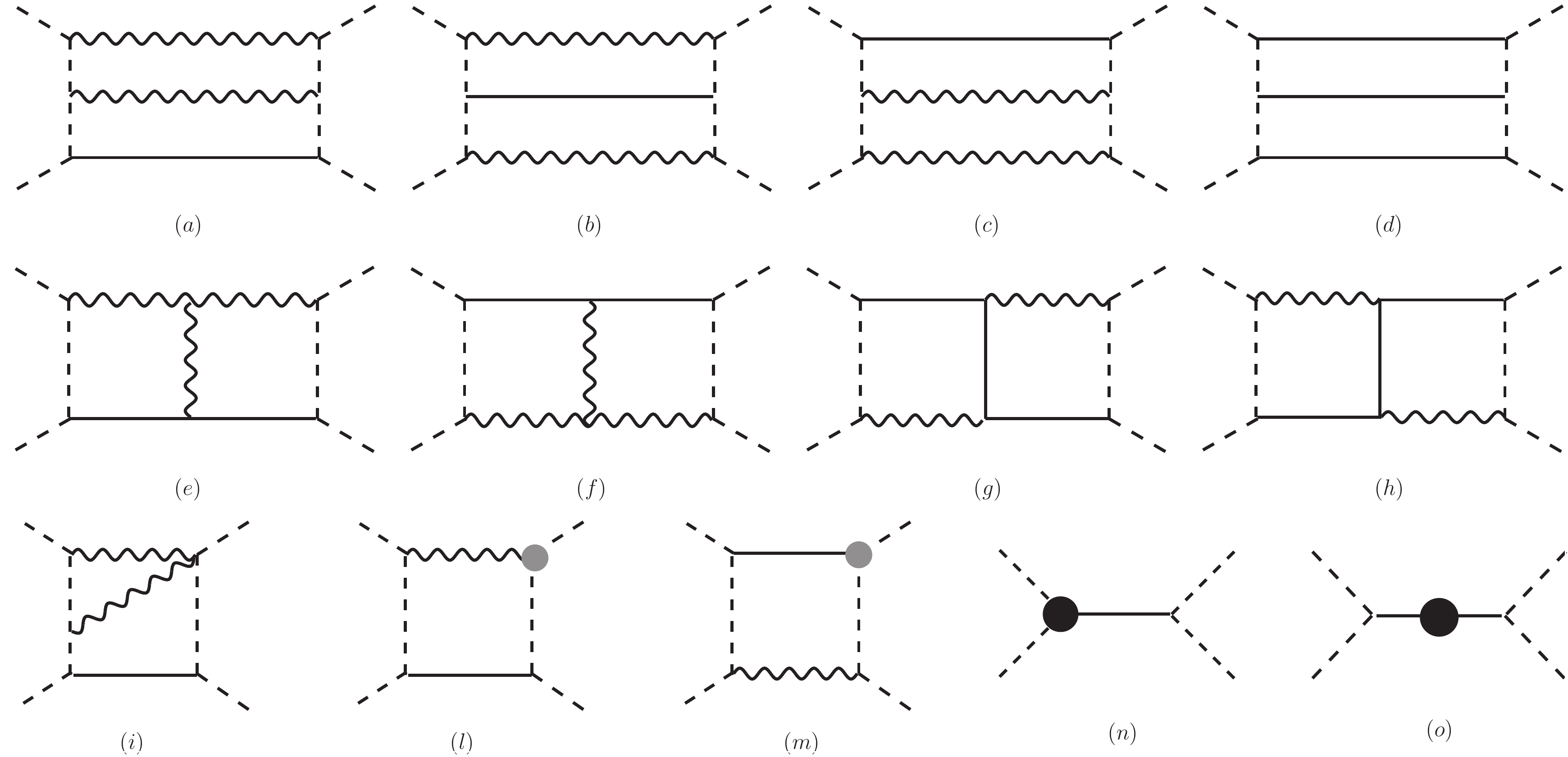} 
\caption{Non vanishing two--loop diagrams contributing to $Q\tilde{Q}\bar{\tilde{Q}}\bar{Q}$ amplitude. Gray and black bullets stand for one-- and two--loop insertions respectively.} \label{pic2loops}
\end{figure}

We consider the process  $(Q \tilde{Q} \bar{\tilde{Q}}\bar Q)$ and compute the two--loop correction  $\mc A^{(2)}_1(q \tilde{q} \bar{\tilde{q}} \bar q)$ for the color structure  $Q(1)_{j} \bar{Q}(4)^{j} \, \bar{\tilde Q}(3)_{i} \tilde Q(2)^{i}$. The diagrams which give a non-vanishing contribution are depicted in Fig.\ref{pic2loops}. 
The diagrams (a)--(d)  in the first line have the topology of vertical double boxes and we found useful to simplify their contributions by combining them properly.  After performing the D-algebras and the projections to the purely scalar component  amplitude we obtain
\begin{align}
(a) &=  g^6 N^2 \begin{minipage}{50px} \includegraphics[width=1.5cm]{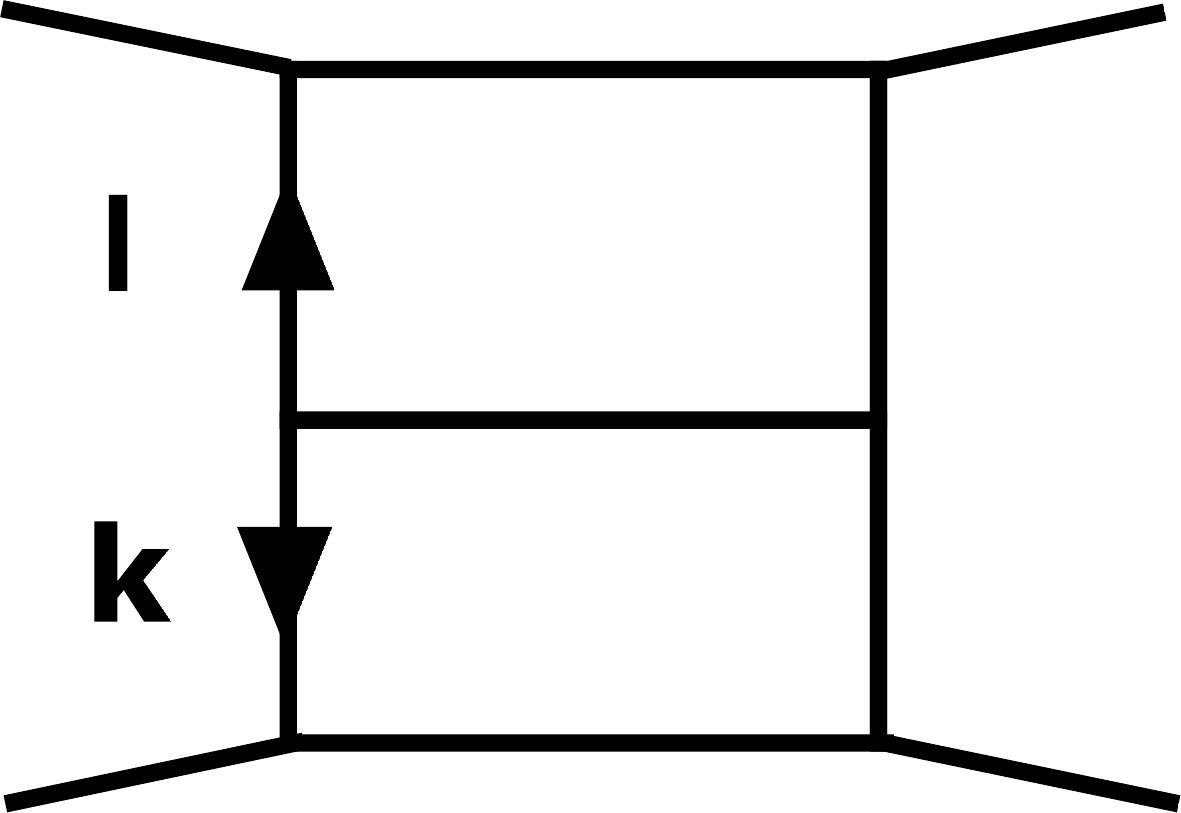} \end{minipage} \bigg( -s k^2 l^2 - l^2 \Tr(p_2 k p_4 p_1) + k^2 \Tr(l p_4 p_1 p_2)  + \Tr(p_2 k (p_3+p_3) l p_4 p_1) \bigg) \non \\
(b) &=  g^6 N^2 \begin{minipage}{50px} \includegraphics[width=1.5cm]{Verdboxmo.pdf} \end{minipage} \bigg( -s k^2 l^2 + k^2 \Tr(p_4 p_1 p_2 l) + l^2 \Tr(k p_3 p_2 p_1) 
+ \Tr(p_4 l k p_3 p_2 p_1) \bigg) \non \\
(c) & =  g^6 N^2 \begin{minipage}{50px} \includegraphics[width=1.5cm]{Verdboxmo.pdf} \end{minipage} \bigg( -s k^2 l^2 + l^2 \Tr(p_1 k p_3 p_2) - k^2 \Tr(l p_3 p_2 p_1) - \Tr(l (p_2+p_3) k p_3 p_2 p_1) \bigg)\non \\
(d) & =  g^6 N^2\begin{minipage}{50px} \includegraphics[width=1.5cm]{Verdboxmo.pdf} \end{minipage} \bigg( -s k^2 l^2 - k^2 \Tr(p_1 l p_3 p_2) - l^2 \Tr(k p_2 p_1 p_4) - \Tr(p_1 l k p_2 p_1 p_4) \bigg) \non 
\end{align}
where we again explicitly write the numerators and pictorially represent the denominators with loop variables $k$ and $l$. As anticipated, the 6-- and 4--gamma traces coming from the different contributions can be nicely combined to produce a simple overall contribution
\begin{align}
(a)+(b)+(c)+(d) & = -2s \,\bigg( \,\begin{minipage}{50px} \includegraphics[width=1.5cm]{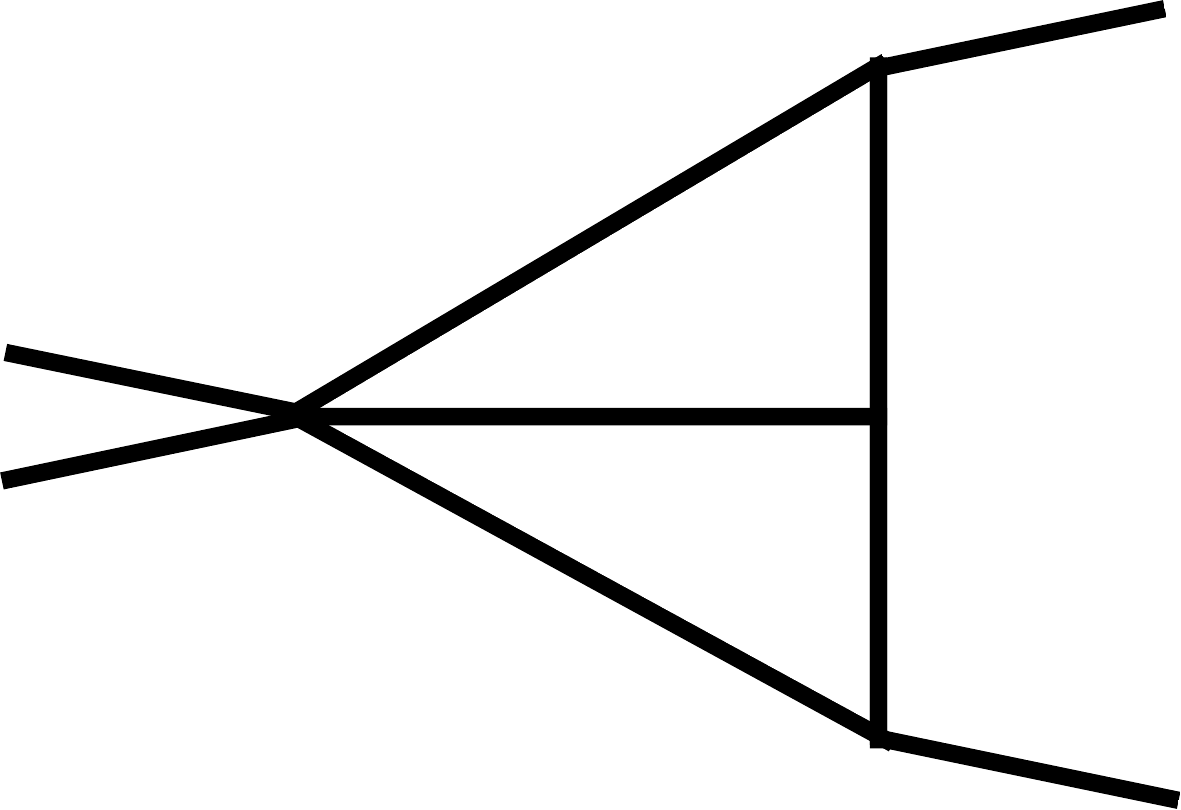} \end{minipage}\!\! + \,\,\begin{minipage}{50px} \includegraphics[width=1.5cm]{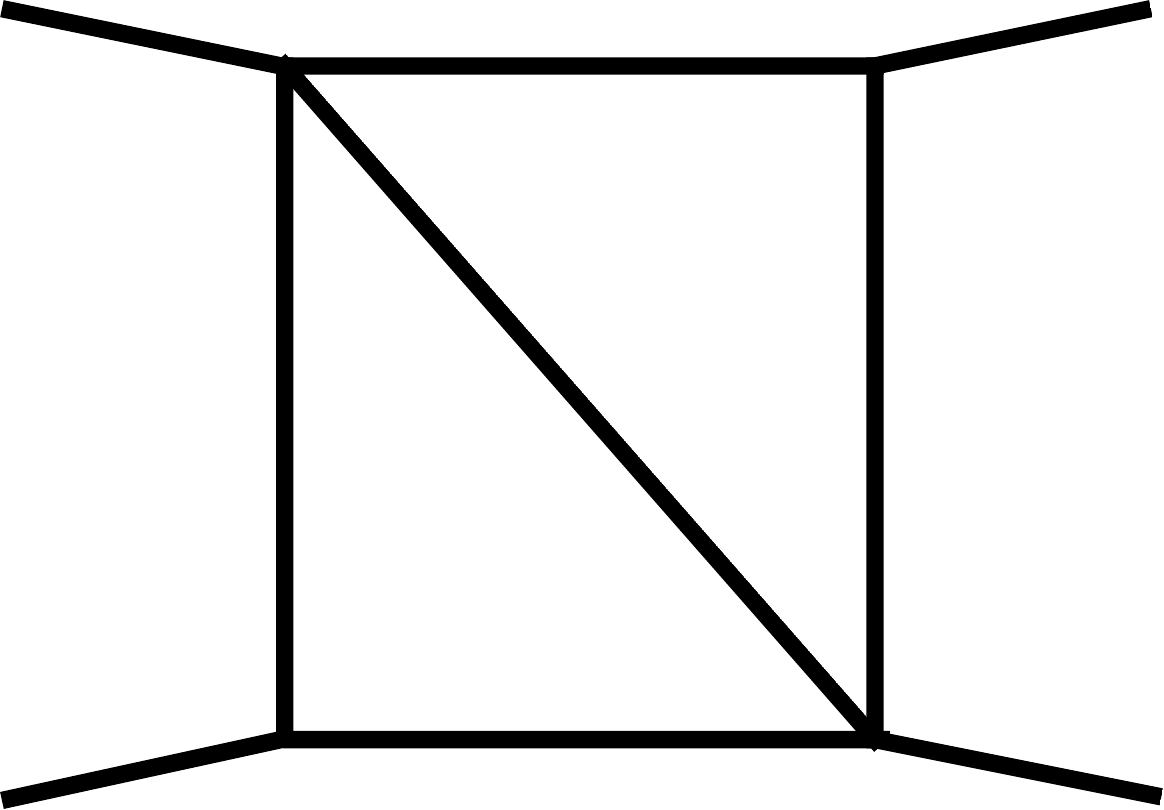} \end{minipage}\bigg) + 4st \hspace{0.3cm} \begin{minipage}{50px} \includegraphics[width=1.5cm]{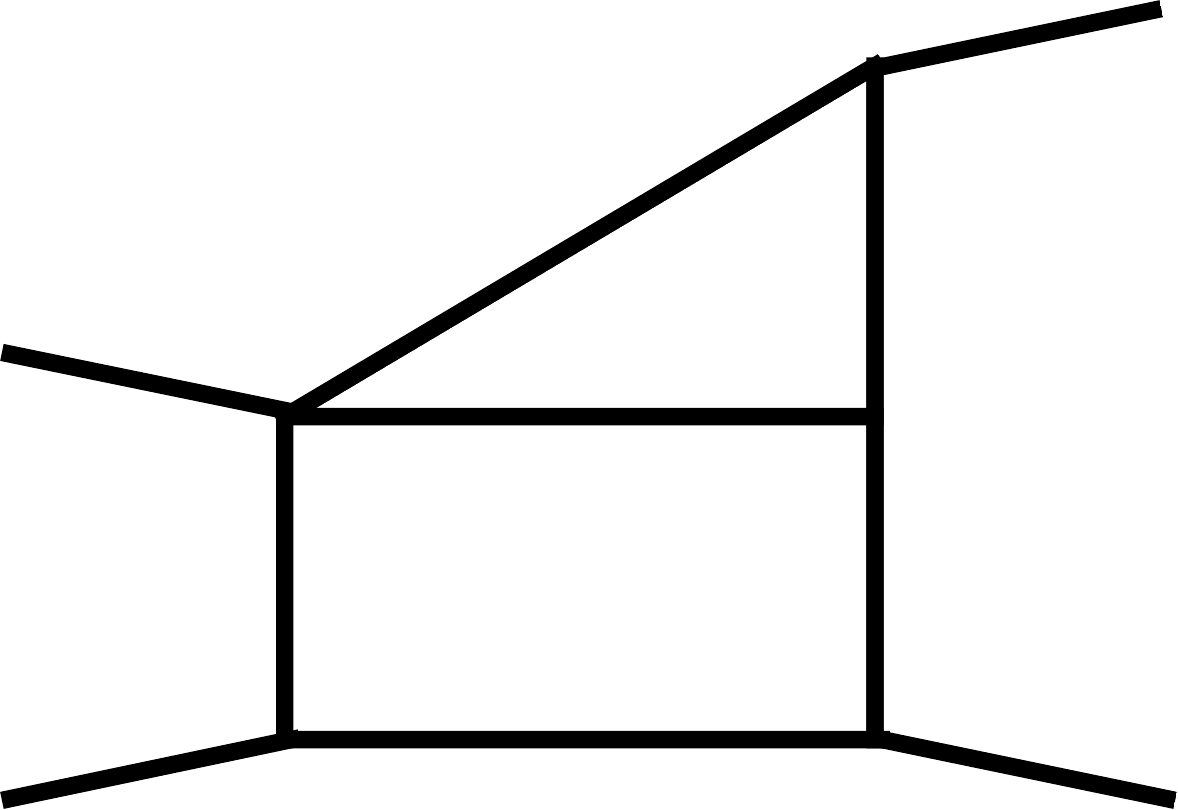} \end{minipage} -s t^2  \hspace{0.3cm} \begin{minipage}{50px} \includegraphics[width=1.5cm]{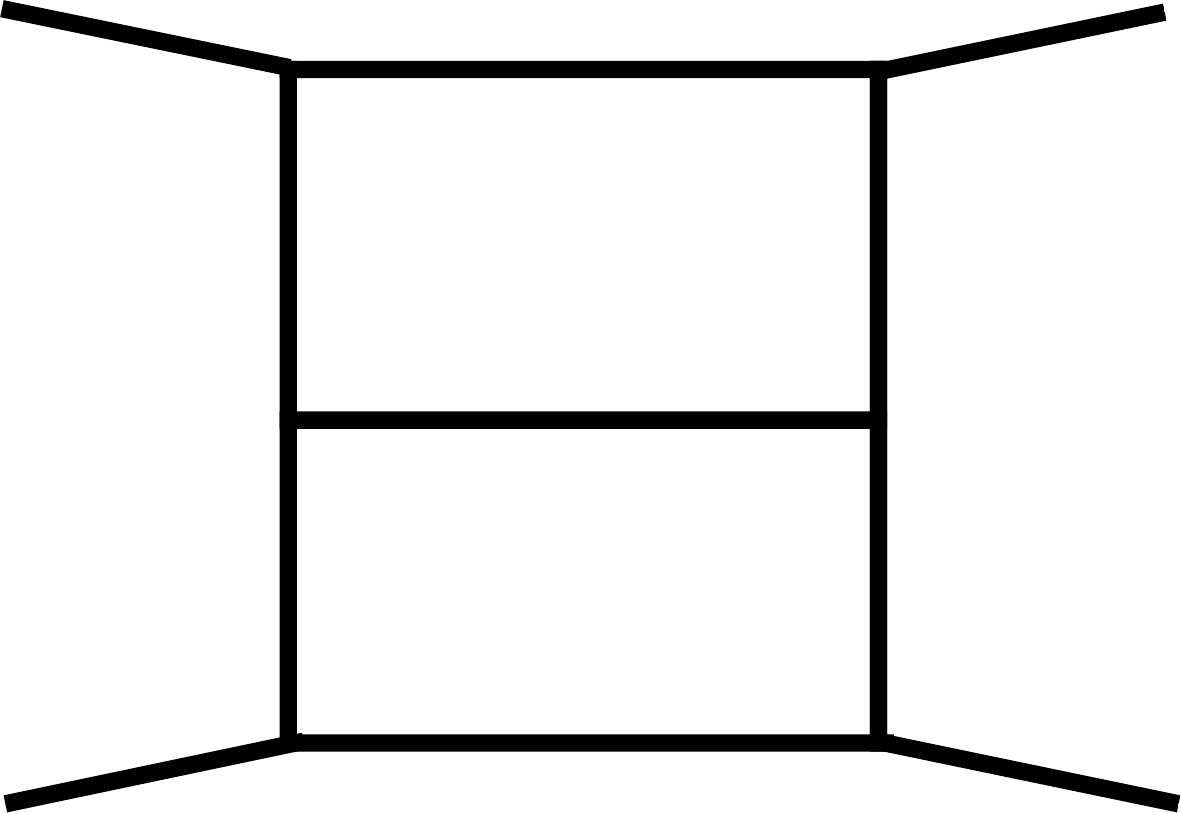} \end{minipage} \non \\[0.1cm] & \label{2init}
\end{align}
where we omitted a  $g^6 N^2$  factor and completed the squares using the symmetries of the integrals to simplify the result. 
The diagram (e)--(h) in the second row of Fig.\ref{pic2loops} have the topology of horizontal double boxes and once again their contribution can be conveniently combined. After D-algebra and projection they give 
\begin{align}
(e) & = \frac{1}{2} g^6 N^2   \,\,\begin{minipage}{50px} \includegraphics[width=1.8cm]{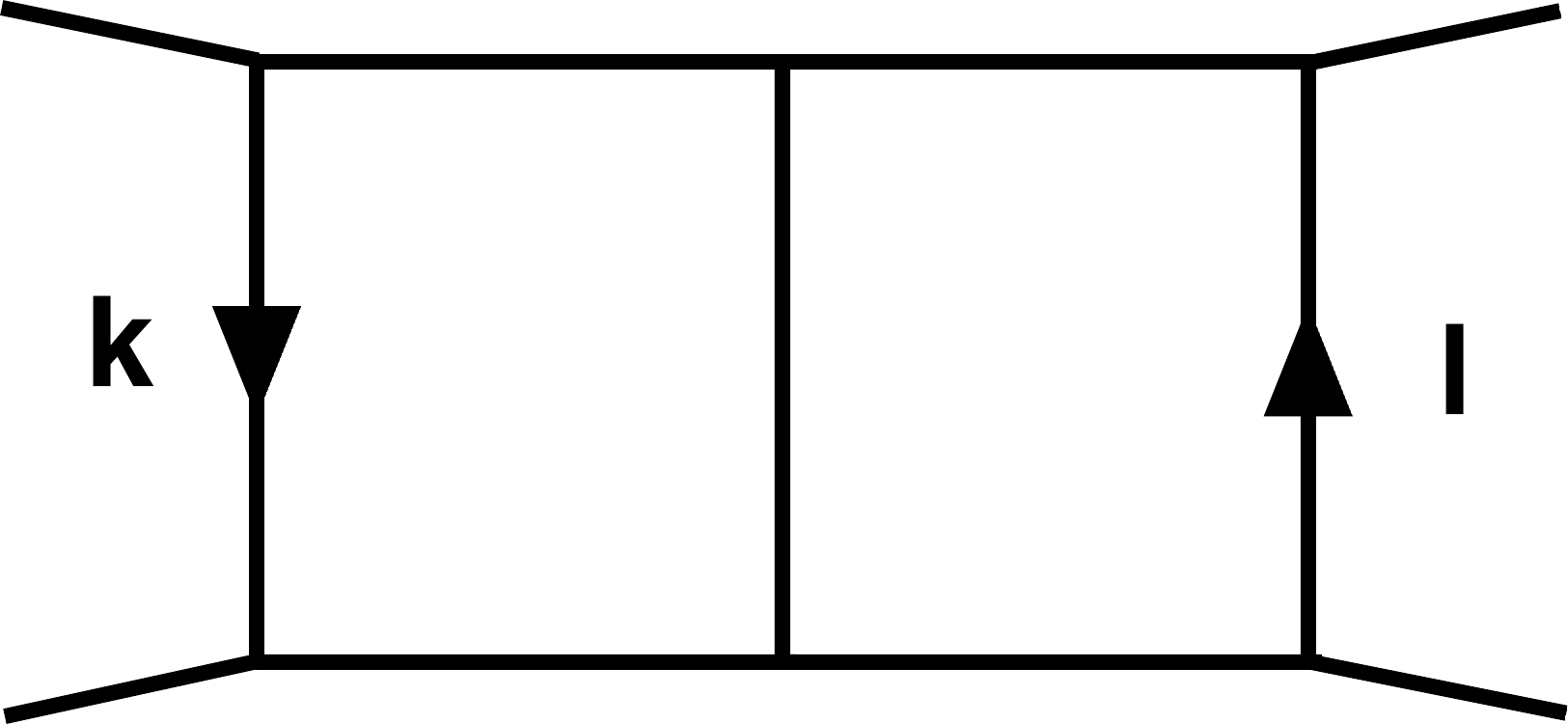} \end{minipage} \,\, \bigg(  -2 k^2 \Tr(l p_1 p_2 p_3) - 2 \Tr(l p_1 k p_2 p_1 p_3) + (k-p2)^2 \Tr(l p_1 p_2 p_3) + \non \\
& \hspace{4cm} -s (k-p_2)^2 (l+p_3)^2 -s(k-p_2)^2 l^2 + (l+p3)^2 \Tr(k p_2 p_1 p_3) \bigg) \non \\
(f) & = \frac{1}{2} g^6 N^2 \,\,\begin{minipage}{50px} \includegraphics[width=1.8cm]{Iladmo.pdf} \end{minipage}\,\, \bigg( 2 k^2 \Tr(l p_2 p_1 p_4) - 2 \Tr(l p_2 k p_1 p_2 p_4) - (k+p1)^2 \Tr(l p_2 p_1 p_4) + \non \\
&  \hspace{4cm} -s (k+p_1)^2 (l-p_4)^2 -s(k+p_1)^2 l^2 -(l-p4)^2 \Tr(k p_1 p_2 p_4) \bigg) \non \\ 
(g) & =  g^6 N^2 \,\,\begin{minipage}{50px} \includegraphics[width=1.8cm]{Iladmo.pdf} \end{minipage} \,\,\bigg(  -s (k+p_1)^2 (l+p_3)^2 + (k+p_1)^2 \Tr(p_2 l p_3 p_1) - \Tr(p_2 l p_3 p_4 k p_1) + \non \\ & \hspace{4cm}- (l+p_3)^2 \Tr(p_4 p_2 p_1 k) 
\bigg) \non \\
(h) & =  g^6 N^2 \,\,\begin{minipage}{50px} \includegraphics[width=1.8cm]{Iladmo.pdf} \end{minipage}\,\, \bigg(  -s (l-p_1)^2 (k-p_2)^2 - (k-p_2)^2 \Tr(p_1 l p_4 p_2)  - \Tr(p_1 l p_4 p_3 k p_2) +\non \\ & \hspace{4cm}+ (l-p_4)^2 \Tr(p_1 p_3 k p_2) 
\bigg)  \non \end{align}
After expanding the traces and completing the squares the overall contribution massively simplifies to
\begin{align}
(e)+(f)+ (g)+(h) &  =  \,\,\big( t-s\big) \,\,\,\begin{minipage}{50px} \includegraphics[width=1.5cm]{Idiag.pdf} \end{minipage} - t \,\,\begin{minipage}{50px} \includegraphics[width=1.5cm]{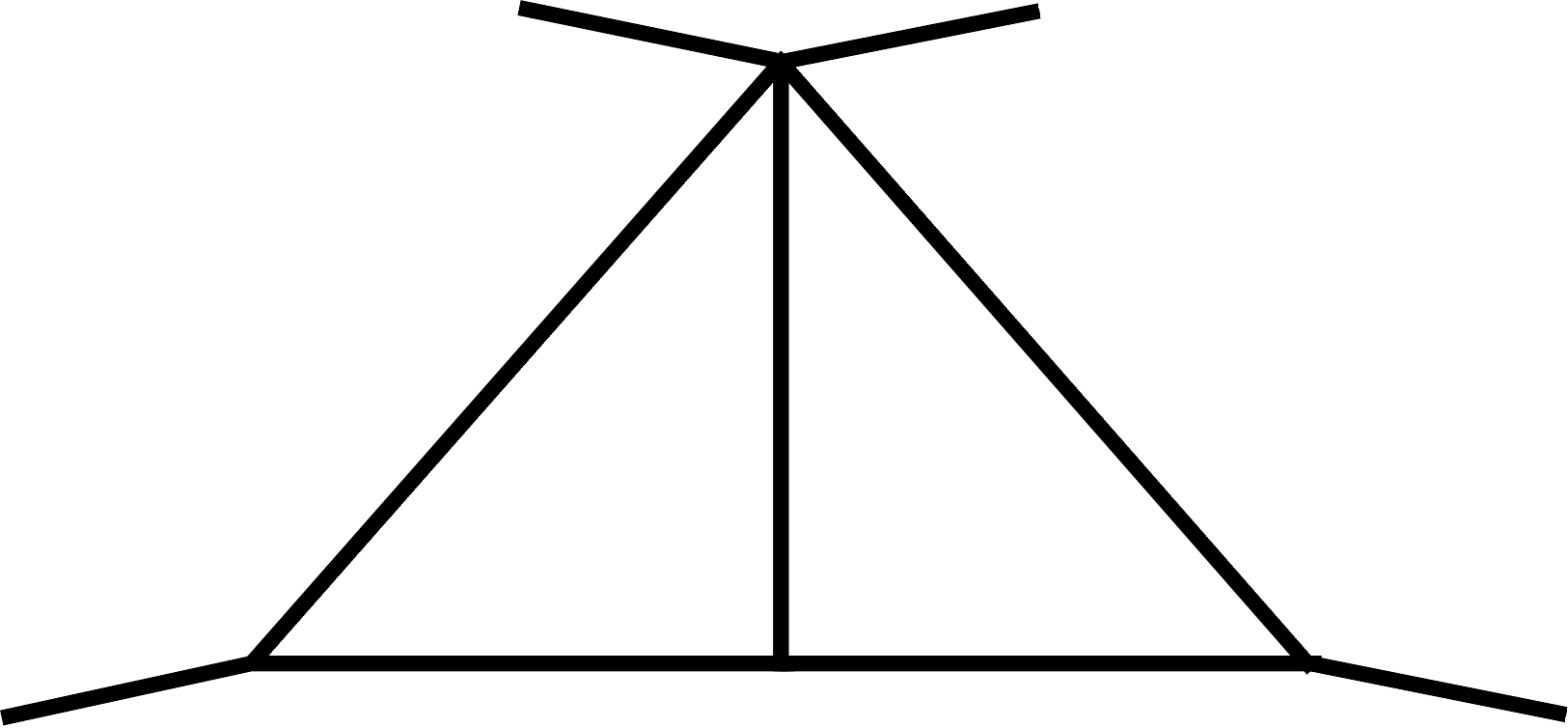} \end{minipage}\!\! - 6 s\,\,\,\begin{minipage}{50px} \includegraphics[width=1.5cm]{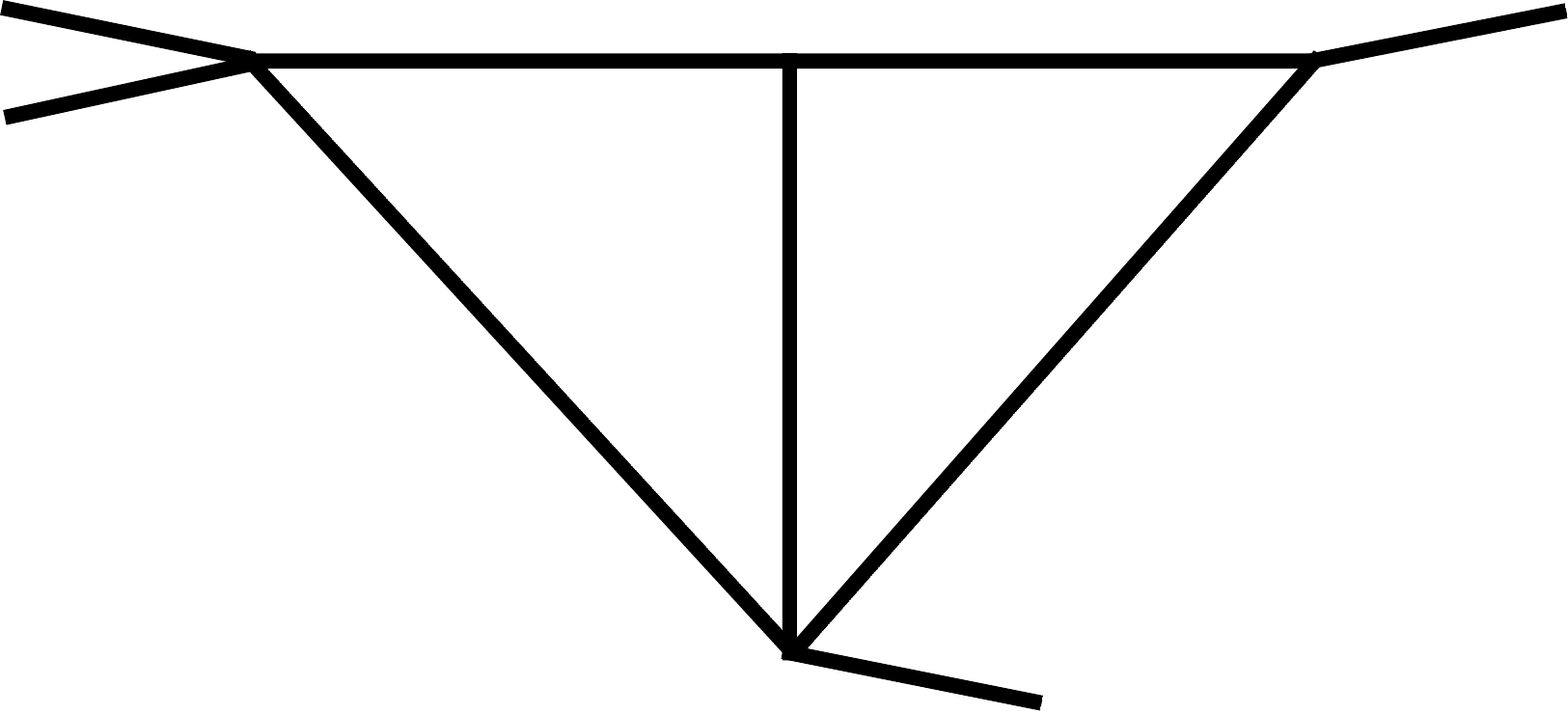} \end{minipage}  \hspace{-0.4cm} \,\, -  s^2 t\,\, \begin{minipage}{50px} \includegraphics[width=1.5cm]{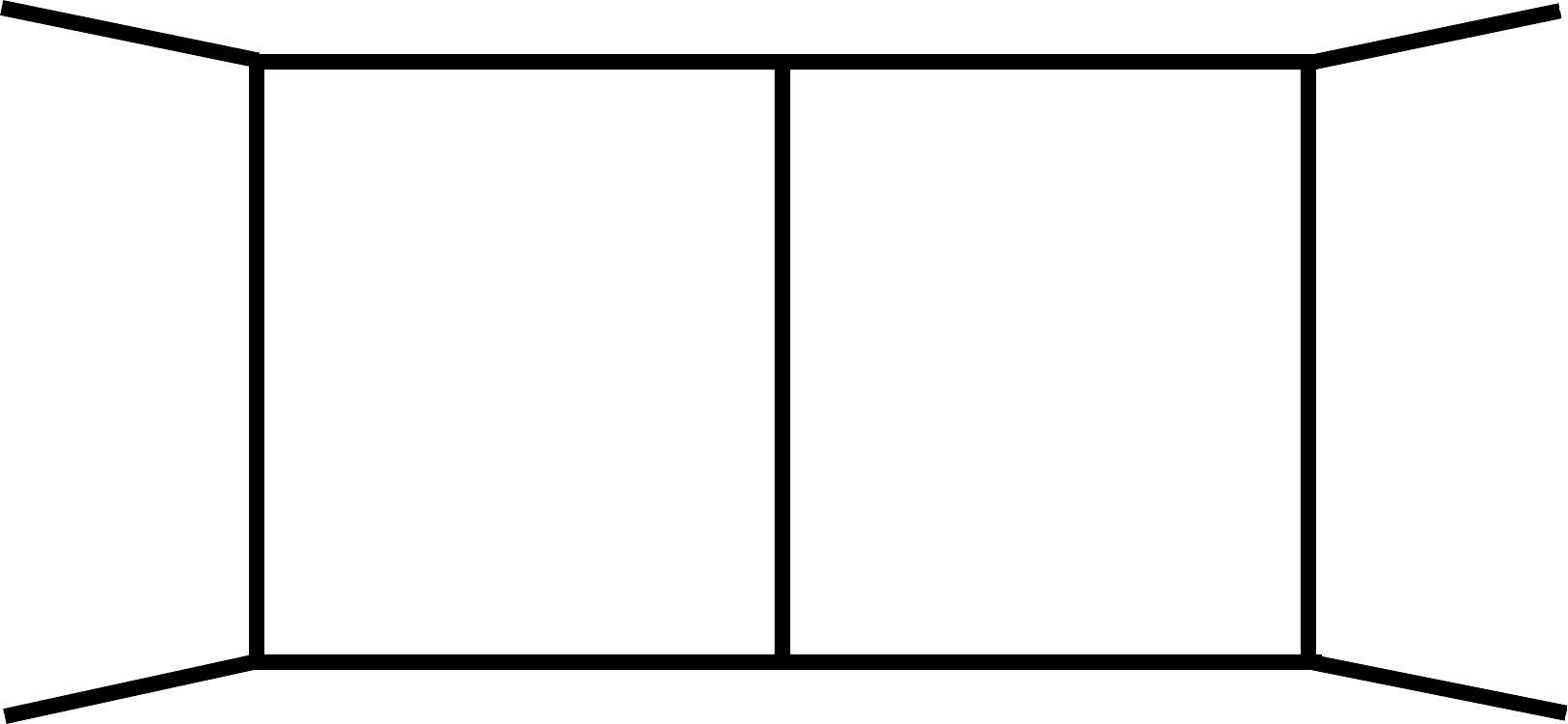} \end{minipage} + \non \\[0.2cm] &- s^2 \,\, \begin{minipage}{50px} \includegraphics[width=1.5cm]{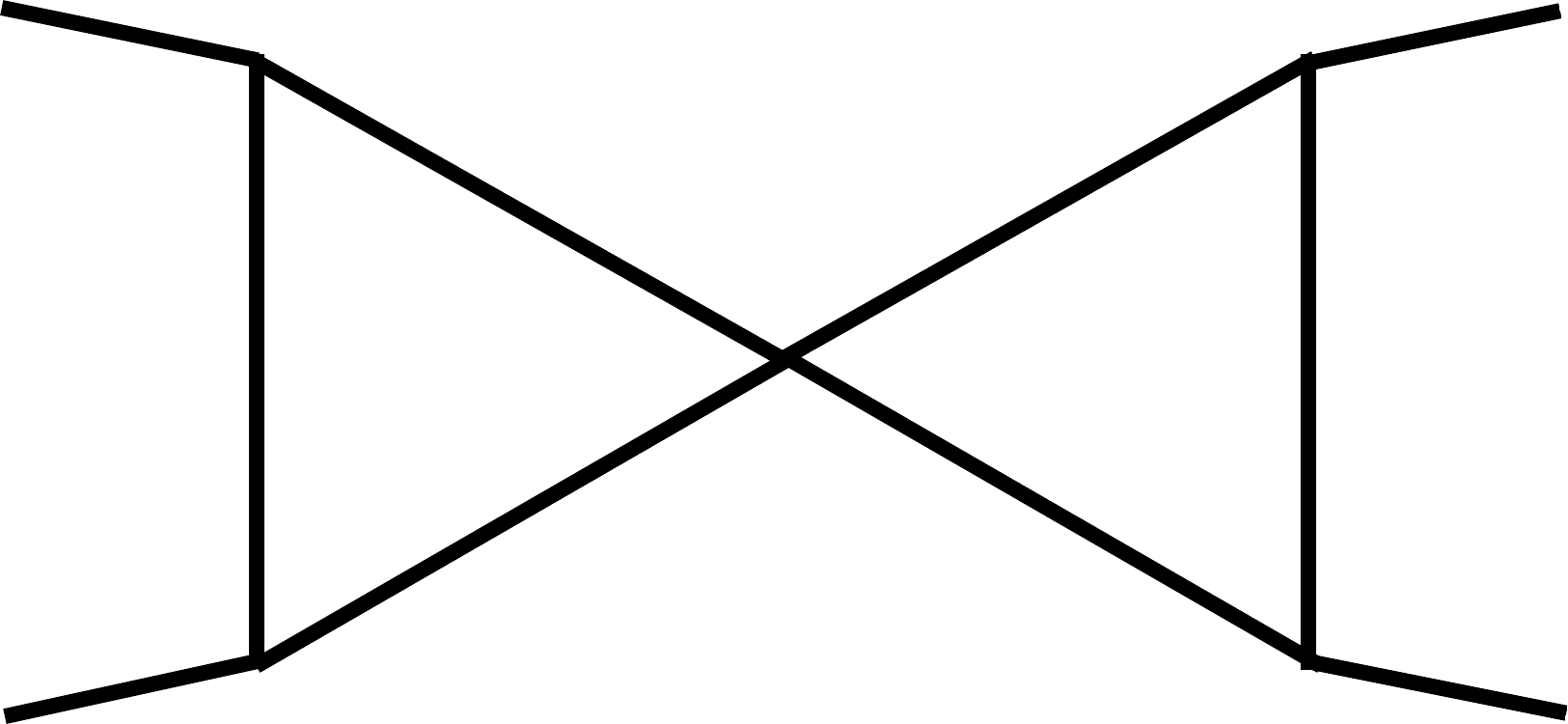} \end{minipage} + 2s^2 \,\, \begin{minipage}{50px} \includegraphics[width=1.5cm]{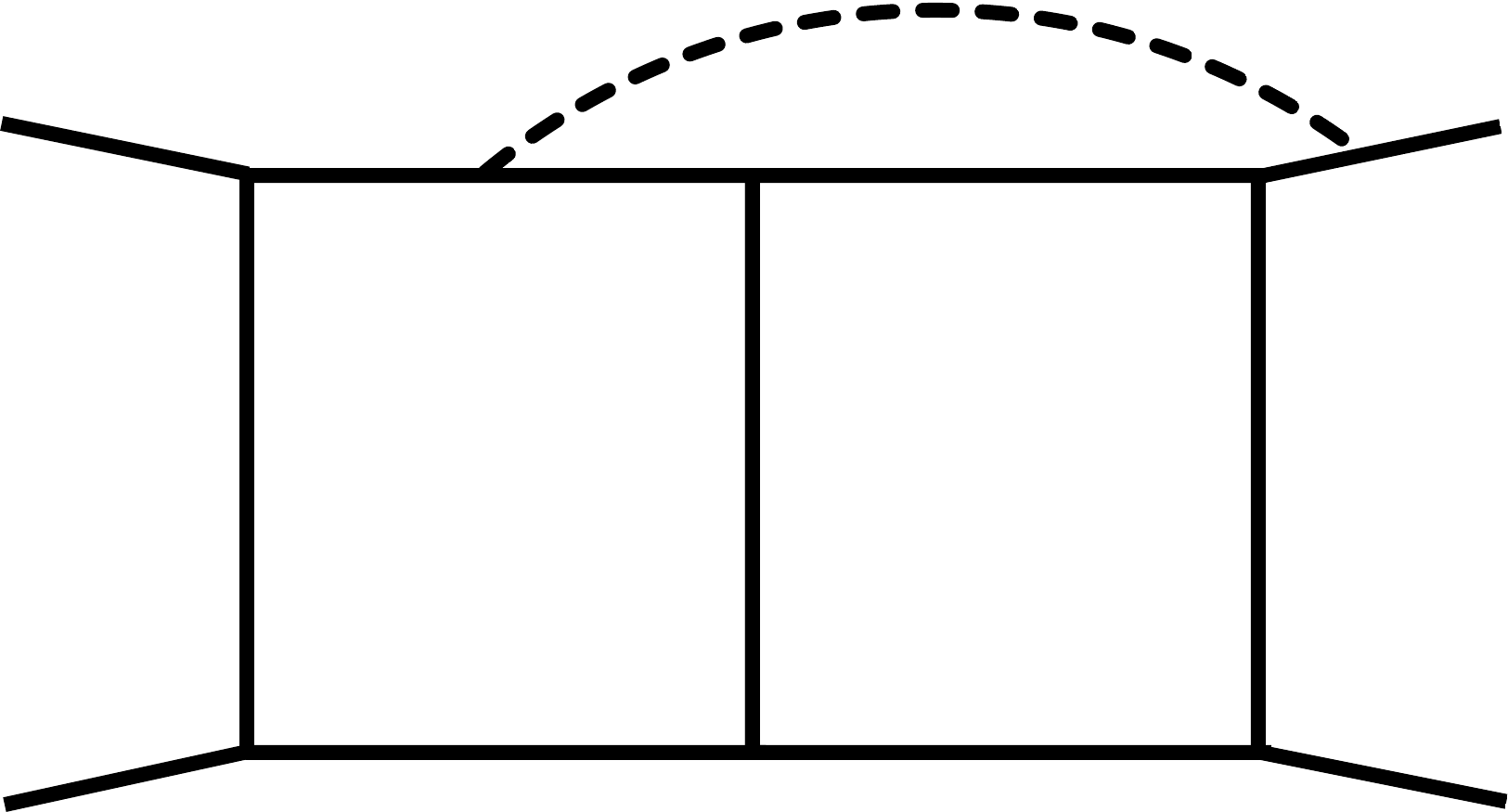} \end{minipage}+ 3 s (k+p_3)^2 \,\,\begin{minipage}{50px} \includegraphics[width=1.5cm]{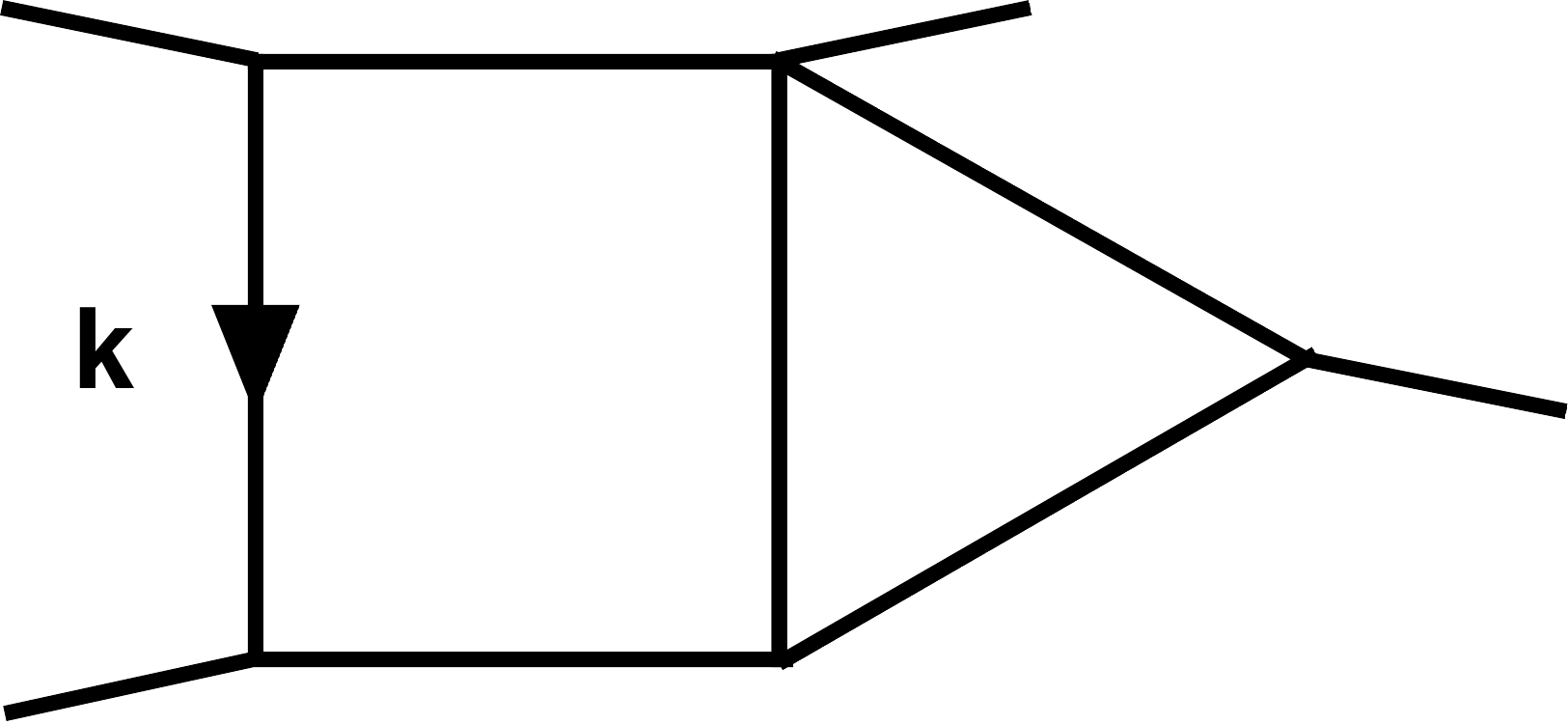} \end{minipage} 
\end{align}
The diagram (i) drawn in Fig.\ref{pic2loops} contributes
\begin{align}
(i)&=  \frac{1}{4} g^6 N^2 \,\,\begin{minipage}{50px} \includegraphics[width=1.8cm]{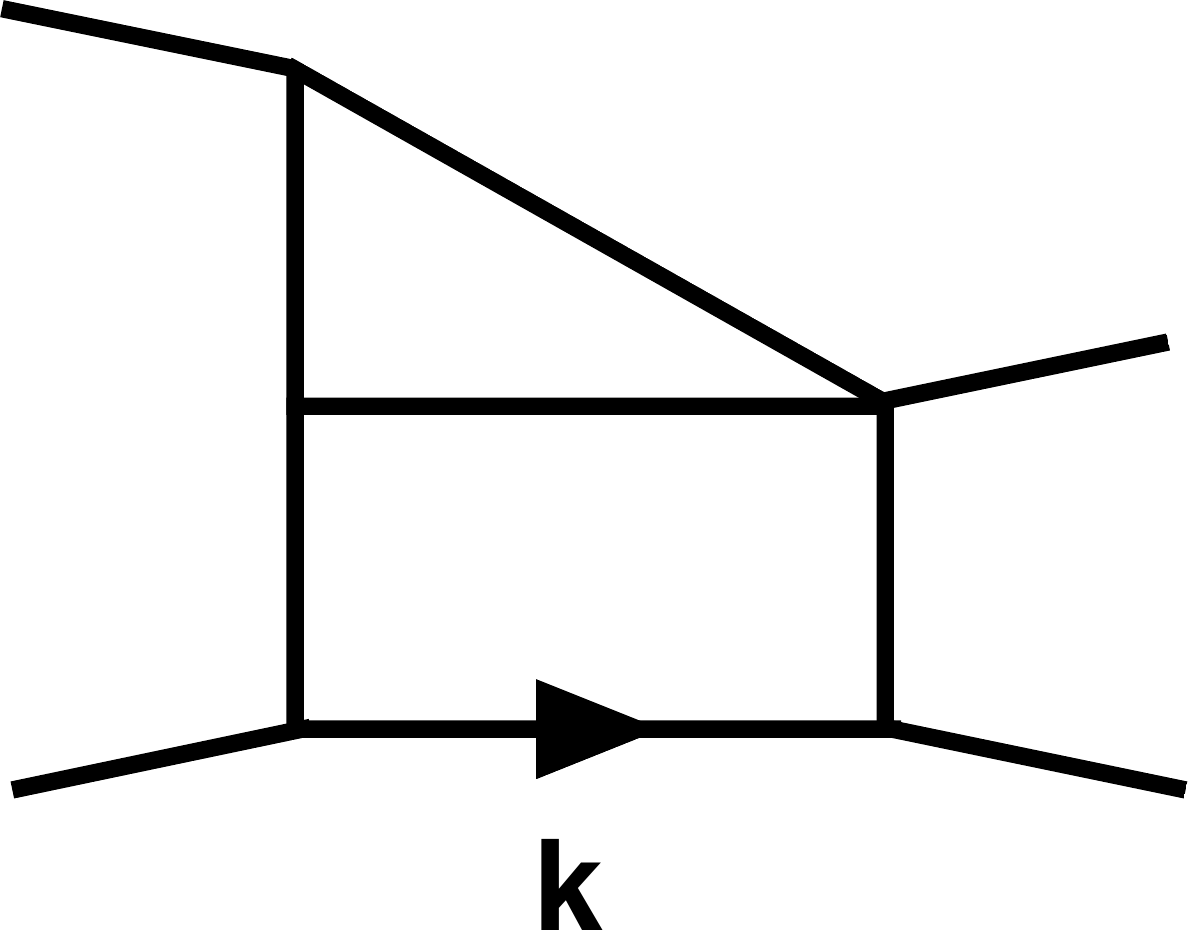} \end{minipage}\,\, \bigg(s k^2-\Tr(k p_3 p_1 p_2)  \bigg) \non
\end{align}
We need to consider four diagrams of type (i) which, after expanding the traces and using symmetries of the integrals, can be combined to give
\begin{align}
(i)  = &  \,\,\big(s+t\big) \,\,\,\begin{minipage}{50px} \includegraphics[width=1.5cm]{Idiag.pdf} \end{minipage} + s\,\,\,\begin{minipage}{50px} \includegraphics[width=1.5cm]{Dtria1.pdf} \end{minipage} -t (k+p_4)^2\,\,\,\begin{minipage}{50px} \includegraphics[width=1.5cm]{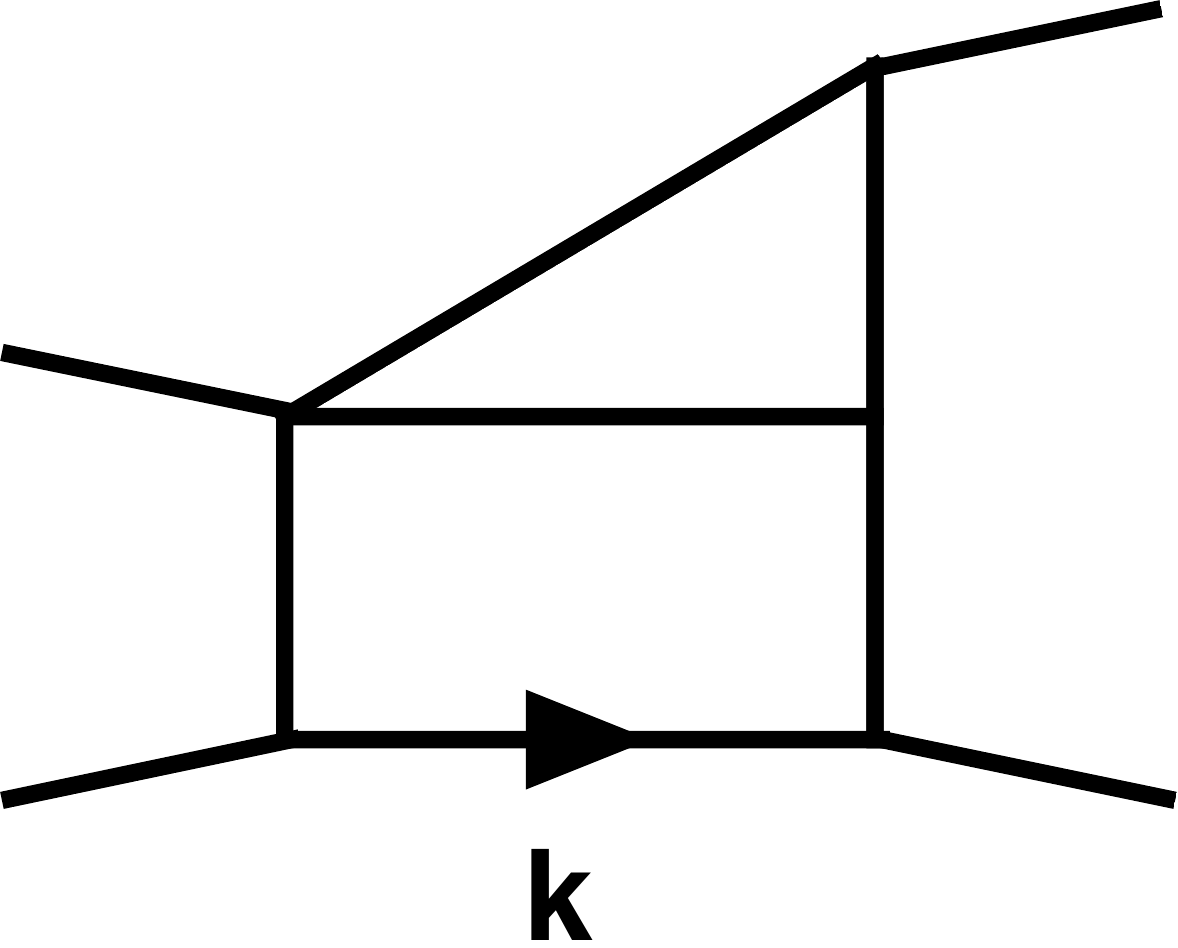} \end{minipage} 
\end{align}
Now we compute one--loop vertex insertions of diagrams (l) and (m). 
\begin{align}
(l)&=  \frac{1}{4} g^6 N^2 \,\,\begin{minipage}{50px} \includegraphics[width=1.8cm]{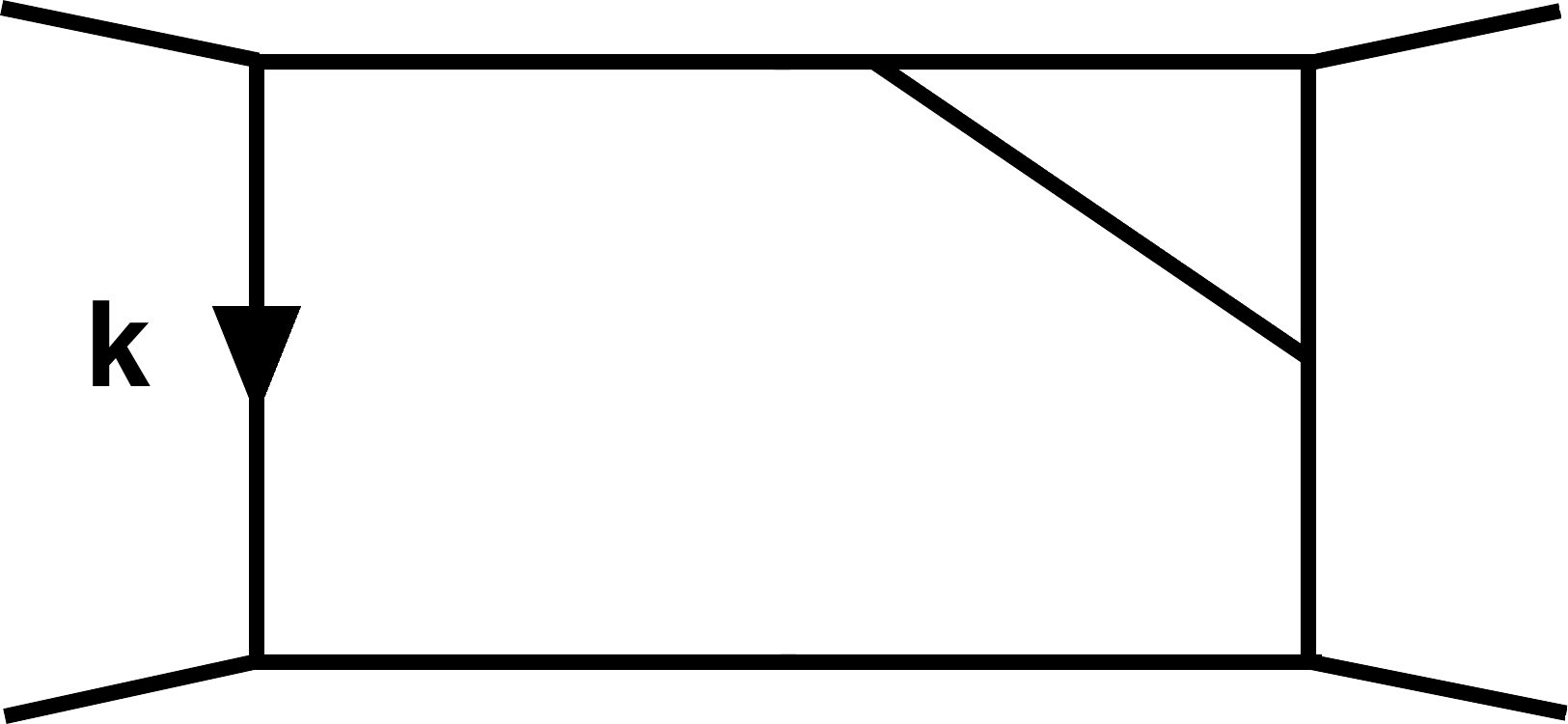} \end{minipage}\,\, \bigg( k^2 \Tr(k p_1 p_2 p_4) + \Tr(k p_1 k p_2 p_1 p_4) - (k+p_1+p_4)^2 \Tr(p_3 p_1 p_2 k) +
\non \\  
& \,\,\,\,\,\,\,\,+ 2 s (k+p_1 +p_4)^2 k \cdot (k-p_2)  + \Tr(p_4 (k+p_1) p_3 p_1 p_2 k) +s \Tr(p_4 (p_1+k)(k-p_2)k) \bigg)  \non \\
(m) & =  g^6 N^2 \,\, \begin{minipage}{50px} \includegraphics[width=1.8cm]{Ivermo2.pdf} \end{minipage}\,\, \bigg( s (k+p_1)^2 (k+p_1 +p_4)^2 + (k+p_1+p_4)^2 \Tr(k p_1 p_2  p_4)\bigg) \non \end{align}
The total contribution coming from one--loop vertex insertions is given by four diagrams of type (l) and four diagrams of type (m). The overall results can be expressed as
\begin{align}
(l)  = &  -t \,\,\,\begin{minipage}{50px} \includegraphics[width=1.5cm]{Dtria3.pdf} \end{minipage} - s\,\,\,\begin{minipage}{50px} \includegraphics[width=1.5cm]{Dtria1.pdf} \end{minipage} + 4s\,\,\,\begin{minipage}{50px} \includegraphics[width=1.5cm]{Dtria4.pdf} \end{minipage} +\non \\ &  + t(k+p_4)^2 \,\,\,\begin{minipage}{50px} \includegraphics[width=1.5cm]{Verhousevec.pdf} \end{minipage} - s(k+p_3)^2 \,\,\,\begin{minipage}{50px} \includegraphics[width=1.5cm]{Horhouse2vec.pdf} \end{minipage}  \\
(m)  = & \,\, 2(s-t)\,\,\,\begin{minipage}{50px} \includegraphics[width=1.5cm]{Idiag.pdf} \end{minipage} +2 t \,\,\,\begin{minipage}{50px} \includegraphics[width=1.5cm]{Dtria3.pdf} \end{minipage} +4s\,\,\,\begin{minipage}{50px} \includegraphics[width=1.5cm]{Dtria4.pdf} \end{minipage} + \non \\ & -2 s(k+p_3)^2 \,\,\,\begin{minipage}{50px} \includegraphics[width=1.5cm]{Horhouse2vec.pdf} \end{minipage} 
\end{align}
The contributions (n) and  (o) come from two--loop insertions of chiral vertex and propagator corrections. They give
\begin{align}
(n) & =  g^6 N^2\,\,\begin{minipage}{50px} \includegraphics[width=1.8cm]{Iladmo.pdf} \end{minipage} \,\,\bigg( s k^2 l^2 +\frac{1}{2} s (k+p_1)^2 l^2 +\frac{1}{2} s (k-p_2)^2 l^2 \bigg) \non \\
(o) & =   0 \nonumber
\end{align}
Combining the two vertex insertion,  we then have an overall 
\begin{align}
(n)  = & \,\, 2s\,\,\,\begin{minipage}{50px} \includegraphics[width=1.5cm]{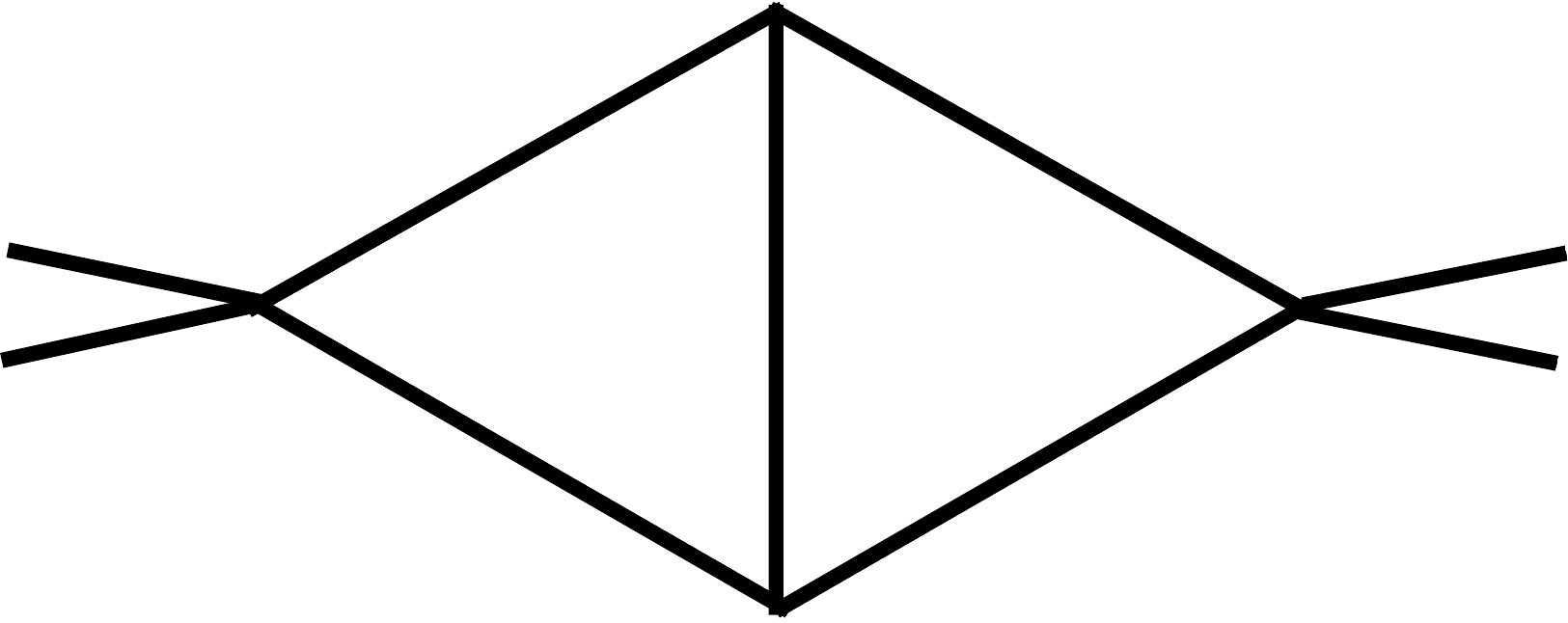} \end{minipage}+2s\,\,\,\begin{minipage}{50px} \includegraphics[width=1.5cm]{Dtria4.pdf} \end{minipage}  \\ 
(o)  = & \,\, 0 \label{2fin} 
\end{align}
It is easy now to sum up pictorially the contributions (\ref{2init})--(\ref{2fin}) and get the final result (we omit the overall $g^6 N^2$ factor)
\begin{align}
\mc A^{(2)}_1(q \tilde{q} \bar{\tilde{q}} \bar q) = & \,-2 s \,\,\,\begin{minipage}{50px} \includegraphics[width=1.5cm]{Dtria1.pdf} \end{minipage}\!\!+4 s \,\,\,\begin{minipage}{50px} \includegraphics[width=1.5cm]{Dtria4.pdf} \end{minipage}  \!\!- s^2 t \begin{minipage}{50px} \includegraphics[width=1.5cm]{Ilad.pdf}\end{minipage}-\!\!s t^2 \begin{minipage}{50px} \includegraphics[width=1.5cm]{Verdbox.pdf} \vspace{0.2cm}\end{minipage} + \non  \\[0.2cm] 
& - s^2\,\, \begin{minipage}{50px} \includegraphics[width=1.5cm]{Ibra.pdf} \end{minipage} + 4st \,\, \begin{minipage}{50px} \includegraphics[width=1.5cm]{Verhouse.pdf}\end{minipage}\!\!\!  + 2s^2 \,\, \begin{minipage}{50px} \includegraphics[width=1.6cm]{Ivlad.pdf} \end{minipage}  + 2 s \,\,\,\begin{minipage}{50px} \includegraphics[width=1.5cm]{Dtria6.pdf} \end{minipage}
\end{align}
We now have expressed the contributions coming from super Feynman diagrams in terms of scalar integrals and scalar integrals with irreducible numerators. Each of these  integrals can now be expanded on the basis of two--loop master integrals using the formulas (\ref{mexpini})--(\ref{mexpfin}) in Appendix \ref{append}. The full amplitude can then be written as the following linear combination on the master integral basis
\begin{align} \label{2master}
\mc A^{(2)}_1(q \tilde{q} \bar{\tilde{q}} \bar q)  = & - s^2 t \, \begin{minipage}{50px} \includegraphics[width=1.8cm]{Ilad.pdf} \end{minipage} - s t^2 \, \begin{minipage}{50px} \includegraphics[width=1.5cm]{Verdbox.pdf} \end{minipage} + 2 s^2  \,\begin{minipage}{50px} \includegraphics[width=1.8cm]{Ivlad.pdf} \end{minipage}\,\,\, \,\, - 24\, a t \,\,\, \begin{minipage}{50px} \includegraphics[angle=90,width=0.95cm]{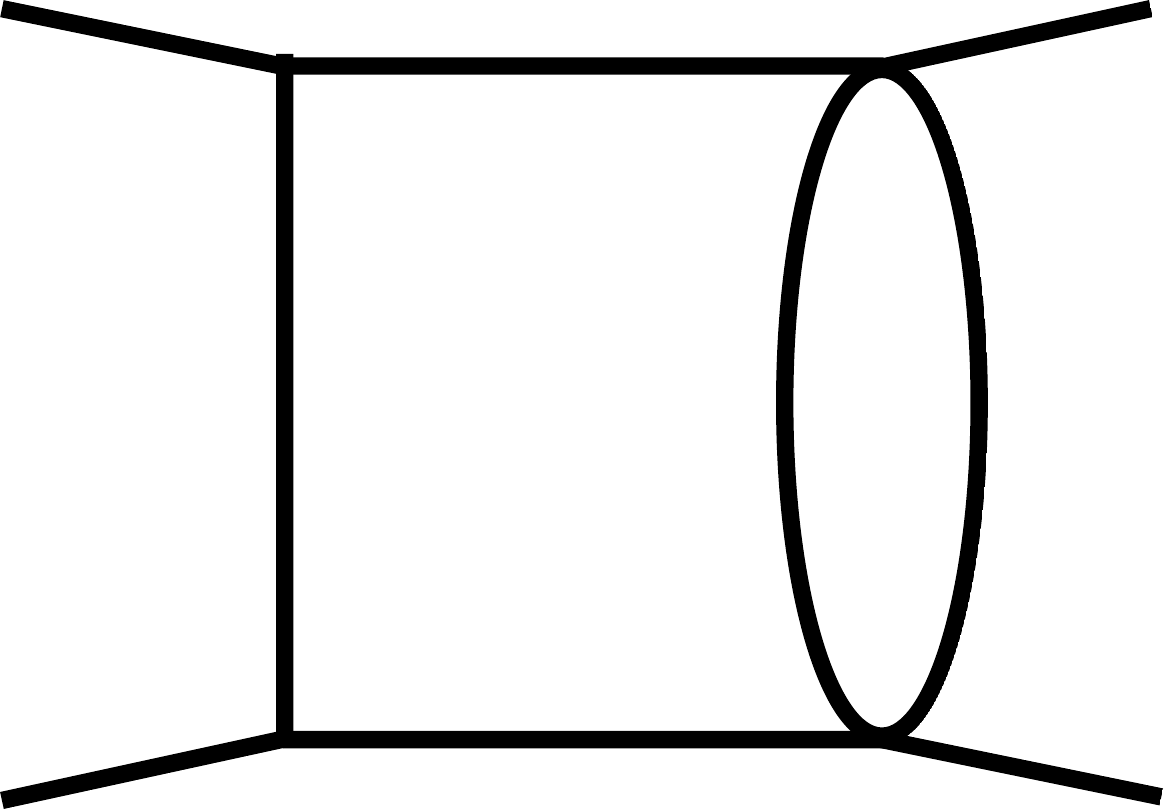} \end{minipage}  \hspace{-0.5cm} + \non \\[0.2cm] &-4\,(a + a^2) \,\, \begin{minipage}{50px} \includegraphics[width=1.7cm]{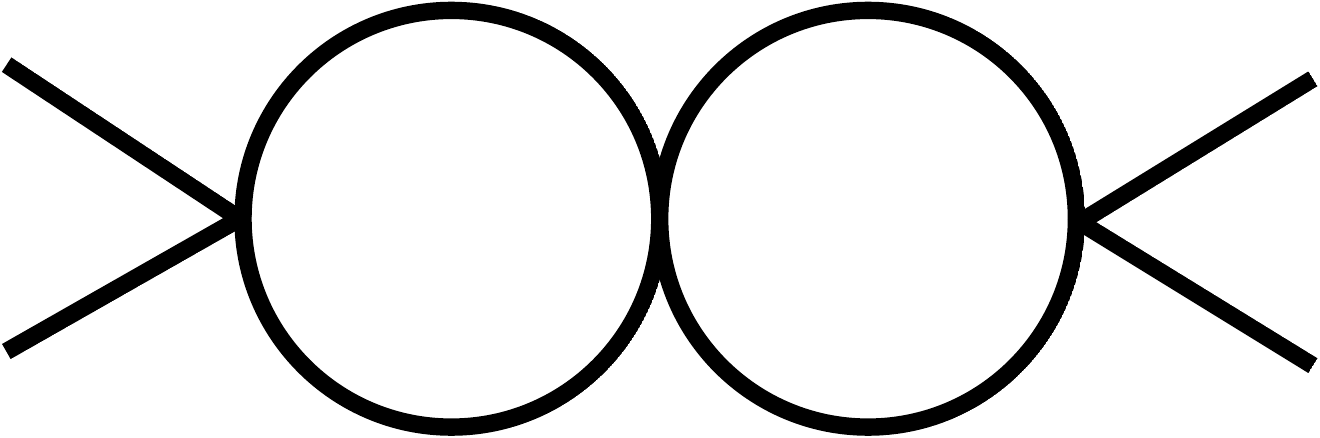} \end{minipage} \,\,\, +  \frac{4c - 18 a c }{as} \,\,\, \begin{minipage}{50px} \includegraphics[width=1.5cm]{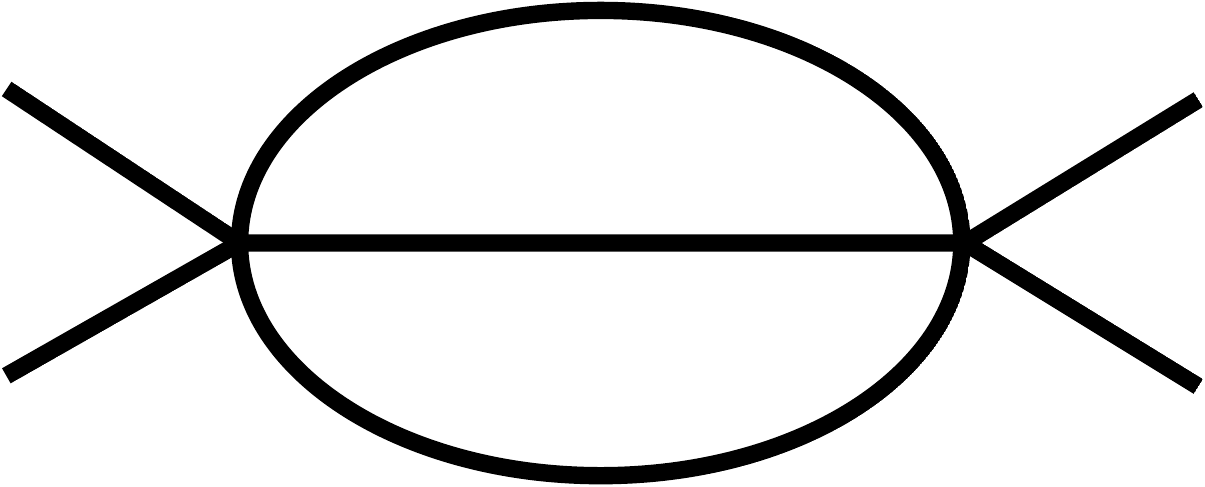} \end{minipage}- \frac{12 c}{t} \,\,\, \begin{minipage}{50px} \includegraphics[angle=90,width=0.5cm]{Isunset.pdf} \end{minipage} \hspace{-0.8cm}+  \non \\[0.2cm]   & + 4 b  \,\,\, \begin{minipage}{50px} \includegraphics[width=1.5cm]{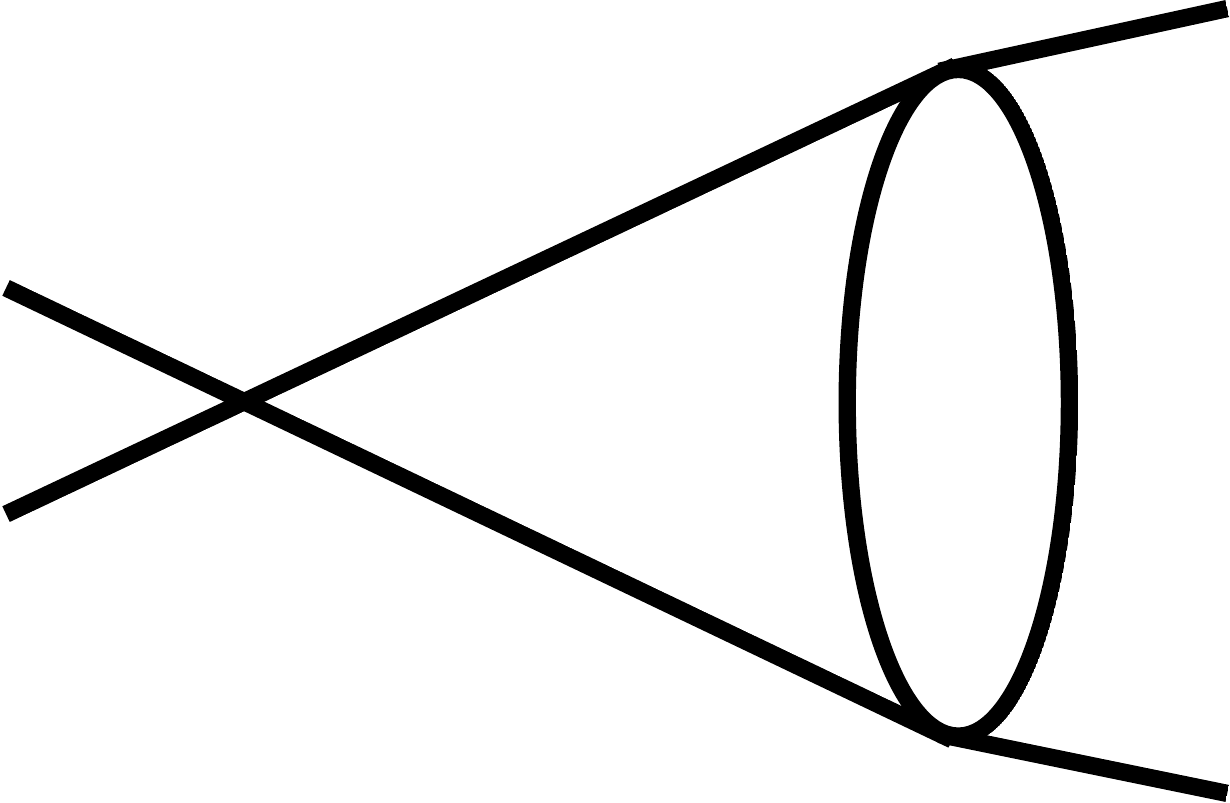} \end{minipage}  - 12 b \,\,\, \begin{minipage}{50px} \includegraphics[angle=90, width=0.95cm]{Itria.pdf} \end{minipage} \hspace{-0.6cm} +\, 12\, (s+t) \,\,\, \begin{minipage}{50px} \includegraphics[width=1.5cm]{Idiag.pdf} \end{minipage}
\end{align}
where for convenience we defined the coefficients
\begin{align} 
a & = - \frac{1-2 \e}{2\e} 
\qquad \,\,\, b  =  \frac{(1-2\e)(1-3\e)}{2\e^2}
\qquad \,\,\, c  = -\frac{(1-2\e)(1-3\e)(2-3\e)}{2\e^3} \label{mastercoeff}
\end{align}
Looking carefully at the final result (\ref{2master}) and more generally at the expressions of the Feynman integrals contributing to each single diagram given in equations (\ref{mexpini})--(\ref{mexpfin}) of Appendix \ref{append} we notice the following remarkable property. With the exception of eq. (\ref{DT6}), a given master integral in the linear combinations comes always multiplied by a fixed coefficient which is a function of the parameter $\e$. Expanding in $\e$ the product between the coefficient and the corresponding master integral it is easy to verify that, even if the master integral itself contains terms of mixed transcendentality, the product always satisfy the maximal transcendentality property.
Take for instance the sunset integral whose expansion is given in (\ref{sunset}). It is clear that to orders which are relevant for the computation it does not preserve maximal transcendentality. Nevertheless in all the expansions (\ref{mexpini})--(\ref{mexpfin}), with the exception of eq. (\ref{DT6}), it comes multiplied by the factor $c$  defined in (\ref{mastercoeff}). Expanding the product we obtain
\begin{equation}
\!-\frac{(1-2\e)(1-3\e)(2-3\e)}{2\e^3} \, \, \begin{minipage}{50px} \includegraphics[width=1.5cm]{Isunset.pdf} \end{minipage} \!\!\!=   \frac{e^{-2 \gamma_{E} \epsilon}}{(4 \pi)^{4-2\epsilon}}\frac{1}{s^{-1+2\e}}\bigg[ \frac{1}{4 \e^4} - \frac{\pi^2}{24 \e^2} - \frac{8 \zeta(3)}{3 \e} -\frac{19 \pi^4 }{480} \,\bigg] \non
\end{equation}
which respects the maximal transcendentality principle. It is clear from this analysis that the contribution to the final result (\ref{2master}) coming from the integral given in (\ref{DT6}) is the only one that breaks the maximum transcendentality principle. 
The horizontal and vertical ladders in the first line of (\ref{2master}) are the only integrals respecting dual conformal symmetry, which is thus broken for the full amplitude as expected.

Inserting in (\ref{2master}) the expansions in $\e$ of the master integrals of Appendix (\ref{2loopmasters}) and dividing by the tree level amplitude, the final result can be cast in the following form
\begin{framed}
\begin{align}  \label{2loopsfinal} 
 & \mc M^{(2)}_1(q \tilde{q} \bar{\tilde{q}} \bar q) \,\, =  \,\,\frac{e^{-2 \gamma_{E} \epsilon}}{(4 \pi)^{4-2\epsilon} t^{2\e}} \,\, \bigg[\,\, \frac{2}{\e^4} \,- \, \frac{1}{\e^2} \bigg(\frac{13 \pi^2}{6} + 2 \ln^2 x \bigg)  - \, \frac{1}{\e} \,\bigg(2\pi^2 \ln(1+x)+ \frac{19}{3}\zeta(3)   \,\, + \non \\[0.2cm] & + \frac{2}{3} \ln^2x(\ln x + 3 \ln(1+x)) + 4 \ln x \Li{2}{-x}-4\Li{3}{-x} \bigg) \,+\,  4 (3 \ln x - \ln (1 + x) ) \zeta(3) \,\,+   \non \\[0.1cm]
& +\frac{23 }{60} \pi^4  +\frac{2}{3}\pi^2\ln x\ln(1+x) - (\pi^2 + \ln^2 x)\ln^2(1+x) + 4 S_{2,2}(-x) - 4 \ln x S_{1,2}(-x)  \,\, + \non \\[0.2cm]
&+ \!4 \ln(1+x) \Li{3}{-x}\!+\frac{2}{3}(\pi^2\!\!-6\ln x\ln(1+x))\Li{2}{-x}\!+\frac{1}{6} (4\pi^2+\ln^2 x) \ln^2 x - 12 \zeta(3) \bigg]  
\end{align}
\end{framed}
In order to get to the compact expression  (\ref{2loopsfinal}) we had to combine (generalized) polylogarithms with ones with inverse arguments using the identities listed in Appendix (\ref{appident}). 
An important consistency check of our result is given by the fact that we exactly reproduce the exponential structure of the infrared poles which is expected for the scattering of massless particles in general gauge theories \cite{Catani:1998bh,Sterman:2002qn}. In fact,  if we read the poles from the following general exponential expression for the amplitude
$$
 \mc M^{(2)} = \frac{f_1(\e)}{2} (\mc M^{(1)}(\e))^2 + \lambda f_2(\e) \mc M^{(1)}(2 \e)
$$
we exactly reproduce our result with the choice
\begin{align} 
f_1 & = 1  \label{scaling1} \\
f_2 & = -2 \zeta(2) - \e 14 \zeta(3) + \mc O (\e^2) \label{scaling2}
\end{align}
which is very reminiscent of the corresponding expansions for the scaling functions  in the $\mc N =4$ SYM, where it was shown \cite{Anastasiou:2003kj} that $f_1=1$ and $f_2= - \zeta(2) - \e \zeta(3) + \mc O (\e^2)$. Nevertheless the finite part of the amplitude does not exhibit exponential behaviour {\it \`a la} BDS as in the $\mc N =4$ SYM case.

\section{Conclusions}
In this paper we have computed four--point  scattering amplitudes in $\mc N=2$ SCQCD up to two loops in the Veneziano limit. 
At one loop we have considered all possible four--point scalar amplitudes, which can be classified into three independent sectors, according to the color representation of the external particles.

In the adjoint sector, namely when the external particles are four scalar fields in the adjoint representation of the gauge group, 
we found, in agreement with \cite{Glover:2008tu}, that the one--loop result (\ref{adj1}) coincides with the one for the  planar $\mc N=4$ SYM gluon scattering amplitude. So in this sector the one--loop result is dual conformal invariant and respects the maximum transcendentality principle.
It would be important to go further and check if this connection with $\mc N=4$ SYM survives at higher loops. 
In fact the difference between the expectation value of light--like Wilson loops evaluated in $\mc N=4$ SYM and in $\mc N=2$ SCQCD was computed and it was found a non vanishing term at three loops \cite{Andree:2010na}. It would be interesting to check if this deviation is present also for scattering amplitudes, in order to understand if the Wilson loop/scattering amplitude duality is valid in this context. We left this computation  for a future work \cite{progress}.

We presented new results outside the adjoint sector. In the mixed sector, with two adjoint scalar fields and a quark/antiquark pair as external particles, we computed the one--loop scattering amplitude given in eq. (\ref{bifu}). 
In the fundamental sector, with only fundamental fields as external particles, we presented results up to two loops, given in eq. (\ref{1fund}) and eq. (\ref{2loopsfinal}). 
In these sectors we found that the loop results are not dual conformal invariant and do not respect the maximum transcendentality principle at two loops. It would be interesting to check the behaviour of higher loop corrections.

To check our two--loop result we analyzed its IR structure in the dimensional regularization scheme. 
We found that the IR structure is in agreement with the exponentiation of IR divergences which is predicted by the general analysis of \cite{Catani:1998bh,Sterman:2002qn}, with scaling functions (\ref{scaling1}) and (\ref{scaling2}) which are reminiscent of those of $\mc N=4$ SYM. 
In contrast with planar scattering amplitudes in $\mc N=4$ SYM, we found that the finite part of our two--loop result does not exponentiate, as suggested by the lack of dual conformal symmetry.

It would be interesting to extend our work to higher point scattering amplitudes. 
Finally, the generalization of our computations to the two parameter family of interpolating superconformal theories which connects $\mc N=2$ SCQCD to the $Z_2$ orbifold of  
$\mc N=4$ SYM through a parameter continuous deformation might lead to important insights into the connection with $\mc N=4$ SYM.

\section{Acknowledgements}
AM thanks Marco Bianchi, Matias Leoni, Silvia Penati and Joao Pires for useful discussions. This work has been supported in part by INFN and MPNS-COST Action MP1210  ``The String Theory Universe".

\appendix

\section{Superspace conventions} \label{appena}

We work in four dimensional Euclidean ${\cal N}=1$ superspace described by coordinates $(x^\mu, \th^\alpha, \thb^{\,\dot{\b}} )$, $\a, \dot{\b} =1,2$  following the conventions of  \cite{Gates:1983nr}. Spinor indices are raised and lowered following  NW-SE conventions
\begin{eqnarray}
  \psi^\alpha=C^{\alpha\beta}\psi_\beta  \qquad \psi_\alpha=\psi^\beta C_{\beta\alpha}\qquad  \bar{\psi}^{\dot\alpha}=C^{\dot\alpha\dot\beta}\bar{\psi}_{\dot\beta}  \qquad \bar{\psi}_{\dot\alpha}=\bar{\psi}^{\dot\beta} C_{\dot\beta\dot\alpha}
\end{eqnarray}
using the matrices 
\begin{eqnarray}
  C^{\alpha\beta} = C^{\dot\alpha\dot\beta} =\left(\begin{array}{cc} 0 & i \\ -i & 0 \end{array}\right) \qquad
  C_{\alpha\beta} = C_{\dot\alpha\dot\beta}=\left(\begin{array}{cc} 0 & -i \\ i & 0 \end{array}\right)
\end{eqnarray}
which obey the relations 
\begin{eqnarray}
  C^{\alpha\beta}\, C_{\gamma\delta}
  &=\delta^\alpha{}_\gamma\, \delta^\beta{}_\delta - \delta^\alpha{}_\delta\, \delta^\beta{}_\gamma \quad .
\end{eqnarray}
Spinors are contracted according to
\begin{eqnarray}
  \psi\chi & =\psi^\alpha\, \chi_\alpha=\chi^\alpha\, \psi_\alpha=\chi\psi
  \qquad
  \psi^2=\frac{1}{2}\, \psi^\alpha\, \psi_\alpha  \\
\bar\psi \bar\chi & =\bar\psi^{\dot\alpha}\, \bar\chi_{\dot\alpha}=\bar\chi^{\dot\alpha}\, \bar\psi_{\dot\alpha}=\bar\chi\bar\psi
  \qquad
  \bar\psi^2=\frac{1}{2}\, \bar\psi^{\dot\alpha}\, \bar\psi_{\dot\alpha} 
\end{eqnarray}
Vector and bispinor indices are exchanged using Pauli matrices $(\sigma_{\mu})^{\alpha\dot\beta}$
\begin{eqnarray}
\begin{array}{lll}
\mathrm{coordinates:}\qquad   &  x^\mu=(\sigma^\mu)_{\alpha\dot\beta}\, x^{\alpha\dot\beta}
    &\qquad
     x^{\alpha\dot\beta} = \frac{1}{2}\, (\sigma_\mu)^{\alpha\dot\beta}\, x^\mu \\
 \mathrm{derivatives:}\qquad    &   \partial_\mu=\frac{1}{2}\, (\sigma_\mu)^{\alpha\dot\beta}\,
    \partial_{\alpha\dot\beta} & \qquad
    \partial_{\alpha\dot\beta}=(\sigma^\mu)_{\alpha\dot\beta}\, \partial_\mu \\
\mathrm{fields:}\qquad   &
      V_\mu=\frac{1}{\sqrt{2}}\, (\sigma_\mu)^{\alpha\dot\beta}\, V_{\alpha\dot\beta}
      & \qquad V_{\alpha\dot\beta}=\frac{1}{\sqrt{2}}\, (\sigma^\mu)_{\alpha\dot\beta}\, V_\mu
\end{array}
\end{eqnarray}
The Pauli matrices satisfy 
\begin{equation}
 \sigma_{\mu} ^{\phantom{\mu}\alpha\dot\beta}\,\sigma^{\nu}_{\phantom{\nu}\alpha\dot\beta} = 2 \,\delta_{\mu}^{\phantom{\mu}\nu} \qquad \qquad \sigma_{\mu}^{\phantom{\mu}\alpha\dot\beta} \,\sigma^{\mu} _{\phantom{\mu}\gamma\dot\eta}= 2 \,\delta^{\alpha}_{\phantom{\alpha}\gamma} \delta^{\dot\beta}_{\phantom{\dot\beta}\dot\eta}
\end{equation}
which imply the following trace identities
\begin{eqnarray}
  &\tr(\sigma^\mu\, \sigma^\nu)
  &\equiv - (\sigma^\mu)^{\alpha \dot\beta}\,(\sigma^\nu)_{\alpha\dot\beta}
  = - 2\,  g^{\mu\nu} \\
  &\tr(\sigma^\mu\, \sigma^\nu\, \sigma^\rho\, \sigma^\eta) &\equiv (\sigma^\mu)^{\alpha \dot\beta}\, (\sigma^\nu)_{\gamma\dot\beta}\,
  (\sigma^\rho)^{\gamma\dot\delta}\, (\sigma^\eta)_{\alpha\dot\delta} = \non\\
  &&= 2\, (g^{\mu\nu}\, g^{\rho\eta}-g^{\mu\rho}\, g^{\nu\eta}+g^{\mu\eta}\, g^{\nu\rho})
\end{eqnarray}
It follows that the scalar product of two vectors can be rewritten as
 \begin{eqnarray}
p\cdot k \, = \, \frac12 \,   p^{\alpha\dot\beta}\,k_{\alpha\dot\beta}
\end{eqnarray}
Superspace covariant derivatives are defined as
\begin{equation}
  D_\a = \pa_\a + \frac{i}{2}\,  \thb^{\dot\b}\,  \pa_{\a\dot\b}
  \qquad , \qquad \Db_{\dot\a} = \bar\pa_{\dot\a}
  + \frac{i}{2}\,  \th^{\b}\,  \pa_{\b\dot\a}
\end{equation}
and satisfy the anticommutator  $\{D_\a ,\,  \Db_{\dot\b}\} = i\,  \pa_{\a\dot\b}$. 

Integration in superspace is defined as $\int d^2 \theta = \frac{1}{2}\partial^\a\partial_\a$, $\int d^2 \bar\theta = \frac{1}{2}\bar\partial^{\dot\a}\bar\partial_{\dot\a}$ and $\int d^4 \theta = d^2 \theta d^2 \bar\theta$, such that we can project to components using
\begin{eqnarray} \label{proje}
 \int d^4x \,d^2 \theta & & = \int    d^4x  \,D^2 |_{\theta=\bar{\theta}=0} \qquad \int d^4x \,d^2 \bar\theta = \int    d^4x \bar D^2 |_{\theta=\bar{\theta}=0} \non \\
\int d^4x \,d^4 \theta & & = \int    d^4x \, \bar D^2 D^2 |_{\theta=\bar{\theta}=0}
\end{eqnarray}
We define the components of the chiral superfields as  
$$
\Phi (x,\theta) = \phi(x) + \theta^\a \psi_\a(x) + \theta^2 F(x) \qquad Q (x,\theta) = q(x) + \theta^\a \lambda_\a(x) + \theta^2 G(x)
$$
with a similar expansion for $\tilde{Q}$ and  corresponding expressions for the conjugated superfields. We will need for our purpose only the lowest components of the scalar multiplets, which can be readily obtained by projections using (\ref{proje}).

The superfields $V$ and $\Phi$ are in the adjoint representation of the gauge group, that is  $V = V_a T^a$ and $\Phi = \Phi_a T^a$, where $T^a$ are the $SU(N)$ generators. When needed, adjoint indices will be denoted by $a,b,c,\dots$. The superfields  $\tilde{Q}$ and $Q$ are respectively in the fundamental and antifundamental representation of $SU(N)$. When needed, (anti)fundamental indices will be denoted by the letters $i,j,k,\dots$ The generators of $SU(N)$ obey
$$
(T^a)_i^{\phantom{i}j} (T^a)_k^{\phantom{k}l} = \delta_i^l \delta_k^j - \frac{1}{N} \delta_i^j \delta_k^l
$$
and are normalized as $\Tr( T^a T^b )= \delta^{ab}$.

\section{Feynman rules}\label{appenb}

We report here the action with interaction terms expanded up to the order needed in the computation
\begin{align}\label{expS} 
S  = & \int\d^4x \d^4\theta \bigg[  \tr \left( \bar{\Phi} \Phi + g \, \bar{\Phi}V \Phi -  g \, \bar{\Phi} \Phi V + \frac{g^2}{2} \bar{\Phi} \Phi V V +\frac{g^2}{2} \bar{\Phi} V V\Phi  - g^2 \bar{\Phi}V \Phi V \right) +\non \\ 
 & \,\,+ \,\, \bar{Q}^{ I}Q_{ I}  + \tilde{Q}^{ I} \bar{\tilde{Q}}_{ I} + g \,\bar{Q}^{ I} V Q_{ I}  - g \,\tilde{Q}^{ I} V \bar{\tilde{Q}}_{ I} + \frac{g^2}{2} \,\bar{Q}^{ I}V VQ_{ I}  +  \frac{g^2}{2} \,\tilde{Q}^{ I}V V\bar{\tilde{Q}}_{ I}  \,\, + \non \\ 
& \,\, +  \tr \left(-\frac{1}{2} V \Box V + \frac{g}{2}\,V\{D^{\alpha} V, \bar{D}^2 D_{\alpha}V\} + \frac{g^2}{8}\,[V, D^{\alpha} V] \bar{D}^2 [V,D_{\alpha}V]  \,\, + \right. \non \\  &  \left. \,\, + \,\,  \bar{c}'c  -  c' \bar{c}  + \frac{g}{2}\, (c'+\bar{c}')[V,c+\bar{c}]+  \frac{g^2}{12}\, (c'+\bar{c}')[V,[V,c-\bar{c}]] \,\right) \,\bigg]  \,\, + \non \\
& \,\, + i g \int\d^4x\d^2\theta  \ \tilde{Q}^{ I} \Phi Q_{ I}
-i g \int\d^4x\d^2\bar\theta  \  \bar Q^{ I}\bar\Phi \bar{\tilde{Q}}_{ I} \,\, + \,\, \dots
\end{align}
The Feynman rules for the propagators are 
\begin{align}
\langle V^aV^b \rangle & = \,
\begin{minipage}{60px}
 \includegraphics[width=1.5cm]{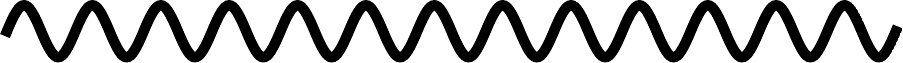}
\end{minipage}\hspace{-0.6cm} =\,  -   \frac{\delta(\theta_1-\theta_2)}{p^2}\, \delta^{a b}  \\
\langle \Phi^a\bar{\Phi}^b \rangle & = \,
\begin{minipage}{60px}
 \includegraphics[width=1.5cm]{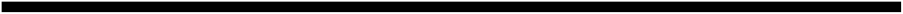}
\end{minipage}\hspace{-0.6cm} = \,\, \frac{\delta(\theta_1-\theta_2)}{p^2}\, \delta^{a b}  \\
\langle Q_{i I} \bar{Q}^{j J}  \rangle = \langle \bar{\Q}_{i I} \tilde{Q}^{j J}  \rangle & = \,
\begin{minipage}{60px}
 \includegraphics[width=1.5cm]{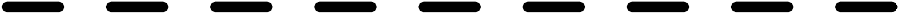}
\end{minipage}\hspace{-0.6cm} = \,\, \frac{\delta(\theta_1-\theta_2)}{p^2}\, \delta_{i}^{j}  \delta_{I}^{J}  \\
\langle \bar{c}'^a c^b \rangle = - \langle c'^a \bar{c}^b \rangle & = 
\,\begin{minipage}{60px}
 \includegraphics[width=1.5cm]{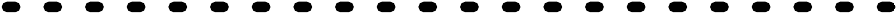}
\end{minipage}\hspace{-0.6cm} =\,  \frac{\delta(\theta_1-\theta_2)}{p^2}\, \delta^{a b} 
\end{align} 
Vertices can be immediately read from the expanded action (\ref{expS}). We work  directly with traces in color space selecting only the Feynman diagrams which contribute to the chosen color configuration (see \cite{Dixon:1996wi} for a review of the method).

\section{Vertex and propagator insertions}\label{appenc}

We discuss here the one-- and two--loop insertions of corrected propagators and vertices which are relevant for our computation. At one loop the vector and matter field propagators receive corrections from the following  diagrams \vspace{0.2cm}
\begin{equation} \label{appc1}
\begin{minipage}{12cm}\hspace{0.2cm}
\includegraphics[scale=0.7]{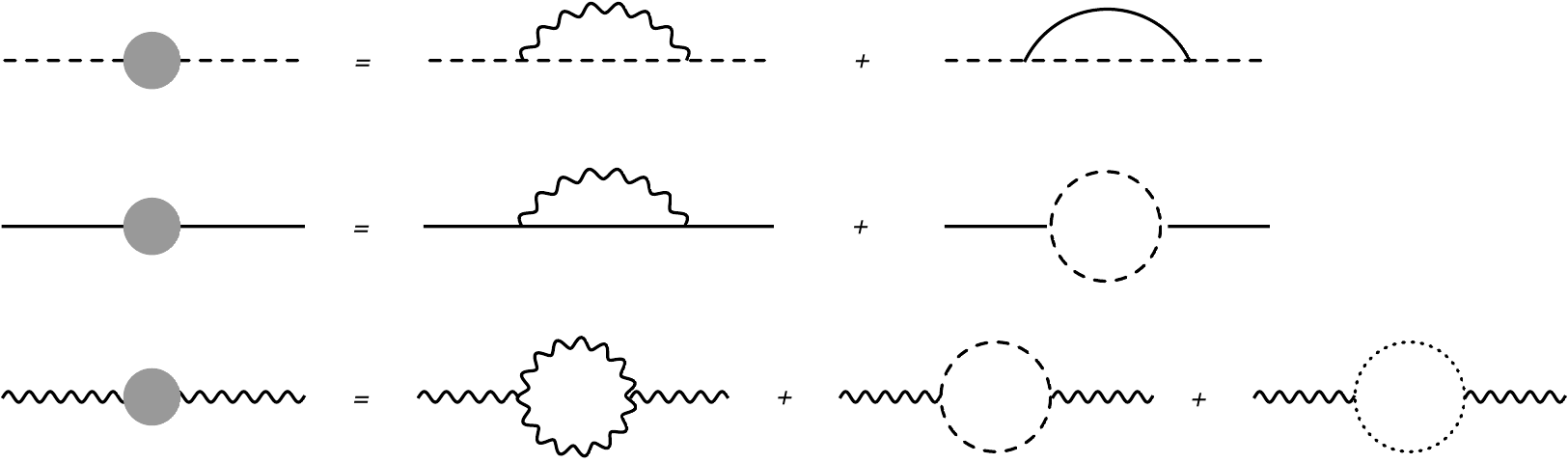}
\end{minipage} \vspace{0.2cm }\end{equation}
It is easy to show that  the two diagrams which correct the $Q$ propagator in the first line of (\ref{appc1}) cancel each other for every value of $N_f$. Also in the case of the $\Phi$ and $V$ propagators the corrections exactly sum up to zero, as can be shown by comparison with $\mathcal{N}=4$ SYM theory. Indeed, the propagators of $\mathcal{N}=2$ SCQCD are corrected by the same diagrams correcting the corresponding propagators of  $\mathcal{N}=4$ SYM, provided we substitute the adjoint matter loops with  fundamental ones. The matter loops in the two theories yield the same result when $N_f=2 N$. 

Due to finiteness theorems for the superpotential, the finiteness of the scalar chiral propagators is enough to ensure conformal invariance at one--loop order, where the condition $N_f=2 N$ has been used non-trivially as shown above. 
At two loops the quantum corrections to the chiral superfield propagators vanish \cite{Pomoni:2011jj}: \vspace{0.3cm}
\begin{equation} \label{2loopschir}
\begin{minipage}{50px}
\includegraphics[width=1.7cm]{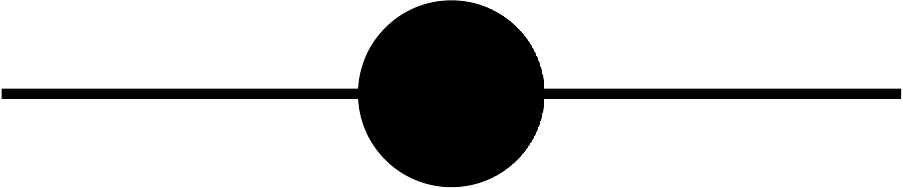}
\end{minipage} = 0
\end{equation}
The finiteness of (\ref{2loopschir}) is enough to ensure that the theory is conformal at two loops. 

In our computations we also need the following one--loop vertex corrections (an overall factor $g^3 N$ is stripped out)
\begin{align}
& \begin{minipage}{12cm} \hspace{0.2cm}
\includegraphics[scale=0.36]{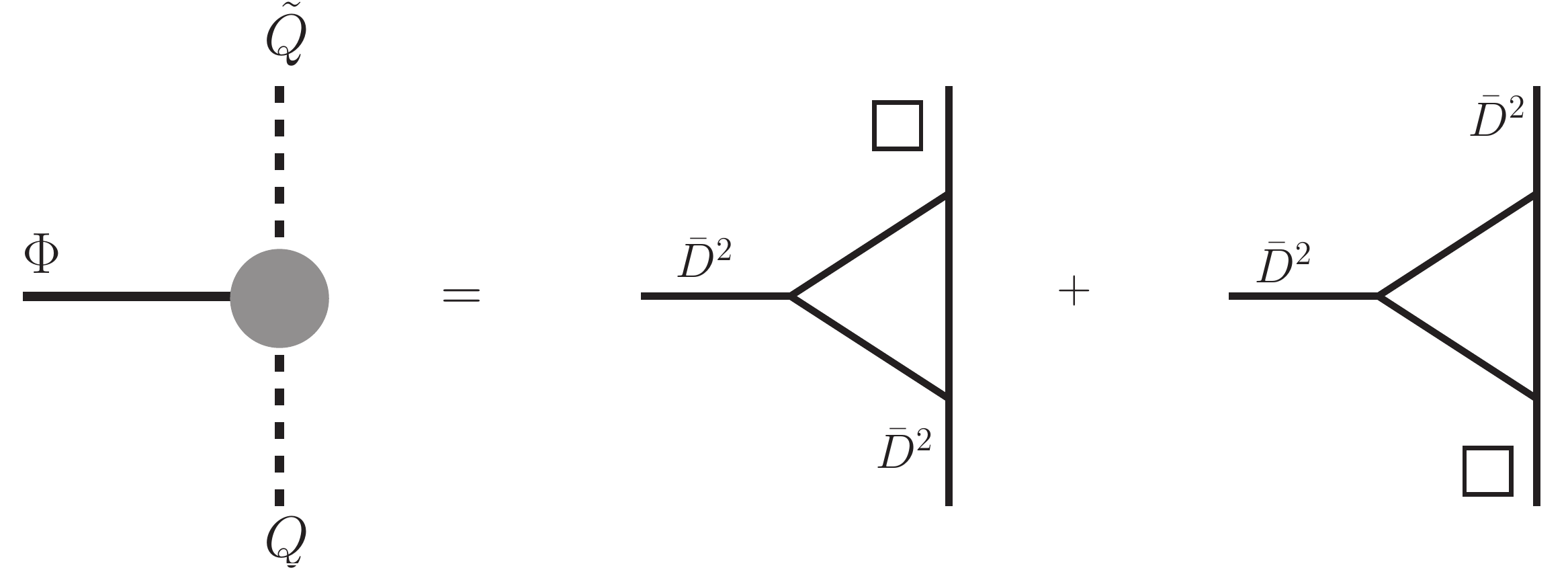}
\end{minipage} \\[0.1cm]
& \begin{minipage}{12cm} \hspace{0.2cm}
\includegraphics[scale=0.36]{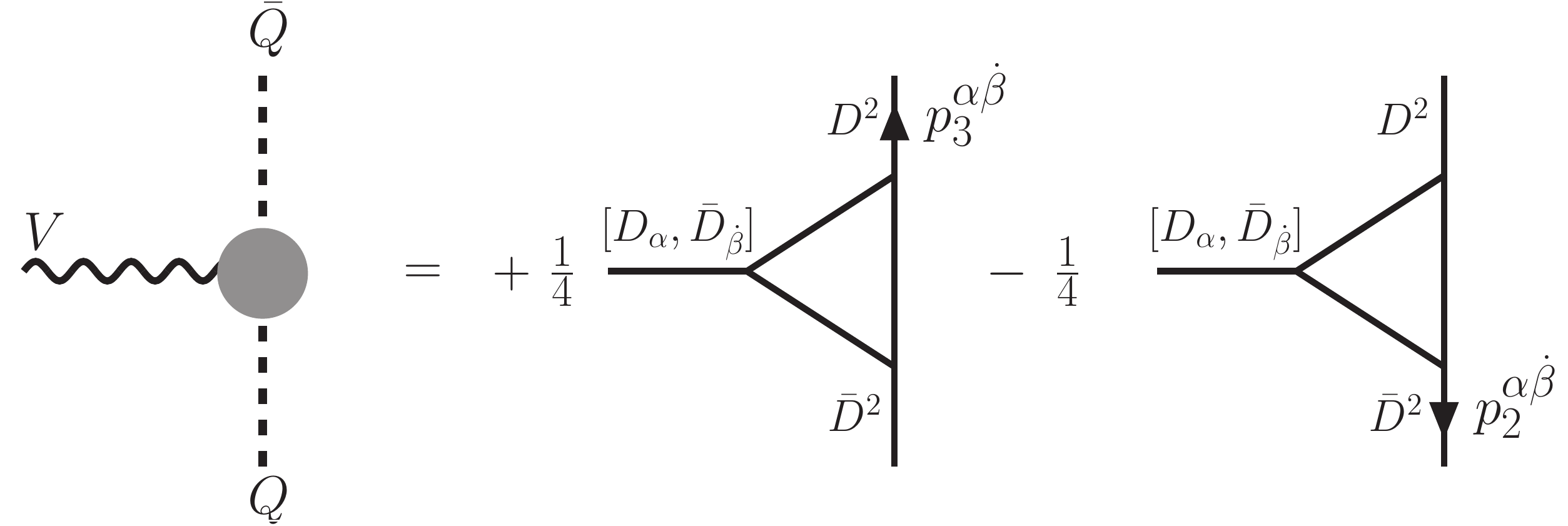}
\end{minipage} \\[0.1cm]
& \begin{minipage}{12cm} \hspace{0.2cm}
\includegraphics[scale=0.36]{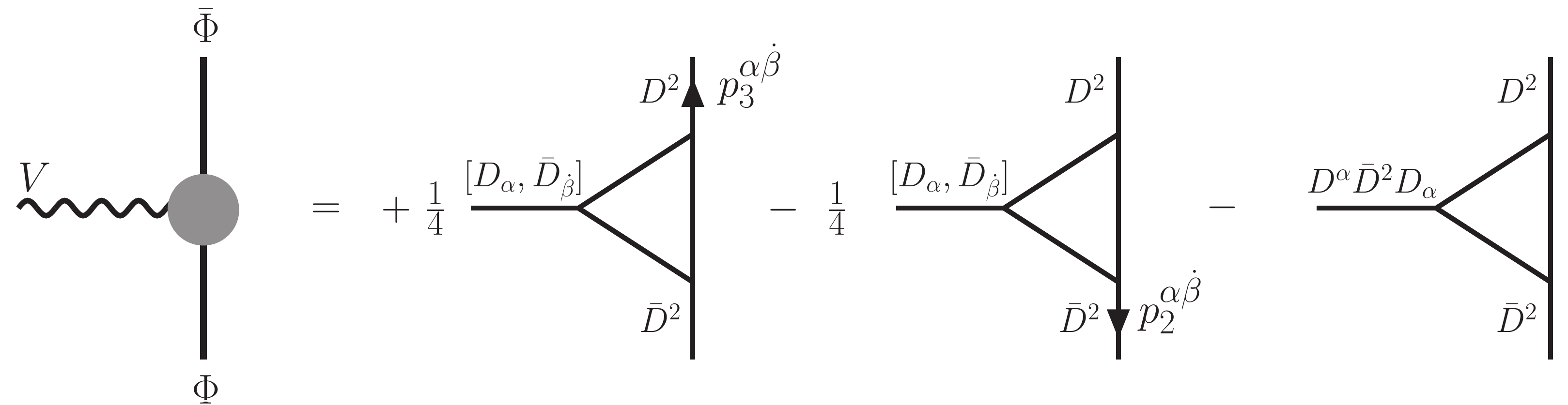}
\label{FigVertex}
\end{minipage} 
\end{align} 
where we depicted only diagrams which contribute at leading color order. In the first correction above we omitted also an overall factor $i$.
We follow here the representation of \cite{Pomoni:2011jj}, where the diagrams are evaluated off-shell and the expansions can be directly inserted in higher loop supergraph structures.  
At two loops we need the chiral vertex correction (we omit an overall $i g^5 N^2$)
\begin{align}
&  \hspace{-0.1cm}
\includegraphics[scale=0.32]{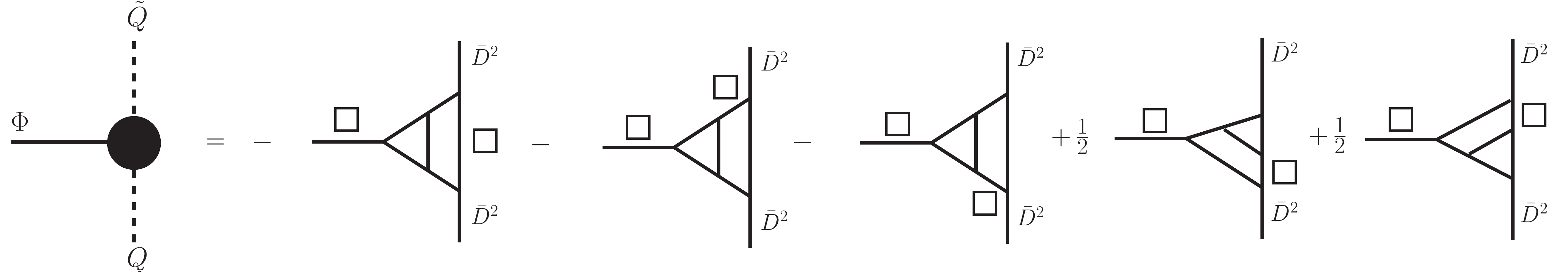}
\end{align}
In this case the full off-shell expansion of the vertex gets lenghty \cite{Pomoni:2011jj}. We report here only the terms giving a non-vanishing contribution to the amplitude in Fig.\ref{pic2loops}, namely the ones which survive after taking on-shell momenta for the external fields $Q$ and $\tilde{Q}$.

\section{Integrals}\label{append}

In this Appendix we discuss how we deal with the Feynman integrals resulting from D-algebras.  At one--loop order  computations are easy enough to directly reduce each integral into a sum of box and triangle scalar integrals by hand.  At two loops, we find convenient to express the integrals in terms of a set of known master integrals by using the {\it Mathematica} package {\bf FIRE} \cite{Smirnov:2008iw}. In (\ref{2loopmasters}) we introduce the two--loop master integral basis and the explicit expressions in dimensional regularization. In (\ref{2lexpans}) we list the expansions of the amplitude integrals on the master basis.  External momenta in the pictures are always labeled counterclockwise starting from the upper left corner of the  and are always put on the mass shell  $p_1^2= p_2^2= p_3^2= p_4^2=0$. 

\subsection{One-loop integrals}

At one--loop order all the tensor and scalar amplitude integrals can be reduced by completing the squares and integration by parts to a combination of the two following integrals: the scalar triangle
\begin{align}
 I_{\mathrm{triangle}}(s) & =  \begin{minipage}{50px} \includegraphics[width=1.5cm]{Itria1.pdf} \end{minipage} = \int \frac{\d^d k}{(2\pi)^d} \frac{1}{k^2 (k-p_3)^2 (k+p_4)^2} = \non \\[0.2cm] &=
 \frac{\G(3-d/2) \G^2(d/2-2)}{s^{3-d/2} (4\pi)^{d/2} \G(d-3)} \label{IntTriangle} = \frac{e^{-\gamma_E \e} }{s^{1+\e}\ (4\pi)^{2-\e}}\left[\frac{1}{\e^2}-
\frac{\pi^2}{12} + \mc O(\e)\right]   \end{align}
and the scalar box integral
\begin{align}
 I_{\mathrm{box}} &= \begin{minipage}{50px} \includegraphics[width=1.5cm]{Ibox1.pdf} \end{minipage} = \int \frac{\d^d k}{(2\pi)^d} \frac{1}{k^2 (k-p_1)^2 (k-p_1-p_4)^2 (k+p_2)^2} 
\label{IntBox} \end{align}
This can be easily evaluated with Mellin-Barnes representations and its  $\e$ expansion reads
\begin{equation}
  I_{\mathrm{box}} = \frac{2 \,e^{-\gamma_E \e}}{s t\ (4\pi)^{2-\e}} \left[ \left(\frac{1}{s^{\e}} + \frac{1}{t^{\e}}\right)\frac{1}{\e^2} -\frac{2}{3}\pi^2 - \frac{1}{2} \ln^2 \frac{t}{s} + \mc O(\e)\right]  \end{equation}
where the dependence on the dimensional regularization mass regulator $\mu$ is understood. At one--loop order, we also need the expressions for triangle and box integrals with numerators. One can directly evaluate the needed tensor integrals  
\begin{align}
& I_{\mathrm{triangle}}^{\alpha \dot\beta} = \begin{minipage}{50px} \includegraphics[width=1.5cm]{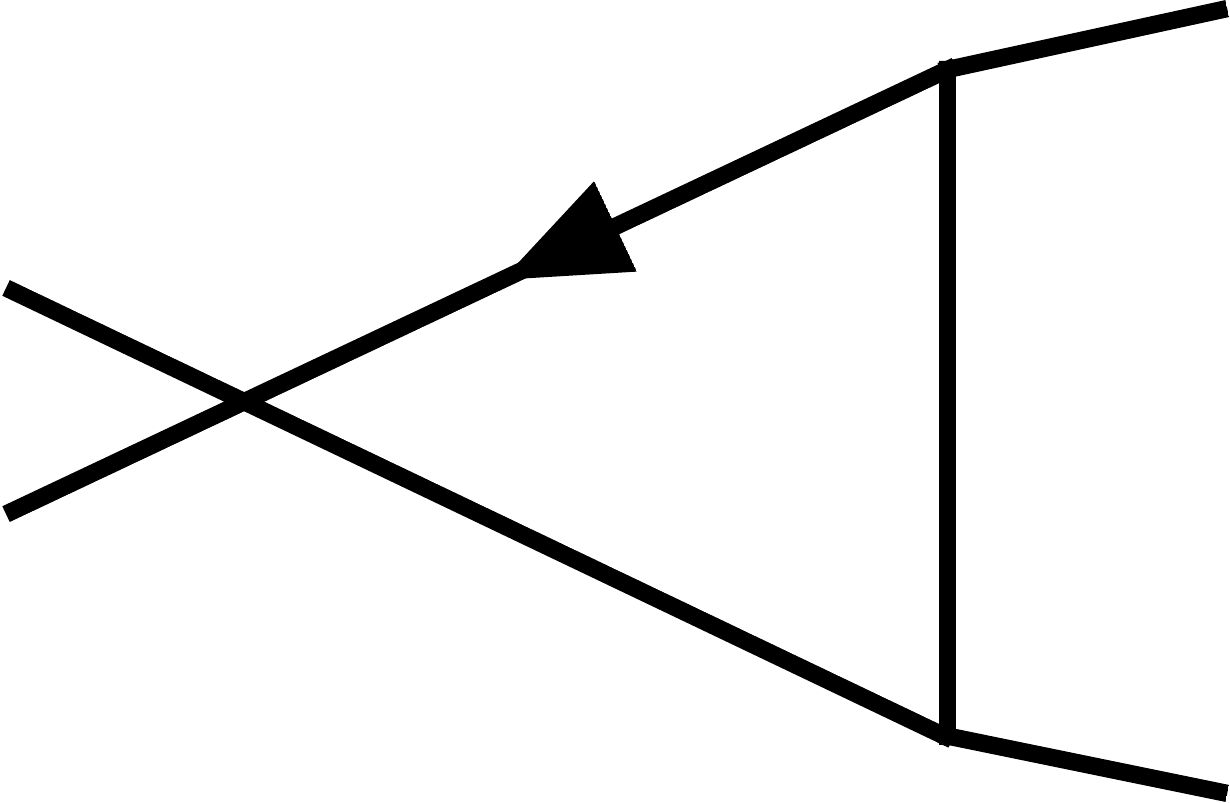} \end{minipage} = \int \frac{\d^d k}{(2\pi)^d} \frac{k^{\alpha \dot\beta}}{k^2 (k+p_4)^2 (k-p_1-p_2)^2} =\non \\[0.2cm]
& = \frac{\G(3-d/2) \G(d/2-2) \G(d/2-1)}{(4\pi)^{d/2} s^{3-d/2} \G(d-2)}(p_1+p_2)^{\alpha \dot\beta} 
 -\frac{\G(3-d/2) \G^2(d/2-2)}{(4\pi)^{d/2} s^{3-d/2} \G(d-2)} p_4^{\alpha \dot\beta} \label{IntVecTriangle} \end{align}
After the expansion in $\e$ \vspace{0.1cm}
\begin{align}
 I_{\mathrm{triangle}}^{\alpha \dot\beta} &= \frac{e^{(2-\gamma_E) \e} }{s^{1+\e}\ (4\pi)^{2-\e}} \left[-\frac{1}{\e} +\mc O(\e) \right] 
 (p_1+p_2)^{\alpha \dot\beta} + \frac{e^{(2-\gamma_E) \e} }{s^{1+\e} (4\pi)^{2-\e}} \left[-\frac{1}{\e^2} - 2 + 
 \frac{\pi^2}{12}+\mc O(\e)\right] p_4^{\alpha \dot\beta}  \end{align}
It is also useful to define the following vector--box  integral \vspace{0.1cm}
\begin{align}
 & I_{\mathrm{box}}^{\alpha \dot\beta} = \begin{minipage}{50px} \includegraphics[width=1.5cm]{Ibox1vec.pdf} \end{minipage} = \int \frac{\d^d k}{(2\pi)^d} \frac{k^{\alpha \dot\beta}}{k^2 (k-p_1)^2 (k-p_1-p_4)^2 (k+p_2)^2} \label{IntVecBox}
\end{align}
which can be evaluated as in the scalar case and expanded in $\e$ \vspace{0.1cm}
\begin{align}
  & I_{\mathrm{box}}^{\alpha \dot\beta} = \left[\frac{e^{-\gamma_E \e}}{s t\ (4\pi)^{2-\e}} \left( \frac{1}{t^{\e}}\,\frac{1}{\e^2} 
  -\frac{\pi^2}{12}\right) - \frac{\pi^2 + \ln^2 \frac{t}{s}}{2 (4\pi)^2 s (s+t)} \right] (p_1-p_2)^{\alpha \dot\beta} \,\, + \non \\
  & + \left[\frac{e^{-\gamma_E \e}}{s t\ (4\pi)^{2-\e}} \left( \frac{1}{s^{\e}}\,\frac{1}{\e^2} -\frac{7}{12}\pi^2 - \frac{1}{2} \ln^2 \frac{t}{s}\right) 
  + \frac{\pi^2 + \ln^2 \frac{t}{s}}{2 (4\pi)^2 s (s+t)} \right] (p_1+p_4)^{\alpha \dot\beta} + \mc O(\e) \non \end{align} 
 
\subsection{Two--loop master integrals} \label{2loopmasters}

At two--loops the integrals are expressed as linear combinations on the following master integral basis
\begin{eqnarray}
I_{spec} (s) & = &\begin{minipage}{50px} \includegraphics[width=1.5cm]{Ispec.pdf} \end{minipage}= \,\, \int \!\frac{\d^d k}{(2\pi)^d}  \frac{\d^d l}{(2\pi)^d} \frac{1}{k^2 (k+p_1+p_2)^2 l^2 (l-p_1-p_2)^2}\vspace{0.2cm}\\
I_{sunset} (s) & = & \begin{minipage}{50px} \includegraphics[width=1.5cm]{Isunset.pdf} \end{minipage}=  \,\, \int \!\frac{\d^d k}{(2\pi)^d}  \frac{\d^d l}{(2\pi)^d} \frac{1}{k^2 l^2 (l-k-p_1-p_2)^2} \vspace{0.2cm} \label{sunset} \\
I_{tri} (s) & = & \begin{minipage}{50px} \includegraphics[width=1.5cm]{Itria.pdf} \end{minipage} = \int \!\frac{\d^d k}{(2\pi)^d}  \frac{\d^d l}{(2\pi)^d} \frac{1}{k^2 (k+p_1+p_2)^2 l^2 (l-k+p_4)^2}  \vspace{0.2cm}\\
I_{mug} (s,t) & = &\begin{minipage}{50px} \includegraphics[width=1.5cm]{Imug.pdf} \end{minipage} = \int \!\frac{\d^d k}{(2\pi)^d}  \frac{\d^d l}{(2\pi)^d} \frac{1}{k^2 (k+p_1)^2 (k-p_2)^2 l^2 (l-k-p_1-p_4)^2} \vspace{0.2cm}\\
I_{diag}  (s,t) & = & \begin{minipage}{50px} \includegraphics[width=1.5cm]{Idiag.pdf} \end{minipage} = \int \!\frac{\d^d k}{(2\pi)^d}  \frac{\d^d l}{(2\pi)^d} \frac{1}{k^2 (k-p_2)^2 l^2 (l-p_4)^2 (l-k-p1-p4)^2}  \vspace{0.2cm}\\
I_{lad} (s,t) & = & \begin{minipage}{50px} \includegraphics[width=1.6cm]{Ilad.pdf} \end{minipage} = \!\!\int \!\frac{\d^d k}{(2\pi)^d}  \frac{\d^d l}{(2\pi)^d} \frac{1}{k^2 (k+p_1)^2 (k-p_2)^2 l^2 (l+p_3)^2 (l-p_4)^2 (l-k-p1-p4)^2}  \nonumber\\ && \nonumber\\& & \\
I_{vlad} (s,t) & = &\begin{minipage}{50px} \includegraphics[width=1.6cm]{Ivlad.pdf} \end{minipage} = \!\!\int \!\frac{\d^d k}{(2\pi)^d}  \frac{\d^d l}{(2\pi)^d} \frac{(k+p_1+p_4)^2}{k^2 (k+p_1)^2 (k-p_2)^2 l^2 (l+p_3)^2 (l-p_4)^2 (l-k-p_1-p_4)^2} \nonumber \\ &&\nonumber \\
&&
\end{eqnarray}  \vspace{0.05cm}
These can be expanded in  dimensional regularization up to the needed order
 \begin{align}
I_{spec} (s) & =  \frac{e^{-2 \gamma_{E} \epsilon}}{(4 \pi)^{4-2\epsilon} s^{2\e}} \!\left[ \frac{1}{\e^2} + \frac{4}{\e} + 12-\frac{\pi^2}{6} +\e \left(32 - \frac{2\pi^2}{3} - \frac{14}{3} \zeta(3)\right) + \right. \non \\ & \left. + \e^2 \!\left( 80 - 2 \pi^2 - \frac{7 \pi^4}{120} - \frac{56}{3} \zeta(3) \right)+  \mathcal{O}(\e^3)\right] \\ 
I_{sunset} (s) & =  \frac{e^{-2 \gamma_{E} \epsilon}}{(4 \pi)^{4-2\epsilon}}\frac{1}{s^{-1+2\e}} \left[ - \frac{1}{4\e} - \frac{13}{8} + \e \left(-\frac{115}{16} +  \frac{\pi^2}{24} \right) + 
 \e^2\left(-\frac{865}{32} + \frac{13 \pi^2}{48} + \frac{8}{3}\zeta(3)\right) +  \right. \vspace{0.2cm} \nonumber\\ &\left. 
 + \e^3 \left(-\frac{5971}{64} + \frac{115\pi^2}{96} + \frac{ 19\pi^4}{480} +  \frac{52}{3} \zeta(3) \right) + \mathcal{O}(\e^4)  \right]\vspace{0.2cm} \\
I_{tri} (s) & =   \frac{e^{-2 \gamma_{E} \epsilon}}{(4 \pi)^{4-2\epsilon}}\frac{1}{s^{2\e}} \left[ \frac{1}{2 \e^2} + \frac{5}{2 \e} + \frac{19}{2} +\frac{\pi^2}{12} + 
 \e \left(\frac{65}{2} + \frac{5 \pi^2}{12} - \frac{13}
{3}\zeta(3)\right) + \right. \vspace{0.2cm} \nonumber\\
& \left. + \e^2 \left(\frac{211}{2} + \frac{19 \pi^2}{12} - \frac{41 \pi^4}{720} - \frac{65}{3}\zeta(3)  \right)  + \mathcal{O}(\e^3) \right]\vspace{0.2cm}  \\
I_{mug} (s,t) & =   \frac{e^{-2 \gamma_{E} \epsilon}}{(4 \pi)^{4-2\epsilon}} \frac{1}{t^{\e}s^{1+\e}}\bigg\{\frac{1}{\e^3} +\frac{2}{\e^2}+\frac{1}{\e}\left(4-\frac{\pi^2}{2} \right) + 8 -\pi^2 -\frac{32}{3}\zeta(3) +\non \\ &  + \Li{3}{-x} - \ln x \Li{2}{-x} - \frac{1}{2}(\pi^2 +\ln^2 x)\ln (1+x) + \frac{1}{2}\pi^2\ln x+\frac{1}{6}\ln^3 x +\non \\ & + \e \bigg[-2 \Li{4}{-x}+(2+\ln x+\ln(1+x))\Li{3}{-x}+ \non \\ & -\left(\frac{\pi^2}{6}+2\ln x+\ln x \ln(1+x)\right)\Li{2}{-x} -S_{1,2}(-x)\ln x + S_{2,2}(-x) + \non \\& +\left(\ln x - \ln(1+x) -\frac{64}{3} \right) \zeta(3)  + 16 - 2 \pi^2 - \frac{31 }{180} \pi^4 + \non \\ & - \frac{5}{24} \ln^4 x+\frac{\pi^2}{4}(-4-2\ln x-\ln (1+x))\ln(1+x)  + \non \\ &+ \frac{1}{6} (2+\ln x +4 \ln(1+x))\ln^3 x+\frac{\pi^2}{6} \ln x (6+3\ln x+5\ln(1+x))+ \non \\ &-\frac{1}{12}(8\pi^2+3\ln(1+x)(4+2\ln x+\ln(1+x))\ln^2 x \bigg] + \mathcal{O}(\e^2) \bigg\} \vspace{0.2cm}\\
I_{diag}  (s,t) & =  - \frac{e^{-2 \gamma_{E} \epsilon}}{(4 \pi)^{4-2\epsilon}}\frac{1}{s+t} \left[-\frac{1}{\e^2} \left(\frac{\ln^2 x}{2} +\frac{\pi^2}{2}  \right) 
+ \frac{1}{\e} \bigg(2 \Li{3}{-x} -2\ln x \Li{2}{-x} +
 \right. \non \\ &  
- (\ln^2 x +\pi^2)  \ln(1+x) +\frac{2}{3}\ln^3 x + \ln s \ln^2 x + \pi^2 \ln t -2\zeta(3) \bigg)  - 4 \Li{4}{-x} +  \non \\ &
+ 4  \bigg(\ln(1+x) -  \ln s \bigg) \Li{3}{-x} 
+ 2  \bigg(\ln^2 x +2 \ln s \ln x-2\ln x \ln(1+x) \bigg) \Li{2}{-x}
+\non \\ &
+ 2 \bigg(\frac{2}{3}\ln^3x + \ln s \ln^2x + \pi^2 \ln t
-2\zeta(3)
\bigg) \ln(1+x)  + 4 (S_{2,2}(-x) - \ln x S_{1,2}(-x))
+\non \\ &
-(\ln^2 x +\pi^2) \ln^2(1+x)
-\frac{1}{2} \ln^4 x-\frac{4}{3}\ln s \ln^3 x
-\left(\ln^2 s + \frac{11}{12}\pi^2\right) \ln^2x +\non \\ & 
-\pi^2 \ln^2s -2\pi^2 \ln s \ln x
+ 4 \zeta(3)\ln t -\frac{\pi^4}{20} +  \mathcal{O}(\e) \bigg]  \\
I_{lad} (s,t) & =  - \frac{e^{-2 \gamma_{E} \epsilon}}{(4 \pi)^{4-2\epsilon}}\frac{1}{t s^{2+2\e}} \left[ -\frac{4}{\e^4} +\frac{5\ln x}{\e^3}
- \frac{1}{\e^2}  \left( 2 \ln^2 x -\frac{5}{2} \pi^2  \right) +  \right.
 \non \\ &
- \frac{1}{\e} \bigg( \frac{2}{3}\ln^3 x +\frac{11}{2}\pi^2 \ln x
-\frac{65}{3} \zeta(3) + 4 \Li{3}{ -x } -4\ln x \Li{2}{ -x }
+
\non \\ & -2\left(\ln^2 x +\pi^2 \right) \ln(1+x) \bigg) 
+\frac{4}{3}\ln^4 x +6 \pi^2 \ln^2 x
-\frac{88}{3} \zeta(3)\ln x +\frac{29}{30}\pi^4 + \non \\ 
&  - 4 \left(S_{2,2}(-x) - \ln x S_{1,2}(-x)  \right)
+ 44 \Li{4}{ -x } - 4 \bigg(\ln(1+x) + 6 \ln x  \bigg) \Li{3}{ -x } +
\non \\ & 
+ 2\left(\ln^2 x +2 \ln x \ln(1+x) +\frac{10}{3}\pi^2\right) \Li{2}{-x} +
\non \\ & \left.
+\left(\ln^2 x +\pi^2 \right) \ln^2(1+x)
-\frac{2}{3} \left(4\ln^3 x +5\pi^2 \ln x -6\zeta(3)\right) \ln(1+x) + \mathcal{O}(\e) \right] \\
I_{vlad} (s,t) & =  \frac{\Gamma[1+\e]^2}{(4 \pi)^{4-2\epsilon}}\frac{1}{s^{2+2\e}} \left[ \frac{9}{4 \e^4} -
\frac{ 2 \ln x}{\e^3}
- \frac{7 \pi^2}{3\e^2} + \frac{1}{\e} \left( \frac{4}{3}\ln^3 x+ \frac{14}{3}\pi^2\ln x +
\right. \right. \non \\ &  -4 (\ln^2 x+\pi^2) \ln (1+x) +8 \Li{3}{-x}-8\ln x \Li{2}{-x}-16\zeta(3) \bigg) + 
\non \\& 
+ 20 S_{2,2}(-x) - 20 \ln x S_{1,2}(-x) 
- 28 \Li{4}{ -x } + \non \\ & 
+\bigg(8\ln x +20\ln (1+x)\bigg)\Li{3}{ -x} +\bigg(6\ln^2 x -20 \ln x\ln (1+x)-\frac{4\pi^2}{3}\bigg)\Li{2}{ -x} +
\non \\  &
-\frac{4}{3}\ln^4 x - \frac{13}{3} \pi^2 \ln^2 x
+ \left(\frac{16}{3}\ln^3 x + \frac{26}{3} \pi^2\ln x \right)\ln(1+ x)+
\non \\ &\left.-5(\ln^2 x +\pi^2)\ln^2 (1+x) +\bigg(28\ln x -20 \ln(1+x)\bigg)\zeta(3) -\frac{7 \pi^4}{45} +  \mathcal{O}(\e)\right]
\end{align}
 with $x=t/s$. Corresponding expressions can be written for t-channel integrals. 

\subsection{Two-loop expansions on master basis}\label{2lexpans}

We list here the expansions of the amplitude integrals on the master integral basis. We also report  the integrals which eventually get canceled in the sum but are still present at the level of single diagrams in order to make manifest the the transcendentality order of each contribution. \vspace{0.2cm}
\begin{align}
 \begin{minipage}{50px} \includegraphics[width=1.5cm]{Dtria1.pdf} \end{minipage}  = & \,\, \frac{c}{s^2} \,\,\, \begin{minipage}{50px} \includegraphics[width=1.5cm]{Isunset.pdf} \end{minipage} \label{mexpini} \\[0.2cm]
 \begin{minipage}{50px} \includegraphics[width=1.5cm]{Dtria4.pdf} \end{minipage}  = &\,\, -\frac{c}{s^2} \,\,\, \begin{minipage}{50px} \includegraphics[width=1.5cm]{Isunset.pdf} \end{minipage} +\frac{b}{ s} \,\,\, \begin{minipage}{50px} \includegraphics[width=1.5cm]{Itria.pdf} \end{minipage}  \\[0.2cm]
\begin{minipage}{50px} \includegraphics[width=1.5cm]{Dtria6.pdf} \end{minipage}  = &\,\,\, \frac{2}{a}\bigg[ \,\frac{c} {s^2} \,\,\, \begin{minipage}{50px} \includegraphics[width=1.5cm]{Isunset.pdf} \end{minipage} -\frac{a^2}{ s} \,\,\, \begin{minipage}{50px} \includegraphics[width=1.7cm]{Ispec.pdf} \end{minipage} \bigg] \label{DT6} \\[0.2cm]
 \begin{minipage}{50px} \includegraphics[width=1.5cm]{Ibra.pdf} \end{minipage}  = &\,\,\, \frac{4 a^2}{s^2}  \,\, \,\begin{minipage}{50px} \includegraphics[width=1.7cm]{Ispec.pdf} \end{minipage}  \\[0.2cm]
 \begin{minipage}{50px} \includegraphics[width=1.5cm]{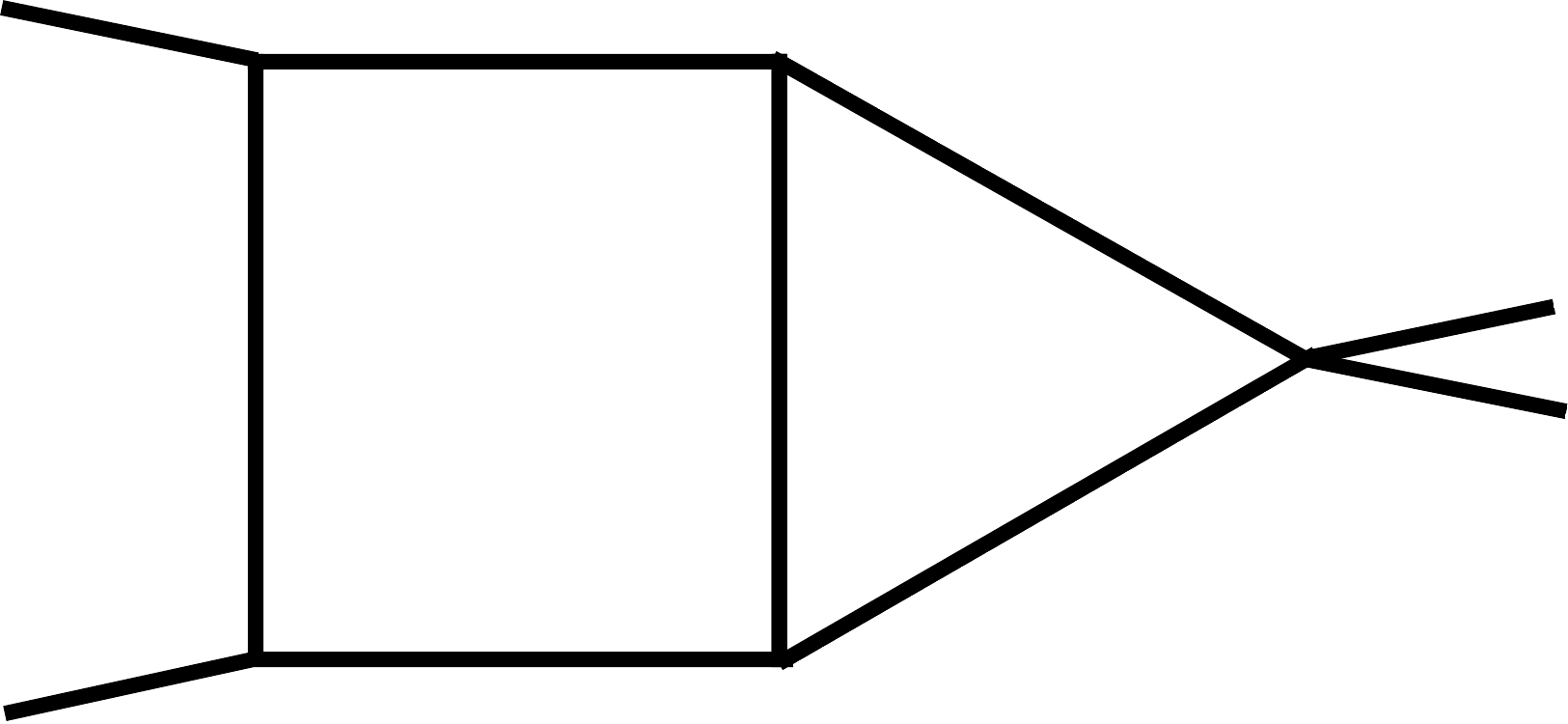} \end{minipage}  =  & -\frac{6c}{s^3 } \,\,\, \begin{minipage}{50px} \includegraphics[width=1.5cm]{Isunset.pdf} \end{minipage} + \, \frac{3b}{ s^2} \,\,\,\begin{minipage}{50px} \includegraphics[width=1.5cm]{Itria.pdf} \end{minipage}  +  \frac{4 a^2}{s^2} \,\,\, \begin{minipage}{50px} \includegraphics[width=1.7cm]{Ispec.pdf} \end{minipage}  \\[0.2cm]
\begin{minipage}{50px} \includegraphics[width=1.5cm]{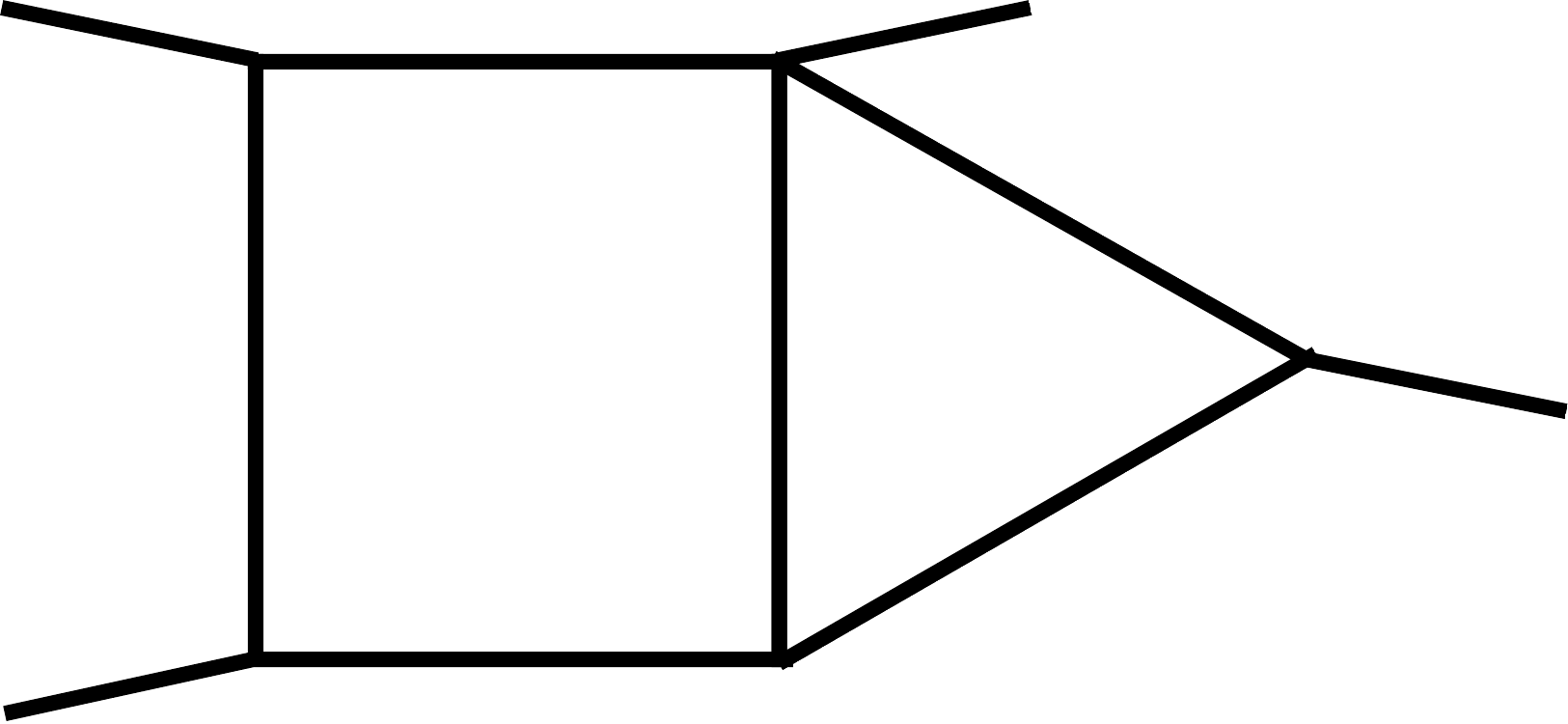} \end{minipage}  =  & -3 c\,\,\,\bigg[ \frac{1}{s^2 t } \,\, \begin{minipage}{50px} \includegraphics[width=1.5cm]{Isunset.pdf} \end{minipage}+ (s \leftrightarrow t)  \,\bigg] 
 - 3 \frac{b}{ st} \,\, \,\begin{minipage}{50px} \includegraphics[width=1.5cm]{Itria.pdf} \end{minipage} + \non \\[0.2cm] & + 3\,\,\frac{s+t}{s t} \,\, \begin{minipage}{50px} \includegraphics[width=1.5cm]{Idiag.pdf} \end{minipage} - 6\,\,\frac{a}{t} \,\,\,   \begin{minipage}{50px} \includegraphics[width=1.5cm]{Imug.pdf} \end{minipage}  \\[0.2cm]
\begin{minipage}{50px} \includegraphics[width=1.5cm]{Horhouse2vec.pdf} \end{minipage} (k &+p_3)^2  = \,\,   - \frac{2c}{s^2 } \,\,\,\, \begin{minipage}{50px} \includegraphics[width=1.5cm]{Isunset.pdf} \end{minipage}-\frac{3c}{s t  } \,\, \,\,\,\begin{minipage}{50px} \includegraphics[angle=90,width=0.5cm]{Isunset.pdf} \end{minipage}   \hspace{-0.9cm} \,\, - \frac{2b}{s} \,\, \,\begin{minipage}{50px} \includegraphics[width=1.5cm]{Itria.pdf} \end{minipage} +\non \\[0.2cm] &+ 3\,\,\frac{s+t}{s} \,\, \begin{minipage}{50px} \includegraphics[width=1.5cm]{Idiag.pdf} \end{minipage}  - 4 a\,\,\, \begin{minipage}{50px} \includegraphics[width=1.5cm]{Imug.pdf} \end{minipage} \label{mexpfin}
\end{align}
where the coefficients $a,b,c$ are defined in (\ref{mastercoeff}). \noindent Corresponding integrals in the t-channel can be obtained by  $s \leftrightarrow t$.

\subsection{Polylogarithm identities}\label{appident}

The two--loop amplitude is a function of standard polylogarithms and Nielsen generalized polylogarithms defined as
\begin{align}
S_{n,p}(z) = \frac{(-1)^{n+p-1}}{(n-1)! p!} \int^{1}_{0} \!\!dt\, \,\frac{(\ln t )^{n-1} \,\,\, (\ln(1-zt))^p}{t} \qquad \,\,\, S_{n,1}(z) = \Li{n+1}{z} 
\end{align}
Following the literature (see e.g. appendix A of \cite{Bern:2005iz}),  the final result can be simplified by using the following set of identities for the polylogarithms with inverse argument 
\begin{align} 
\Li{2}{-1/x} & = -\Li{2}{-x} -\frac{\pi^2}{6}-\frac{1}{2}\,\ln^2 x \non \\
\Li{3}{-1/x} & =  \Li{3}{-x} +\frac{\pi^2}{6}\,\ln x +\frac{1}{6}\,\ln^3 x \non \\
\Li{4}{ -1/x} &= -\Li{4}{-x} - \frac{1}{24}\ln^4 x - 
   \frac{\pi^2}{12} \ln^2 x - \frac{7 \pi^4}{360} \label{invpoly} \\
S_{1,2}(-1/x) & = -S_{1,2}(-x) + \Li{3}{-x} - \ln x \Li{2}{-x} +\zeta(3)-\frac{1}{6}\,\ln^3 x \non \\
S_{2,2}(-1/x) & = S_{2,2}(-x) -2 \Li{4}{-x} + \ln x \Li{3}{-x} -\ln x \zeta(3)+\frac{1}{24}\,\ln^4 x -\frac{7\pi^4}{360} \non
\end{align}
In our case $x=t/s$ is a positive real number, thus the above identities hold for the whole domain of x.

 \end{document}